\documentclass[a4paper]{article}
\usepackage{amsmath,amsfonts,epsfig,color}

\numberwithin{equation}{section}

\newcommand{\hs}[1]{\hspace*{#1cm}}
\definecolor{light}{gray}{.95}

\newcommand{\barr}{\begin{array}}
\newcommand{\earr}{\end{array}}
\newcommand{\nn}{\nonumber}

\newcommand{\et}[1]{e^{\mbox{\footnotesize $#1$}}}
\newcommand{\bld}[1]{\mbox{\boldmath $#1$}}

\newcommand{\dt}{\! \cdot \!}
\newcommand{\wdg}{\! \wedge \!}
\newcommand{\crs}{\! \times \!}
\newcommand{\scp}{\! \ast \!}
\newcommand{\la}{\langle}
\newcommand{\ra}{\rangle}
\newcommand{\lra}{\leftrightarrow}

\newcommand{\half}{{\textstyle \frac{1}{2}}}
\newcommand{\third}{{\textstyle \frac{1}{3}}}
\newcommand{\tthird}{{\textstyle \frac{2}{3}}}

\newcommand{\qrt}{{\textstyle \frac{1}{4}}}

\newcommand{\tld}{\sim}

\newcommand{\etc}{\textit{etc.}}

\newcommand{\btimes}{\mbox{\boldmath $\times$}}
\newcommand{\eqv}{=}

\newcommand{\alp}{\alpha}
\newcommand{\bet}{\beta}
\newcommand{\gam}{\gamma}
\newcommand{\del}{\delta}
\newcommand{\eps}{\epsilon}
\newcommand{\kap}{\kappa}
\newcommand{\lam}{\lambda}
\newcommand{\sig}{\mbox{\boldmath $\sigma$}}
\newcommand{\om}{\omega}
\newcommand{\Gam}{\Gamma}

\newcommand{\Lam}{\Lambda}

\newcommand{\Om}{\Omega}

\newcommand{\ba}{\mbox{\boldmath $a$}}
\newcommand{\bb}{\mbox{\boldmath $b$}}

\newcommand{\bx}{\mbox{\boldmath $x$}}

\newcommand{\bA}{\mbox{\boldmath $A$}}
\newcommand{\bB}{\mbox{\boldmath $B$}}

\newcommand{\bD}{\mbox{\boldmath $D$}}
\newcommand{\bE}{\mbox{\boldmath $E$}}

\newcommand{\bH}{\mbox{\boldmath $H$}}

\newcommand{\bJ}{\mbox{\boldmath $J$}}

\newcommand{\bsig}{\mbox{\boldmath $\sig$}}

\newcommand{\rdot}{\dot{r}}

\newcommand{\tdot}{\dot{t}}
\newcommand{\vdot}{\dot{v}}
\newcommand{\xdot}{\dot{x}}

\newcommand{\thetadot}{\dot{\theta}}

\newcommand{\cla}{{\mathcal{A}}}

\newcommand{\cld}{{\mathcal{D}}}

\newcommand{\clf}{{\mathcal{F}}}
\newcommand{\clg}{{\mathcal{G}}}

\newcommand{\clj}{{\mathcal{J}}}
\newcommand{\cll}{{\mathcal{L}}}

\newcommand{\clr}{{\mathcal{R}}}
\newcommand{\cls}{{\mathcal{S}}}
\newcommand{\clt}{{\mathcal{T}}}

\newcommand{\clw}{{\mathcal{W}}}

\newcommand{\dif}[1]{\partial_{#1}}
\newcommand{\da}{\partial_a}
\newcommand{\db}{\partial_b}
\newcommand{\dc}{\partial_c}
\newcommand{\dd}{\partial_d}

\newcommand{\dr}{\partial_r}

\newcommand{\dx}{\partial_x}
\newcommand{\dX}{\partial_X}

\newcommand{\dift}{\partial_t}

\newcommand{\dmu}{\partial_{\mu}}
\newcommand{\dnu}{\partial_{\nu}}
\newcommand{\dpsi}{\partial_{\psi}}

\newcommand{\si}{\bsig_1}
\newcommand{\sj}{\bsig_2}
\newcommand{\sk}{\bsig_3}
\newcommand{\sr}{\bsig_r}
\newcommand{\isi}{I \hspace{-1pt} \bsig_1}
\newcommand{\isj}{I \hspace{-1pt} \bsig_2}
\newcommand{\isk}{I \hspace{-1pt} \bsig_3}

\newcommand{\sigk}{\bsig_k}

\newcommand{\gi}{\gamma_{1}}
\newcommand{\gj}{\gamma_{2}}
\newcommand{\gk}{\gamma_{3}}
\newcommand{\go}{\gamma_{0}}

\newcommand{\gamum}{\gamma^\mu}
\newcommand{\gamdm}{\gamma_\mu}

\newcommand{\gamdn}{\gamma_\nu}

\newcommand{\hgi}{\hat{\gamma}_{1}}
\newcommand{\hgj}{\hat{\gamma}_{2}}
\newcommand{\hgk}{\hat{\gamma}_{3}}
\newcommand{\hgo}{\hat{\gamma}_{0}}
\newcommand{\hgamum}{\hat{\gamma}^\mu}
\newcommand{\hgamdm}{\hat{\gamma}_\mu}
\newcommand{\hgam}{\hat{\gamma}}

\newcommand{\sigr}{\bsig_r}
\newcommand{\sigth}{\bsig_\theta}
\newcommand{\sigph}{\bsig_\phi}

\newcommand{\Rrev}{\tilde{R}}

\newcommand{\psirev}{\tilde{\psi}}
\newcommand{\psidot}{\dot{\psi}}
\newcommand{\psidag}{\psi^{\dagger}}
\newcommand{\psibar}{\bar{\psi}}
\newcommand{\phirev}{\tilde{\phi}}
\newcommand{\phidot}{\dot{\phi}}
\newcommand{\phidag}{\phi^{\dagger}}

\newcommand{\lif}{\mathsf{f}}
\newcommand{\lbf}{\bar{\mathsf{f}}}
\newcommand{\lih}{\mathsf{h}}
\newcommand{\lbh}{\bar{\mathsf{h}}}
\newcommand{\liR}{\mathsf{R}}

\newcommand{\ho}{\lbh}
\newcommand{\hu}{\lih}
\newcommand{\fo}{\lbf}
\newcommand{\fu}{\lif}

\newcommand{\grad}{\nabla}

\newcommand{\dgrad}{\dot{\grad}}
\newcommand{\bgrad}{\mbox{\boldmath $\grad$}}

\newcommand{\pardot}{\dot{\partial}}
\newcommand{\deriv}[2]{\frac{\partial #1}{\partial #2}}

\newcommand{\csthet}{\cos\!\theta\,}

\newcommand{\dhi}{\det(\hu)^{-1}} 
\renewcommand{\dh}{\det(\hu)}
\newcommand{\dhua}{\partial_{\hu(a)}}
\newcommand{\dhoa}{\partial_{\ho(a)}}
\newcommand{\hint}{\int |d^4x| \dhi \,}

\newcommand{\ric}{{\cal R}}

\newcommand{\cldd}{\dot{\cld}}

\newcommand{\dho}{\dot{\lbh}}
\newcommand{\dhu}{\dot{\lih}}

\newcommand{\phht}{{\hat{\phi}}}
\newcommand{\thht}{{\hat{\theta}}}
\newcommand{\dL}{\dot{L}}
\newcommand{\rhodot}{\dot{\rho}}

\newcommand{\Hdt}{\dot{H}}
\newcommand{\dOm}{\dot{\Om}}
\newcommand{\slI}{\textsl{I}}
\newcommand{\slII}{\textsl{II}}
\newcommand{\slIII}{\textsl{III}}
\newcommand{\slD}{\textsl{D}}
\newcommand{\slN}{\textsl{N}}
\newcommand{\urev}{\tilde{u}}

\newcounter{bean}
\newenvironment{romanlist}%
{\begin{list}%
{(\roman{bean})}{\topsep 0in \usecounter{bean}}}%
{\end{list}}

\begin{document}


\thispagestyle{empty}

\vspace*{0.4cm}

\noindent
\textsf{\textbf{\large GRAVITY, GAUGE THEORIES AND \\ GEOMETRIC ALGEBRA}}

\vspace{0.4 cm}

\noindent
{\large Anthony Lasenby\footnote{E-mail: a.n.lasenby@mrao.cam.ac.uk},
Chris Doran\footnote{E-mail: c.doran@mrao.cam.ac.uk} and Stephen
Gull\footnote{E-mail: steve@mrao.cam.ac.uk}}

\vspace{0.4cm}
\noindent
Astrophysics Group, Cavendish Laboratory, Madingley Road, \\
Cambridge CB3 0HE, UK.

\vspace{0.4cm}

\begin{abstract}
A new gauge theory of gravity is presented.  The theory is constructed
in a flat background spacetime and employs gauge fields to ensure that
all relations between physical quantities are independent of the
positions and orientations of the matter fields.  In this manner all
properties of the background spacetime are removed from physics, and
what remains are a set of `intrinsic' relations between physical
fields.  For a wide range of phenomena, including all present
experimental tests, the theory reproduces the predictions of general
relativity.  Differences do emerge, however, through the first-order
nature of the equations and the global properties of the gauge fields,
and through the relationship with quantum theory.  The properties of
the gravitational gauge fields are derived from both classical and
quantum viewpoints.  Field equations are then derived from an action
principle, and consistency with the minimal coupling procedure selects
an action that is unique up to the possible inclusion of a
cosmological constant.  This in turn singles out a unique form of
spin-torsion interaction.  A new method for solving the field
equations is outlined and applied to the case of a time-dependent,
spherically-symmetric perfect fluid.  A gauge is found which reduces
the physics to a set of essentially Newtonian equations.  These
equations are then applied to the study of cosmology, and to the
formation and properties of black holes.  Insistence on finding global
solutions, together with the first-order nature of the equations,
leads to a new understanding of the role played by time reversal.
This alters the physical picture of the properties of a horizon around
a black hole.  The existence of global solutions enables one to
discuss the properties of field lines inside the horizon due to a
point charge held outside it.  The Dirac equation is studied in a
black hole background and provides a quick derivation of the Hawking
temperature.  Some applications to cosmology are also discussed, and a
study of the Dirac equation in a cosmological background reveals that
the only models consistent with homogeneity are spatially flat.  It is
emphasised throughout that the description of gravity in terms of
gauge fields, rather than spacetime geometry, leads to many simple and
powerful physical insights.  The language of `geometric algebra' best
expresses the physical and mathematical content of the theory and is
employed throughout.  Methods for translating the equations into other
languages (tensor and spinor calculus) are given in appendices.

\end{abstract}

\newpage


\noindent
This paper first appeared in \textit{Phil.\/ Trans.\/ R.\/ Soc.\/ Lond.\/ A}
(1998) \textbf{356}, 487--582.  The present version has been updated
with some corrections and improvements in notation.  Some extra
citations have been added, but these are not intended to be an
exhaustive catalogue of work since the paper was completed in 1996.

\vspace{0.5cm}

\noindent
April 2004

\tableofcontents

\newpage

\vspace*{5cm}
\begin{center}
{\LARGE Part I --- Foundations}
\end{center}

\newpage


\section{Introduction} \label{Intro}

In modern theoretical physics particle interactions are described by
gauge theories.  These theories are constructed by demanding that
symmetries in the laws of physics should be local, rather than global,
in character.  The clearest expositions of this principle are
contained in quantum theory, where one initially constructs a
Lagrangian containing a global symmetry.  In order to promote this to
a local symmetry, the derivatives appearing in the Lagrangian are
modified so that they are unchanged in form by local transformations.
This is achieved by the introduction of fields with certain
transformation properties (`gauge fields'), and these fields are then
responsible for inter-particle forces.  The manner in which the gauge
fields couple to matter is determined by the `minimal coupling'
procedure, in which partial (or directional) derivatives are replaced
by covariant derivatives.  This is the general framework that has been
applied so successfully in the construction of the standard model of
particle physics, which accounts for the strong, weak and
electromagnetic forces.

But what of gravity: can general relativity be formulated as a gauge
theory?  This question has troubled physicists for many
years~\cite{uti56,kib61,iva83}.  The first work that recovered
features of general relativity from a gauging argument was due to
Kibble~\cite{kib61}, who elaborated on an earlier, unsuccessful
attempt by Utiyama~\cite{uti56}.  Kibble used the 10-component
Poincar\'{e} group of passive infinitesimal coordinate transformations
(consisting of four translations and six rotations) as the global
symmetry group.  By gauging this group and constructing a suitable
Lagrangian density for the gauge fields, Kibble arrived at a set of
gravitational field equations --- though not the Einstein equations.
In fact, Kibble arrived at a slightly more general theory, known as a
`spin-torsion' theory.  The necessary modifications to Einstein's
theory to include torsion were first suggested by Cartan~\cite{car22},
who identified torsion as a possible physical field.  The connection
between quantum spin and torsion was made
later~\cite{kib61,wey50,sci64}, once it had become clear that the
stress-energy tensor for a massive fermion field must be
asymmetric~\cite{weys47,bea63}.  Spin-torsion theories are sometimes
referred to as Einstein--Cartan--Kibble--Sciama  theories.
Kibble's use of passive transformations was criticised by Hehl {\em et
al.}~\cite{heh76}, who reproduced Kibble's derivation from the
standpoint of active transformations of the matter fields.  Hehl {\em
et al.} also arrived at a spin-torsion theory, and it is now generally
accepted that torsion is an inevitable feature of a gauge theory based
on the Poincar\'{e} group.

The work of Hehl {\em et al.}~\cite{heh76} raises a further issue.  In
their gauge theory derivation Hehl {\em et al.} are clear that `{\em
coordinates and frames are regarded as fixed once and for all, while
the matter fields are replaced by fields that have been rotated or
translated}'.  It follows that the derivation can only affect the
properties of the matter fields, and not the properties of spacetime
itself.  Yet, once the gauge fields have been introduced, the authors
identify these fields as determining the curvature and torsion of a
Riemann--Cartan spacetime.  This is possible only if it is assumed from
the outset that one is working in a Riemann--Cartan spacetime, and not
in flat Minkowski spacetime.  But the idea that spacetime is curved is
one of the cornerstone principles of general relativity.  That this
feature must be introduced {\em a priori\/}, and is not derivable from
the gauge theory argument, is highly undesirable --- it shows that the
principle of local gauge invariance must be supplemented with further
assumptions before general relativity is recovered.  The conclusions
are clear: classical general relativity must be modified by the
introduction of a spin-torsion interaction if it is to be viewed as a
gauge theory, and the gauge principle alone fails to provide a
conceptual framework for general relativity as a theory of gravity.

In this paper we propose an alternative theory of gravity that is
derived from gauge principles alone.  These gauge fields are functions
of position in a single Minkowski vector space.  But here we
immediately hit a profound difficulty.  Parameterising points with
vectors implies a notion of a Newtonian `absolute space' (or
spacetime) and one of the aims of general relativity was to banish this idea.  
So can we possibly retain the idea of representing points with
vectors without introducing a notion of absolute space?  The answer to
this is yes --- we must construct a theory in which points are
parameterised by vectors, but the physical relations between fields
are independent of where the fields are placed in this vector space.
We must therefore be free to move the fields around the vector space
in an arbitrary manner, without in any way affecting the physical
predictions.  In this way our abstract Minkowski vector space will
play an entirely passive role in physics, and what will remain are a
set of `intrinsic' relations between spacetime fields at the same
point.  Yet, once we have chosen a particular parameterisation of
points with vectors, we will be free to exploit the vector space
structure to the full, secure in the knowledge that any physical
prediction arrived at is ultimately independent of the
parameterisation.

The theory we aim to construct is therefore one that is invariant
under arbitrary field displacements.  It is here that we make contact
with gauge theories, because the necessary modification to the
directional derivatives requires the introduction of a gauge field.
But the field required is not of the type usually obtained when
constructing gauge theories based on Lie-group symmetries.  The gauge
field coupling is of an altogether different, though very natural,
character.  However, this does not alter the fact that the theory
constructed here is a gauge theory in the broader sense of being
invariant under a group of transformations.  The treatment presented
here is very different from that of Kibble~\cite{kib61} and Hehl {\em
et al.}~\cite{heh76}.  These authors only considered infinitesimal
translations, whereas we are able to treat arbitrary finite field
displacements.  This is essential to our aim of constructing a theory
that is independent of the means by which the positions of fields are
parameterised by vectors.

Once we have introduced the required `position-gauge' field, a further
spacetime symmetry remains.  Spacetime fields are not simply scalars,
but also consist of vectors and tensors.  Suppose that two spacetime
vector fields are equated at some position.  If both fields are then
rotated at a point, the same intrinsic physical relation is obtained.
We therefore expect that all physical relations should be invariant
under local rotations of the matter fields, as well as displacements.
This is necessary if we are to achieve complete freedom from the
properties of the underlying vector space --- we cannot think of the
vectors representing physical quantities as having direction defined
relative to some fixed vectors in Minkowski spacetime, but are only
permitted to consider relations between matter fields.  Achieving
invariance under local rotations introduces a further gauge field,
though now we are in the familiar territory of Yang--Mills type
interactions (albeit employing a non-compact Lie group).

There are many ways in which the gauge theory presented here offers
both real and potential advantages over traditional general
relativity.  As our theory is a genuine gauge theory, the status of
physical predictions is always unambiguous --- any physical prediction
must be extracted from the theory in a gauge-invariant manner.
Furthermore, our approach is much closer to the conventional theories
of particle physics, which should ease the path to a quantum theory.
One further point is that discarding all notions of a curved spacetime
makes the theory conceptually much simpler than general relativity.
For example, there is no need to deal with topics such as
differentiable manifolds, tangent spaces or fibre
bundles~\cite{egu80}.

The theory developed here is presented in the language of `{\em
geometric algebra}'~\cite{hes-gc,DGL93-notreal}.  Any physical theory
can be formulated in a number of different mathematical languages, but
physicists usually settle on a language which they feel represents the
`optimal' choice.  For quantum field theory this has become the
language of abstract operator commutation relations, and for general relativity it is
Riemannian geometry.  For our gauge theory of gravity there seems
little doubt that geometric algebra is the optimal language available
in which to formulate the theory.  Indeed, it was partly the desire to
apply this language to gravitation theory that led to the development
of the present theory.  (This should not be taken to imply that
geometric algebra cannot be applied to standard general relativity --- it certainly
can~\cite{hes-gc,hes86,hes-sta,sob81}.  It has also been used to
elaborate on Utiyama's approach~\cite{hes-sta}.)  To us, the use of
geometric algebra is as central to the theory of gravity presented
here as tensor calculus and Riemannian geometry were to Einstein's
development of general relativity.  It is the language that most clearly exposes the
structure of the theory.  The equations take their simplest form when
expressed in geometric algebra, and all reference to coordinates and
frames is removed, achieving a clean separation between physical
effects and coordinate artefacts.  Furthermore, the geometric algebra
development of the theory is entirely self-contained.  All problems
can be treated without ever having to introduce concepts from other
languages, such as differential forms or the Newman--Penrose formalism.

We realise, however, that the use of an unfamiliar language may deter
some readers from exploring the main physical content of our theory
--- which is of course independent of the language chosen to express
it.  We have therefore endeavoured to keep the mathematical content of
the main text to a minimum level, and have included appendices
describing methods for translating our equations into the more
familiar languages of tensor and spinor calculus.  In addition, many
of the final equations required for applications are simple scalar
equations.  The role of geometric algebra is simply to provide the
most efficient and transparent derivation of these equations.  It is
our hope that physicists will find geometric algebra a simpler and
more natural language than that of differential geometry and tensor
calculus.

This paper starts with an introduction to geometric algebra and its
spacetime version -- the spacetime algebra.  We then turn to the
gauging arguments outlined above and find mathematical expressions of
the underlying principles.  This leads to the introduction of two
gauge fields.  At this point the discussion is made concrete by
turning to the Dirac action integral.  The Dirac action is formulated
in such a way that internal phase rotations and spacetime rotations
take equivalent forms.  Gauge fields are then minimally coupled to the
Dirac field to enforce invariance under local displacements and both
spacetime and phase rotations.  We then turn to the construction of a
Lagrangian density for the gravitational gauge fields.  This leads to
a surprising conclusion.  The demand that the gravitational action be
consistent with the derivation of the minimally-coupled Dirac equation
restricts us to a single action integral.  The only freedom that
remains is the possible inclusion of a cosmological constant, which
cannot be ruled out on theoretical grounds alone.  The result of this
work is a set of field equations that are completely independent of
how we choose to label the positions of fields with a vector $x$.  The
resulting theory is conceptually simple and easier to calculate with
than the metric-based theory of general relativity.  We call this
theory `\textit{gauge theory gravity}' (GTG).  Having derived the
field equations, we turn to a discussion of measurements, the
equivalence principle and the Newtonian limit in GTG.  We end Part~I
with a discussion of symmetries, invariants and conservation laws.

In Part~II we turn to applications, concentrating mainly on
time-dependent spherically-symmetric systems.  We start by studying
perfect fluids and derive a simple set of first-order equations that
describe a wide range of physical phenomena.  The method of derivation
of these equations is new and offers many advantages over conventional
techniques.  The equations are then studied in the contexts of black
holes, collapsing matter and cosmology.  We show how a gauge can be
chosen that affords a clear, global picture of the properties of these
systems.  Indeed, in many cases one can apply simple, almost
Newtonian, reasoning to understand the physics.  For some of these
applications the predictions of GTG and general relativity are
identical, and these cases include all present experimental tests of
general relativity.  However, on matters such as the role of horizons
and topology, the two theories differ.  For example, we show that the
black-hole solutions admitted in GTG fall into two distinct
time-asymmetric gauge sectors, and that one of these is picked out
uniquely by the formation process.  This is quite different to general
relativity, which admits eternal time-reverse symmetric solutions.  In
discussing differences between GTG and general relativity, it is not
always clear what the correct general relativistic viewpoint is.  We
should therefore be explicit in stating that what we intend when we
talk about general relativity is the full, modern formulation of the
subject as expounded by, for example, Hawking~\&
Ellis~\cite{haw-large} and D'Inverno~\cite{inv-rel}.  This includes
ideas such as worm-holes, exotic topologies and distinct `universes'
connected by black holes~\cite{kauf-front,haw-black}.

After studying some solutions for the gravitational fields we turn to
the properties of electromagnetic and Dirac fields in gravitational
backgrounds.  For example, we give field configurations for a charge
held at rest outside a black hole.  We show how these field lines
extend smoothly across the horizon, and that the origin behaves as a
polarisation charge.  This solution demonstrates how the global
properties of the gravitational fields are relevant to physics outside
the horizon, a fact that is supported by an analysis of the Dirac
equation in a black-hole background.  This analysis also provides a
quick, though physically questionable, derivation of a particle
production rate described by a Fermi--Dirac distribution with the
correct Hawking temperature.  We end with a discussion of the
implications of gauge-theory gravity for cosmology.  A study of the
Maxwell and Dirac equations in a cosmological background reveals a
number of surprising features.  In particular, it is shown that a
non-spatially-flat universe does not appear homogeneous to Dirac
fields --- fermionic matter would be able to detect the `centre' of
the universe if $k\neq 0$.  Thus the only homogeneous cosmological
models consistent with GTG are those that are spatially flat.  (This
does not rule out spatially-flat universes with a non-zero
cosmological constant.)  A concluding section summarises the
philosophy behind our approach, and outlines some future areas of
research.

\section{An outline of geometric algebra}
\label{Outline}

There are many reasons for preferring geometric algebra to other
languages employed in mathematical physics.  It is the most powerful
and efficient language for handling rotations and boosts; it
generalises the role of complex numbers in two dimensions, and
quaternions in three dimensions, to a scheme that efficiently handles
rotations in arbitrary dimensions.  It also exploits the advantages of
labelling points with vectors more fully than either tensor calculus
or differential forms, both of which were designed with a view to
applications in the intrinsic geometry of curved spaces.  In addition,
geometric algebra affords an entirely {\em real\/} formulation of the
Dirac equation~\cite{hes75,DGL93-states}, eliminating the need for
complex numbers.  The advantage of the real formulation is that
internal phase rotations and spacetime rotations are handled in an
identical manner in a single unifying framework.  A wide class of
physical theories have now been successfully formulated in terms of
geometric algebra.  These include classical
mechanics~\cite{hes-nf1,hes74a,vol93}, relativistic
dynamics~\cite{hes74}, Dirac
theory~\cite{hes75,DGL93-states,DGL93-paths,DGL95-elphys},
electromagnetism and
electrodynamics~\cite{DGL93-notreal,DGL93-paths,vol93a}, as well as a
number of other areas of modern mathematical
physics~\cite{DGL93-gras,dor93-spin,DGL93-lft,DGL-polmvl,DGL-polspin}.
In every case, geometric algebra has offered demonstrable advantages
over other techniques and has provided novel insights and unifications
between disparate branches of physics and mathematics.

This section is intended to give only a brief introduction to the
ideas and applications of geometric algebra.  A fuller introduction,
including a number of results relevant to this paper, is set out in
the series of
papers~\cite{DGL93-notreal,DGL93-states,DGL93-paths,DGL93-lft} written
by the present authors.  Elsewhere, the books by Doran~\&
Lasenby~\cite{gap}, Hestenes~\cite{hes-sta,hes-nf1} and Hestenes~\&
Sobczyk~\cite{hes-gc} cover the subject in detail.  A number of other
helpful introductory articles can be found, including those by
Hestenes~\cite{hes-unified,hes91}, Vold~\cite{vol93,vol93a}, and Doran
\& Lasenby~\cite{DL-course}.  The conference
proceedings~\cite{cliffconf1,cliffconf2,cliffconf3} also contain some
interesting and useful papers.

Geometric algebra arose from Clifford's attempts to generalise
Hamilton's quaternion algebra into a language for vectors in arbitrary
dimensions~\cite{cli1878}.  Clifford discovered that both complex
numbers and quaternions are special cases of an algebraic framework in
which vectors are equipped with a single associative product that is
distributive over addition\footnote{The same generalisation was also
found by Grassmann~\cite{gra1877}, independently and somewhat before
Clifford's work.  This is one of many reasons for preferring
Clifford's name (`{\em geometric algebra}') over the more usual `{\em
Clifford algebra}'.}.  With vectors represented by lower-case Roman
letters ($a$, $b$), Clifford's `geometric product' is written simply
as $ab$.  A key feature of the geometric product is that the square of
any vector is a scalar.  Now, rearranging the expansion
\begin{equation}
(a + b)^2 = (a+b)(a+b) = a^2 + (ab + ba) + b^2
\end{equation}
to give
\begin{equation}
ab + ba = (a + b)^2 - a^2 - b^2,
\label{1eq1}
\end{equation}
where the right-hand side of~\eqref{1eq1} is a sum of squares and by
assumption a scalar, we see that the symmetric part of the geometric
product of two vectors is also a scalar.  We write this `inner' or
`dot' product between vectors as
\begin{equation}
a \dt b \eqv \half (ab + ba).
\label{1dot}
\end{equation}
The remaining antisymmetric part of the the geometric product
represents the directed area swept out by displacing $a$ along $b$.
This is the `outer' or `exterior' product introduced by
Grassmann~\cite{gra1877} and familiar to all who have studied the
language of differential forms.  The outer product of two vectors is
called a {\em bivector\/} and is written with a wedge:
\begin{equation}
a \wdg b \eqv \half (ab - ba).
\label{1wdg}
\end{equation}

On combining~\eqref{1dot} and~\eqref{1wdg} we find that the geometric product
has been decomposed into the sum of a scalar and a bivector part,
\begin{equation}
ab = a \dt b + a \wdg b.
\label{1gprod}
\end{equation}
The innovative feature of Clifford's product~\eqref{1gprod} lies in its
mixing of two different types of object: scalars and bivectors.  This
is not problematic, because the addition implied by~\eqref{1gprod} is
precisely that which is used when a real number is added to an
imaginary number to form a complex number.  But why might we want to
add these two geometrically distinct objects?  The answer emerges from
considering reflections and rotations.  Suppose that the vector $a$ is
reflected in the (hyper)plane perpendicular to the unit vector $n$.
The result is the new vector
\begin{equation}
a - 2 (a \dt n) n = a -(an + na)n = - nan.
\label{1reflec1} 
\end{equation}
The utility of the geometric algebra form of the resultant vector,
$-nan$, becomes clear when a second reflection is performed.  If this
second reflection is in the hyperplane perpendicular to the unit
vector $m$, then the combined effect is
\begin{equation}
a \mapsto mnanm.
\end{equation}
But the combined effect of two reflections is a rotation so, defining
the geometric product $mn$ as the scalar-plus-bivector quantity $R$,
we see that rotations are represented by
\begin{equation}
a \mapsto Ra\Rrev.
\label{1rotn}
\end{equation}
Here the quantity $\Rrev=nm$ is called the `{\em reverse}' of $R$ and
is obtained by reversing the order of all geometric products between
vectors:
\begin{equation}
(ab \dots c)^\tld \eqv c \dots ba.
\end{equation}
The object $R$ is called a {\em rotor\/}.  Rotors can be written as an
even (geometric) product of unit vectors, and satisfy the relation
$R\Rrev=1$.  The representation of rotations in the form~\eqref{1rotn} has
many advantages over tensor techniques.  By defining $\csthet\eqv
m\dt n$ we can write
\begin{equation}
R = mn = \exp \Bigl( \frac{m \wdg n}{|m \wdg n|} \theta/2 \Bigr),
\label{O4}
\end{equation}
which relates the rotor $R$ directly to the plane in which the
rotation takes place.  Equation~\eqref{O4} generalises to arbitrary
dimensions the representation of planar rotations afforded by complex
numbers.  This generalisation provides a good example of how the full
geometric product, and the implied sum of objects of different types,
can enter geometry at a very basic level.  The fact that
equation~\eqref{O4} encapsulates a simple geometric relation should also
dispel the notion that Clifford algebras are somehow intrinsically
`quantum' in origin.  The derivation of~\eqref{1rotn} has assumed nothing
about the signature of the space being employed, so that the formula
applies equally to boosts as well as rotations.  The two-sided formula
for a rotation~\eqref{1rotn} will also turn out to be compatible with the
manner in which observables are constructed from Dirac spinors, and
this is important for the gauge theory of rotations of the Dirac
equation which follows.

Forming further geometric products of vectors produces the entire
geometric algebra.  General elements are called `multivectors' and
these decompose into sums of elements of different grades (scalars are
grade zero, vectors grade one, bivectors grade two {\em etc.}).
Multivectors in which all elements have the same grade are termed {\em
homogeneous} and are usually written as $A_r$ to show that $A$
contains only grade-$r$ components.  Multivectors inherit an
associative product, and the geometric product of a grade-$r$
multivector $A_r$ with a grade-$s$ multivector $B_s$ decomposes into
\begin{equation} 
A_{r}B_{s} = \la AB \ra_{r+s} +  \la AB \ra_{r+s-2}
\ldots  + \la AB \ra_{|r-s|} ,
\label{1mvlprod}
\end{equation}
where the symbol $\la M\ra_r$ denotes the projection onto the
grade-$r$ component of $M$.  The projection onto the grade-0 (scalar)
component of $M$ is written $\la M\ra$.  The `$ \cdot $' and `$
\wedge$' symbols are retained for the lowest-grade and highest-grade
terms of the series~\eqref{1mvlprod}, so that
\begin{align}
A_r \dt B_s &\eqv  \la AB \ra_{|r-s|} \label{1intprod} \\
A_r \wdg B_s &\eqv  \la AB \ra_{r+s} , 
\end{align}
which are called the interior and exterior products respectively.  
The exterior product is associative, and satisfies the
symmetry property
\begin{equation}
A_r \wdg B_s = (-1)^{rs} B_s \wdg A_r.
\end{equation}

Two further products can also be defined from the geometric
product.  These are the scalar product 
\begin{equation}
A \scp B \eqv \la AB \ra
\label{1scl}
\end{equation}
and the commutator product
\begin{equation}
A \crs B \eqv \half (AB - BA).
\label{1comm}
\end{equation}
The scalar product~\eqref{1scl} is commutative and satisfies the
cyclic reordering property 
\begin{equation}
\la A \ldots B C \ra = \la C A \ldots B\ra.
\end{equation}
The scalar product~\eqref{1scl} and the interior product~\eqref{1intprod}
coincide when acting on two homogeneous multivectors of the same
grade.  The associativity of the geometric product ensures that the
commutator product \eqref{1comm} satisfies the Jacobi identity
\begin{equation}
A \crs (B \crs C) + B \crs (C \crs A) + C \crs (A \crs B) =0.
\end{equation}

Finally we introduce some further conventions.  Throughout we employ
the operator ordering convention that, {\em in the absence of brackets,
inner, outer, commutator and scalar products take precedence over
geometric products}.  Thus $a \dt b\, c$ means $(a \dt b)c$, not $a
\dt( b c)$.  This convention helps to eliminate unwieldy numbers of
brackets.  Summation convention is employed throughout, except for
indices that denote the grade of a multivector, which are not summed
over.  Natural units ($\hbar=c=4\pi\epsilon_0=G=1$) are used except where
explicitly stated.  Throughout we refer to a Lorentz transformation
(\textit{i.e.} a spatial rotation and/or boost) simply as a `rotation'.

\subsection{The spacetime algebra}
\label{O-STA} 

Of central importance to this paper is the geometric algebra of
spacetime, the {\em spacetime algebra}~\cite{hes-sta}.  To describe
the spacetime algebra it is helpful to introduce a set of four
orthonormal basis vectors $\{\gamdm\}$, $\mu = 0 \ldots 3$, satisfying
\begin{equation}
\gamdm \dt \gamdn = \eta_{\mu \nu} 
= \mbox{diag($+$\ $-$\ $-$\ $-$)}.
\end{equation}
The vectors $\{\gamdm\}$ satisfy the same algebraic relations as
Dirac's $\gamma$-matrices, but they now form a set of four independent
basis vectors for spacetime, not four components of a single vector in
an internal `spin-space'.  The relation between Dirac's matrix algebra
and the spacetime algebra is described in more detail in
Appendix~\ref{app-dirac}, which gives a direct translation of the
Dirac equation into its spacetime algebra form.

A frame of timelike bivectors $\{\sig_k\}$, $k = 1 \ldots 3$ is
defined by
\begin{equation}
\sig_k \eqv  \gam_k\go,
\label{1defsig}
\end{equation}
and forms an orthonormal frame of vectors in the space relative to the
$\go$ direction.  The algebraic properties of the $\{\sig_k\}$ are the
same as those of the Pauli spin matrices, but they now represent an
orthonormal frame of vectors in space, and not three components of a
vector in spin-space.  The highest-grade element (or `pseudoscalar')
is denoted by $I$ and is defined as:
\begin{equation}
I \eqv \go \gi \gj \gk = \si\sj\sk .
\end{equation}
The pseudoscalar satisfies $I^2=-1$.  Since we are in a space of even
dimension, $I$ {\em anti\/}commutes with odd-grade elements, and
commutes only with even-grade elements.  With these definitions, a
basis for the 16-dimensional spacetime algebra is provided by
\begin{equation}
\begin{array}{ccccc}
1 &  \{ \gamdm \} & \{ \sigk, \; I\sigk \} & \{ I \gamdm \} & I \\ 
\text{1 scalar} & \text{4 vectors} & \text{6 bivectors} & 
\text{4 trivectors} & \text{1 pseudoscalar}  .
\end{array} 
\end{equation}

Geometric significance is attached to the above relations as follows.
An observer's rest frame is characterised by a future-pointing
timelike (unit) vector. If this is chosen to be the $\go$ direction
then the $\go$-vector determines a map between spacetime vectors
$a=a^\mu\gamdm$ and the even subalgebra of the spacetime algebra via
\begin{align}
a \go &= a_0 + \ba, \label{1sptsplt} \\
\intertext{where}
a_0 &= a \dt \go \\
\ba &= a \wdg \go.
\end{align}
The `relative vector' $\ba$ can be decomposed in the $\{\sig_k\}$
frame and represents a spatial vector as seen by an observer in the
$\go$-frame.  Since a vector appears to an observer as a line segment
existing for a period of time, it is natural that what an observer
perceives as a vector should be represented by a spacetime bivector.
Equation~\eqref{1sptsplt} embodies this idea, and shows that the algebraic
properties of vectors in relative space are determined entirely by the
properties of the fully relativistic spacetime algebra.

The split of the six spacetime bivectors into relative vectors and
relative bivectors is a frame-dependent operation --- different
observers see different relative spaces.  This fact is clearly
illustrated with the Faraday bivector $F$.  The `space-time
split'~\cite{hes74a} of $F$ into the $\go$-system is made by
separating $F$ into parts that anticommute and commute with $\go$.
Thus
\begin{align}
F &= \bE + I \bB, \label{1Fsplit} \\
\intertext{where}
\bE &= \half(F - \go F \go) \\
I\bB  &= \half(F + \go F \go).
\end{align}
Both $\bE$ and $\bB$ are spatial vectors in the $\go$-frame, and
$I\bB$ is a spatial bivector.  Equation~\eqref{1Fsplit} decomposes $F$
into separate electric and magnetic fields, and the explicit
appearance of $\go$ in the formulae for $\bE$ and $\bB$ shows how this
split is observer-dependent.

The identification of the algebra of three-dimensional space with the
even subalgebra of the spacetime algebra necessitates a convention
that articulates smoothly between the two algebras.  For this reason
relative (or spatial) vectors in the $\go$-system are written in bold
type to record the fact that they are in fact spacetime bivectors.
This distinguishes them from spacetime vectors, which are left in
normal type.   Further conventions are introduced where necessary.

\subsection{Geometric calculus}
\label{O-GC}

Many of the derivations in this paper employ the vector and
multivector derivatives~\cite{hes-gc,DGL93-lft}.  Before defining
these, however, we need some simple results for vector frames.
Suppose that the set $\{e_k \}$ form a vector frame.  The reciprocal
frame is determined by~\cite{hes-gc}
\begin{equation}
e^j = (-1)^{j-1} e_1 \wdg e_2 \wdg \cdots \wdg \check{e}_j \wdg \cdots
\wdg e_n \, E^{-1}  
\end{equation}
where
\begin{equation}
E \eqv e_1 \wdg e_2 \wdg \cdots \wdg e_n
\end{equation}
and the check on $\check{e}_j$ denotes that this term is missing from
the expression.  The $\{e_k \}$ and $\{e^k\}$ frames are related by
\begin{equation}
e_j \dt e^k=\del_j^k .
\end{equation}
An arbitrary multivector $B$ can be decomposed in terms of the $\{e^k\}$
frame into
\begin{align}
B &= \sum_{i<\cdots <j} B_{i \cdots j} \, e^i \wdg \cdots \wdg e^j \\
\intertext{where}
B_{i\cdots j} &= B \dt ( e_j \wdg \cdots \wdg e_i).
\end{align}

Suppose now that the multivector $F$ is an arbitrary function of some
multivector argument $X$, $F=F(X)$.  The derivative of $F$ with
respect to $X$ in the $A$ direction is defined by
\begin{equation}
A \scp \dX F(X) \eqv \lim_{\tau \mapsto 0} \frac{F(X+ \tau A)
-F(X)}{\tau}.
\label{O-GC1}
\end{equation}
From this the multivector derivative $\dX$ is defined by
\begin{equation}
\dX \eqv \sum_{i < \cdots < j} e^i \wdg \cdots \wdg e^j  (e_j \wdg
\cdots \wdg e_i) \scp \dX .
\label{O-GC2}
\end{equation}
This definition shows how the multivector derivative $\dX$ inherits
the multivector properties of its argument $X$, as well as a calculus
from equation~\eqref{O-GC1}.

Most of the properties of the multivector derivative follow from the
result that
\begin{equation} 
\dX \la X A \ra = P_{X}(A) ,
\end{equation}
where $ P_{X}(A)$ is the projection of $A$ onto the grades contained
in $X$.  Leibniz' rule is then used to build up results for more
complicated functions (see Appendix~\ref{app-results}).  The
multivector derivative acts on the next object to its right unless
brackets are present; for example in the expression $\dX AB$ the $\dX$
acts only on $A$, but in the expression $\dX (AB)$ the $\dX$ acts on
both $A$ and $B$.  If the $\dX$ is intended to act only on $B$ then
this is written as $\dot{\partial}_X A\dot{B}$, the overdot
denoting the multivector on which the derivative acts.  As an
illustration, Leibniz' rule can be written in the form
\begin{equation}
\dX (AB) =  \dot{\partial}_X \dot{A} B + \dot{\partial}_X A \dot{B}.
\end{equation}
The overdot notation neatly encodes the fact that, since $\dX$ is a
multivector, it does not necessarily commute with other multivectors
and often acts on functions to which it is not adjacent.

We frequently make use of the derivative with respect to a general
vector to perform a range of linear algebra operations.  For such
operations the following results are useful:
\begin{align}
\da a \dt A_r &= r A_r \\
\da a \wdg A_r &= (n-r) A_r \\
\da A_r a &= (-1)^r (n-2r) A_r, \label{O-gc5}
\end{align} 
where $n$ is the dimension of the space ($n=4$ for all the
applications considered here).  Further results are given in
appendix~\ref{app-results}.

The derivative with respect to spacetime position $x$ is called the
{\em vector derivative}, and is of particular importance.  Suppose
that we introduce an arbitrary set of spacetime coordinates $x^\mu$.
These define a coordinate frame
\begin{equation}
e_\mu = \deriv{}{x^\mu} x.
\end{equation}
The reciprocal frame is denoted $e^\mu$, and from this we define the
vector derivative $\grad$ by 
\begin{equation}
\grad = e^\mu  \deriv{}{x^\mu} = e^\mu \dmu.
\end{equation}
(The importance of the vector derivative means that it is sensible to
give it a unique symbol.)  The vector derivative inherits the
algebraic properties of a vector in the spacetime algebra.  The
usefulness of the geometric product for the vector derivative is
illustrated by electromagnetism.  In tensor notation, Maxwell's
equations are
\begin{equation}
\dmu F^{\mu\nu} = J^{\nu}, \qquad \partial_{[\alp} F_{\mu\nu]} = 0.
\end{equation}
These have the spacetime algebra equivalents~\cite{hes-sta} 
\begin{equation}
\grad \dt F = J, \qquad \grad \wdg F = 0 .
\end{equation}
But we can utilise the geometric product to combine these into the
single equation
\begin{equation}
\grad F = J.
\label{Imax}
\end{equation}
The great advantage of the $\grad$ operator is that it possesses an
inverse, so a first-order propagator theory can be developed for
it~\cite{hes-gc,DGL93-paths}.  This is not possible for the separate 
$\grad \cdot$ and $\grad \wedge$ operators.

\subsection{Linear algebra}
\label{O-Linalg}

Geometric algebra offers many advantages over tensor calculus in
developing the theory of linear
functions~\cite{hes-gc,gap,hes91}.  A linear function mapping
vectors to vectors is written in a sans-serif font, $\fu(a)$, where
$a$ is an arbitrary arguments and plays the role of a placeholder.
The argument of a linear function is usually assumed to be a constant
vector, unless stated otherwise.  The adjoint function is written with
an overbar, $\fo(a)$, and is defined such that
\begin{equation}
a \dt \fu(b) = \fo(a) \dt b.
\end{equation}
It follows that
\begin{equation}
\fo(a) = \db \la \fu(b) a \ra = e_\mu \la \fu(b) e^\mu \ra .
\label{O-L2}
\end{equation}
We will frequently employ the derivative with respect to the vectors
$a$, $b$ {\em etc.} to perform algebraic manipulations of linear
functions, as in equation~\eqref{O-L2}.  The advantage is that all
manipulations are then frame-free.  Of course, the $\da$ and $a$
vectors can easily be replaced by the sum over a set of frame vectors
and their reciprocals, if desired.

A symmetric function is one for which $\fu(a)=\fo(a)$.  For such
functions 
\begin{equation}
\da \wdg \fu(a) = \da \wdg \db \la a \fu(b) \ra = \fu(b) \wdg \db.
\end{equation}
It follows that for symmetric functions
\begin{equation}
\da \wdg \fu(a) = 0,
\end{equation}
which is equivalent to the statement that $\fu(a)=\fo(a)$.

Linear functions extend to act on multivectors via
\begin{equation}  
\fu (a \wdg b \wdg \ldots \wdg c) \eqv \fu(a) \wdg \fu(b) \ldots \wdg
\fu(c),
\label{1defouter}  
\end{equation} 
so that $\fu$ is now a grade-preserving linear function mapping
multivectors to multivectors.  In particular, since the pseudoscalar
$I$ is unique up to a scale factor, we can define
\begin{equation}
\det(\fu) = \fu(I) I^{-1}.
\end{equation}

Viewed as linear functions over the entire geometric algebra, $\fu$
and $\fo$ are related by the fundamental formulae
\begin{equation}
\begin{aligned}
A_{r} \dt {\lbf}(B_{s}) &= {\lbf}( {\fu}(A_{r}) \dt B_{s}) 
\qquad r \leq s  \\
{\fu}(A_{r}) \dt B_{s} &= {\fu}( A_{r} \dt {\lbf}(B_{s}))
\qquad r \geq s ,
\end{aligned} 
\label{1fnadj}
\end{equation}
which are easily derived~ \cite[Chapter 3]{hes-gc}.  The formulae for the
inverse functions are found as special cases of~\eqref{1fnadj},
\begin{equation}
\begin{aligned}
{\fu}^{-1}(A) &= \det(\fu)^{-1} \, {\lbf}(A I) I^{-1}  \\
{\lbf}^{-1}(A) &= \det(\fu)^{-1} \, I^{-1} {\fu}(I A) .
\label{1fninv}
\end{aligned}
\end{equation}
A number of further results for linear functions are contained in
Appendix~\ref{app-results}.  These include a coordinate-free
formulation of the derivative with respect to a linear function, which
proves to be very useful in deriving stress-energy tensors from action
integrals.


\section{Gauge principles for gravitation}
\label{Gauge}

In this section we identify the dynamical variables that will describe
gravitational interactions.  We start by reviewing the arguments
outlined in the introduction.  The basic idea is that all physical
relations should have the generic form $a(x)=b(x)$, where $a$ and $b$
are spacetime fields representing physical quantities, and $x$ is the
spacetime position vector.  An equality such as this can certainly
correspond to a clear physical statement.  But, considered as a
relation between fields, the physical relationship expressed by this
statement is completely independent of where we choose to think of $x$
as lying in spacetime.  In particular, we can associate each position
$x$ with some new position $x'=f(x)$ and rewrite the relation as
$a(x')=b(x')$, and the equation still has precisely the same content.
(A proviso, which will gain significance later, is that the map $f(x)$
should be non-singular and cover all of spacetime.)

A similar argument applies to rotations.  The \textit{intrinsic}
content of a relation such as $a(x)=b(x)$ at a given point $x_0$ is
unchanged if we rotate each of $a$ and $b$ by the same amount.  That
is, the equation $Ra(x_0)\Rrev=Rb(x_0)\Rrev$ has the same physical
content as the equation $a(x_0)=b(x_0)$.  For example scalar product
relations, from which we can derive angles, are unaffected by this
change.  These arguments apply to any physical relation between any
type of multivector field.  The principles underlying gauge theory
gravity can therefore be summarised as follows:

\begin{romanlist}
\setcounter{bean}{0}
\item The physical content of a field equation must be invariant under
arbitrary local displacements of the fields. (This is called
position-gauge invariance.)
\item The physical content of a field equation must be invariant under
arbitrary local rotations of the fields.  (This is called
rotation-gauge invariance.)
\end{romanlist}

In this theory predictions for all measurable quantities, including
distances and angles, must be derived from gauge-invariant relations
between the field quantities themselves, not from the properties of
the background spacetime.  On the other hand, quantities that depend
on a choice of `gauge' are not predicted absolutely and cannot be
defined operationally.

It is necessary to indicate how this approach differs from the one
adopted in gauge theories of the Poincar\'{e} group.  (This is a point
on which we have been confused in the past~\cite{DGL-grav-bel}.)
Poincar\'{e} transformations for a multivector field $M(x)$ are
defined by
\begin{equation}
M(x) \mapsto M' = R M(x') \Rrev
\label{grav1-poi1}
\end{equation}
where
\begin{equation}
x' = \Rrev x R + t,
\label{grav1-poi2}
\end{equation} 
$R$ is a constant rotor and $t$ is a constant vector.  Transformations
of this type mix displacements and rotations, and any attempt at a
local gauging of this spacetime symmetry fails to decouple the
two~\cite{kib61,heh76}.  Furthermore, the fact that the rotations
described by Poincar\'{e} transformations include the
displacement~\eqref{grav1-poi2} (with $t=0$) means that the rotations
discussed under point~(ii) above are not contained in the Poincar\'{e}
group.

As a final introductory point, while the mapping of fields onto
spacetime positions is arbitrary, the fields themselves must be
well-defined.  The fields cannot be singular except at a few special
points.  Furthermore, any remapping of the fields in the must be
one-to-one, else we would cut out some region of physical
significance.  In later sections we will see that general relativity
allows operations in which regions of spacetime are removed.  These
are achieved through the use of singular coordinate transformations
and are the origin of some potential differences between GTG and general
relativity.

\subsection{The position-gauge field}

We now examine the consequences of the local symmetries we have just
discussed.  As in all gauge theories we must study the effects on
\textit{derivatives}, since all non-derivative relations already
satisfy the correct requirements.

We start by considering a scalar field $\phi(x)$ and form its vector
derivative $\grad \phi(x)$.  Suppose now that from $\phi(x)$ we define
the new field $\phi'(x)$ by
\begin{align}
\phi'(x) &\eqv \phi(x'), \\
\intertext{where}
x' &= f(x)
\end{align}
and $f(x)$ is an arbitrary (differentiable) map between spacetime
position vectors.  The map $f(x)$ should not be thought of as a map
between manifolds, or as moving points around; rather, the function
$f(x)$ is merely a rule for relating one position vector to another
within a single vector space.  Note that the new function $\phi'(x)$
is given by the old function $\phi$ evaluated at $x'$.  We could have
defined things the other way round, so that $\phi'(x')$ is given by
$\phi(x)$, but the form adopted here turns out to be easier to
implement in practice.

If we now act on the new scalar field $\phi'$ with $\grad$ we form the
quantity $\grad\phi(x')$.  To evaluate this we first look at the
directional derivatives of $\phi'$,
\begin{align}
a \dt \grad \phi(x') &= \lim_{\eps \rightarrow 0} \frac{1}{\eps}
\Bigl( \phi \bigl(f(x + \eps a) \bigr) - \phi\bigl( f(x) \bigr) \Bigr) \nn \\
&=  \lim_{\eps \rightarrow 0} \frac{1}{\eps}
\Bigl( \phi \bigl(x' + \eps \fu(a)\bigr) - \phi(x') \Bigr) \nn \\
&= \fu(a) \dt \grad_{x'} \phi(x'),
\label{G1a}
\end{align}
where 
\begin{equation}
\fu(a) \eqv a \dt \grad f(x)
\end{equation} 
and the subscript on $\grad_{x'}$ records that the derivative is now
with respect to the new vector position variable $x'$.  The function
$\fu(a)$ is a linear function of $a$ and an arbitrary function of $x$.
If we wish to make the position-dependence explicit we write this as
$\fu(a;x)$.  In general, any position-dependent linear
function with its position-dependence suppressed is to be taken as a
function of $x$.  Also --- as stated previously --- the
argument of a linear function should be assumed to be constant unless
explicitly stated otherwise.

From~\eqref{G1a} we see that
\begin{equation}
\grad_x = \fo(\grad_{x'})
\end{equation}
and it follows that 
\begin{equation}
\grad \phi'(x) = \fo\bigl(\grad_{x'} \phi(x') \bigr).
\label{G1b}
\end{equation} 
The bracketed term on the right-hand side, $\grad_{x'} \phi(x')$, is
the old gradient vector $\grad\phi$ evaluated at $x'$ instead of $x$.
This tells us how to modify the derivative operator $\grad$: we must
introduce a new linear function that assembles with $\grad$ in such a
way that the $\fo$ field is removed when the full object is displaced.
The resulting object will then have the desired property of just
changing its position dependence under arbitrary local displacements.
We therefore introduce the \textit{position-gauge field} $\ho(a;x)$,
which is a linear function of $a$ and an arbitrary function of
position $x$.  As usual this is abbreviated to $\ho(a)$ when the
position-dependence is taken as a function of $x$.  Under the
displacement $x\mapsto x'=f(x)$, $\ho(a)$ is defined to transform to
the new field $\ho'(a;x)$, where
\begin{equation}
\ho'(a;x) \eqv \ho \bigl(\fo^{-1}(a); f(x) \bigr) =
\ho\bigl(\fo^{-1}(a); x' \bigr).
\label{G3}
\end{equation}
This ensures that $\ho(\grad)$ transforms as
\begin{equation}
\ho(\grad_x ; x) \mapsto \ho\bigl( \fo^{-1} (\grad_x); x'\bigr) =
\ho(\grad_{x'};x'). 
\end{equation}
This transformation law ensures that, if we define a vector $A(x)$ by
\begin{equation}
A(x) \eqv \ho(\grad \phi(x)),
\end{equation}
then $A(x)$ transforms simply as $A(x)\mapsto A'(x) = A(x')$ under {\em
arbitrary\/} displacements.  This is the type of behaviour we seek.
The vector $A(x)$ can now be equated with other (possibly
non-differentiated) fields and the resulting equations are unchanged
in form under arbitrary repositioning of the fields in spacetime.

Henceforth, we refer to any quantity that transforms under arbitrary
displacements as
\begin{equation}
M(x) \mapsto M'(x) = M(x')
\end{equation}
as behaving {\em covariantly\/} under displacements.  The position
gauge field $\lih$ enables us to form derivatives of covariant objects
which are also covariant under displacements.  When we come to
calculate with this theory, we will often fix a gauge by choosing a
labelling of spacetime points with vectors.  In this way we remain
free to exploit all the advantages of representing points with
vectors.  Of course, all physical predictions of the theory will
remain independent of the actual gauge choice.

The $\lih$-field is not a connection in the conventional Yang--Mills
sense.  The coupling to derivatives is different, as is the
transformation law~\eqref{G3}.  This is unsurprising, since the group of
arbitrary translations is infinite-dimensional (if we were considering
maps between manifolds then this would form the group of
diffeomorphisms).  Nevertheless the $\lih$-field embodies the idea of
replacing directional derivatives with covariant derivatives, so
clearly deserves to be called a gauge field.

A remaining question is to find the conditions under which the
$\lih$-field can be transformed to the identity.  This would be the
case if $\lih$ was derived entirely from a displacement, which would
imply that
\begin{equation}
\ho(a) = \fo^{-1}(a),
\end{equation}
and hence
\begin{equation}
\ho^{-1}(a) = \fo(a).
\label{G4}
\end{equation}
But, from the definition of $\fu(a)$, it follows that
\begin{equation}
\fo(a) = \db \la a \, b \dt \grad f(x) = \grad \la f(x) a \ra
\end{equation} 
and hence that
\begin{equation}
\grad \wdg \fo(a) = \grad \wdg \grad \la f(x) a \ra = 0.
\label{G4.2}
\end{equation}
So, if the $\ho(a)$ field can be transformed to the identity, it
must satisfy
\begin{equation}
\grad \wdg \ho^{-1}(a) =0.
\label{G4.4}
\end{equation}
An arbitrary $\lih$-field will not satisfy this equation, so in general
there is no way to assign position vectors so that the effects of the
$\lih$-field vanish.  In the light of equations~\eqref{G4.2} and~\eqref{G4.4}
it might seem more natural to introduce the gauge field as
$\ho^{-1}(\grad)$, instead of $\ho(\grad)$.  There is little to choose
between these conventions, though our choice is partially justified by
our later implementation of the variational principle.

\subsection{The rotation-gauge field}
\label{FRGF}

We now examine how the derivative must be modified to allow rotational
freedom from point to point, as described in point~(ii) at the start
of this section.  Here we give an analysis based on the properties of
classical fields.  An analysis based on spinor fields is given in the
following section.  We have already seen that the gradient of a scalar
field is modified to $\ho(\grad\phi)$ to achieve covariance under
displacements.  But objects such as temperature gradients are
certainly physical, and can be equated with other physical quantities.
Consequently vectors such as $\ho(\grad\phi)$ must transform under
rotations in the same manner as all other physical fields.  It follows
that, under local spacetime rotations, the $\lih$-field must transform as
\begin{equation}
\ho(a) \mapsto R \ho(a) \Rrev .
\end{equation} 

Now consider an equation such as Maxwell's equation, which we saw in
Section~\ref{O-GC} takes the simple spacetime algebra form $\grad
F=J$.  Once the position-gauge field is introduced, this equation
becomes
\begin{equation}
\ho(\grad)\clf=\clj, 
\label{Gmx1}
\end{equation}
where
\begin{equation}
\clf \eqv \ho(F) \quad \mbox{and} \quad \clj \eqv \dh \, \hu^{-1} (J).
\end{equation} 
(The reasons behind these definitions will be explained in
Section~\ref{MAX}.  The use of a calligraphic letter for certain
covariant fields is a convention we have found very useful.)
The definitions of $\clf$ and $\clj$ ensure that under local rotations
they transform as 
\begin{equation}
\clf \mapsto R\clf\Rrev \quad \mbox{and} \quad  \clj \mapsto
R\clj\Rrev.
\end{equation}
Any (multi)vector that transforms in this manner under rotations and
is covariant under displacements is referred to as a
\textit{covariant} (multi)vector.

Equation~\eqref{Gmx1} is covariant under arbitrary displacements, and we
now need to make it covariant under local rotations as well.  To
achieve this we replace $\ho(\grad)$ by $\ho(e^\mu)\dmu$ and focus
attention on the term $\dmu \clf$.  Under a position-dependent
rotation we find that
\begin{equation}
\dmu (R\clf\Rrev) = R \dmu \clf \Rrev + \dmu R \,
\clf\Rrev +R \clf \dmu \Rrev.
\end{equation} 
Since the rotor $R$ satisfies $R\Rrev=1$ we find that
\begin{gather}
\dmu R \Rrev + R \dmu \Rrev = 0 \\
\implies \hs{0.5} \dmu R \Rrev = - R \dmu \Rrev = -(
\dmu R \Rrev)^\tld .
\end{gather}
Hence $\dmu R \Rrev$ is equal to minus its reverse and so must be a
bivector.  (In a geometric algebra the bivectors form a representation
of the Lie algebra of the rotation group~\cite{dor93-spin}.)  We can
therefore write
\begin{equation}
\dmu (R\clf\Rrev) = R \dmu \clf \Rrev +2 (\dmu R \Rrev) \crs
(R\clf\Rrev). 
\end{equation} 
To construct a covariant derivative we must therefore add a
connection term to $\dmu$ to construct the operator
\begin{equation}
\cld_\mu \eqv \dmu + \Om_\mu \times.
\label{G5}
\end{equation} 
Here $\Om_\mu$ is a bivector-valued linear function.  To make
the linear argument explicit we write
\begin{equation}
\Om(e_\mu) = \Om_\mu
\end{equation}
so that $\Om(a)=\Om(a;x)$ is a linear function of $a$ with an
arbitrary $x$-dependence.  The commutator product of a multivector
with a bivector is grade-preserving so, even though it contains
non-scalar terms, $\cld_\mu$ preserves the grade of the multivector on
which it acts.

Under local rotations the $\dmu$ term in $\cld_\mu$ cannot change,
and we also expect that the $\cld_\mu$ operator be unchanged in form
(this is the essence of `minimal coupling').  We should therefore have
\begin{equation}
\cld_\mu' =  \dmu + \Om_\mu' \times.
\label{Grdv1}
\end{equation}
But the property that the covariant derivative must satisfy is
\begin{equation}
\cld_\mu' (R\clf\Rrev) = R \cld_\mu \clf \Rrev
\end{equation}
and, substituting~\eqref{Grdv1} into this equation, we find that $\Om_\mu$
transforms as
\begin{equation}
\Om_\mu \mapsto \Om_\mu' = R \Om_\mu \Rrev - 2 \dmu R \Rrev.
\end{equation}
In coordinate-free form we can write this transformation law as
\begin{equation}
\Om(a) \mapsto \Om'(a) = R \Om(a) \Rrev - 2 a \dt \grad R \Rrev.
\label{G6}
\end{equation} 
Of course, since $\Om(a)$ is an arbitrary function of position, it
cannot in general be transformed away by the application of a rotor.
We finally reassemble the derivative~\eqref{G5} with the $\ho(\da)$ term
to form the equation
\begin{equation}
\ho(e^\mu)\cld_\mu \clf = \clj.
\end{equation}
The transformation properties of $\ho(a)$, $\clf$, $\clj$ and $\Om(a)$
ensure that this equation is now covariant under rotations as well as
displacements.

To complete the set of transformation laws, we note that under
displacements $\Om(a)$ must transform in the same way as $a\dt\grad
R\Rrev$, so that 
\begin{equation}
\Om(a;x) \mapsto \Om(\fu(a); f(x)) =  \Om(\fu(a); x').
\label{GOmtrf} 
\end{equation} 
It follows that
\begin{align}
\ho(\da) \, \Om(a) \crs \clf(x) &\mapsto \ho(\fo^{-1}(\da);x') \,
\Om(\fu(a);x') \crs \clf(x') \nn \\
&= \ho(\da;x') \,  \Om(a;x') \crs \clf(x'),
\end{align}
as required for covariance under local translations.

General considerations have led us to the introduction of two new
gauge fields: vector-valued linear function $\ho(a;x)$ and the
bivector-valued linear function $\Om(a;x)$, both of which are
arbitrary functions of position $x$.  This gives a total of $4\times
4+4\times 6=40$ scalar degrees of freedom.  The $\ho(a)$ and $\Om(a)$
fields are incorporated into the vector derivative to form the
operator $\ho(e^\mu)\cld_\mu$, which acts covariantly on multivector
fields.  Thus we can begin to construct equations whose intrinsic
content is free of the manner in which we represent spacetime
positions with vectors.  We next see how these fields arise in the
setting of the Dirac theory.  This enables us to derive the properties
of the $\cld_\mu$ operator from more primitive considerations of the
properties of spinors and the means by which observables are
constructed from them.  First, though, let us compare the fields that
we have defined with the fields used conventionally in general
relativity .  To do this, we first construct the vectors
\begin{equation}
g_\mu \eqv \hu^{-1}(e_\mu), \qquad
g^\mu \eqv \ho(e^\mu).
\end{equation}
From these, the metric is defined by
\begin{equation}
g_{\mu \nu} \eqv g_\mu \dt g_\nu.
\end{equation}
Note that the inner product here is the standard spacetime algebra
inner product and is not related to a curved space metric.  Similarly,
the $\lih$-field can be used to construct a \textit{vierbein}, as
discussed in Appendix~C.  A vierbein in general relativity relates a
coordinate frame to an orthonormal frame.  But, while the $\lih$-field
can be used to construct such a vierbein, it should be clear that it
serves a different purpose in GTG --- it ensures covariance under
arbitrary displacements.  This was the motivation for the introduction
of a vierbein in Kibble's work~\cite{kib61}, although only
infinitesimal transformations could be considered there.  Furthermore,
the $\lih$-field is essential to enable a clean separation between
field rotations and displacements, which again is not achieved in
other approaches.

\subsection{Gauge fields for the Dirac action}
\label{G-dirac}

We now rederive the gravitational gauge fields from symmetries of the
Dirac action.  The point here is that, once the $\lih$-field is
introduced, spacetime rotations and phase rotations couple to the
Dirac field in essentially the same way.  To see this, we start with
the Dirac equation and Dirac action in a slightly unconventional
form~\cite{hes75,DGL93-states,DGL93-lft}.  We saw in
Section~\ref{Outline} that rotation of a multivector is performed by
the double-sided application of a rotor.  The elements of a linear
space that is closed under single-sided action of a representation of
the rotor group are called \textit{spinors}.  In conventional
developments a matrix representation for the Clifford algebra of
spacetime is introduced, and the space of column vectors on which
these matrices act defines the spin-space.  But there is no need to
adopt such a construction.  For example, the even subalgebra of the
spacetime algebra forms a linear space that is closed under
single-sided application of the rotor group.  The even subalgebra is
also an eight-dimensional linear space, the same number of real
dimensions as a Dirac spinor, and so it is not surprising that a
one-to-one map between Dirac spinors and the even subalgebra can be
constructed.  Such a map is given in Appendix~A.  The essential
details are that the result of multiplying the column spinor
$|\psi\ra$ by the Dirac matrix $\hgamum$ is represented in the
spacetime algebra as $\psi\mapsto\gamum\psi\go$, and that
multiplication by the scalar unit imaginary is represented as
$\psi\mapsto\psi\isk$.  It is easily seen that these two operations
commute and that they map even multivectors to even multivectors.  By
replacing Dirac matrices and column spinors by their spacetime algebra
equivalents the Dirac equation can be written in the form
\begin{equation}
\grad \psi \isk -eA\psi = m \psi \go,
\label{G-d1}
\end{equation}
which is now representation-free and coordinate-free.  Using the same
replacements, the free-particle Dirac action becomes
\begin{equation}
S= \int |d^4 x| \la \grad \psi I\gk \psirev - m \psi \psirev \ra
\label{G-d2} 
\end{equation}
and, with the techniques of Appendix~\ref{app-results}, it is simple
to verify that variation of this action with respect to $\psi$ yields
equation~\eqref{G-d1} with $A$=0.  

It is important to appreciate that the fixed $\go$ and $\gk$ vectors
in~\eqref{G-d1} and~\eqref{G-d2} do not pick out preferred directions in
space.  These vectors can be rotated to new vectors $R_0\go\Rrev_0$
and $R_0\gk\Rrev_0$, and replacing the spinor by $\psi\Rrev_0$ recovers
the same equation~\eqref{G-d1}.  This point will be returned to when we
discuss forming observables from the spinor $\psi$.

Our aim now is to introduce gauge fields into the action~\eqref{G-d2} to
ensure invariance under arbitrary rotations and displacements.  
The first step is to introduce the $\lih$-field.  Under a displacement
$\psi$ transforms covariantly,
\begin{equation}
\psi(x) \mapsto \psi'(x) = \psi(x'),
\end{equation}
where $x'=f(x)$.  We must therefore replace the $\grad$ operator by
$\ho(\grad)$ so that $\ho(\grad)\psi$ is also covariant under
translations.  But this on its own does not achieve complete
invariance of the action integral~\eqref{G-d2} under displacements.
The action consists of the integral of a scalar over some region.  If
the scalar is replaced by a displaced quantity, then we must also
transform the measure and the boundary of the region if the resultant
integral is to have the same value.  Transforming the boundary is
easily done, but the measure does require a little work.  Suppose that
we introduce a set of coordinates $x^\mu$.  The measure $|d^4x|$ is
then written
\begin{equation}
|d^4x| = -I e_0 \wdg e_1 \wdg e_2 \wdg e_3 \, dx^0 \, dx^1\, dx^2\, dx^3,
\end{equation}
where
\begin{equation}
e_\mu \eqv \deriv{x}{x^\mu}.
\end{equation}
By definition, $|d^4x|$ is already independent of the choice of
coordinates, but it must be modified to make it position-gauge
invariant.  To see how, we note that under the displacement $x\mapsto
f(x)$, the $\{e_\mu\}$ frame transforms to
\begin{equation}
e_\mu' (x) = \deriv{f(x)}{x^\mu} = \fu(e_\mu).
\end{equation}
It follows that to ensure invariance of the action integral we must
replace each of the $e_\mu$ by $\hu^{-1}(e_\mu)$.  Thus the invariant
scalar measure is
\begin{equation}
-I \, \hu^{-1}(e_0) \wdg \cdots \wdg
\hu^{-1}(e_3) \, dx^0 \cdots dx^3 = \dhi |d^4x|.
\end{equation}
These results lead us to the action
\begin{equation}
S= \int |d^4 x| \dhi \la \ho(\grad) \psi I\gk \psirev - m
\psi \psirev \ra ,
\label{G-d4} 
\end{equation}
which is unchanged in value if the dynamical fields are replaced by 
\begin{align}
\psi'(x) &\eqv \psi(x') \\
\ho'(a;x)  &\eqv \ho(\fo^{-1}(a);x') ,
\end{align}
and the boundary is also transformed.

Having arrived at the action in the form of~\eqref{G-d4} we can now
consider the effect of rotations applied at a point.  The
representation of spinors by even elements is now particularly
powerful because it enables both internal phase rotations and
rotations in space to be handled in the same unified framework.
Taking the electromagnetic coupling first, we see that the
action~\eqref{G-d4} is invariant under the global phase rotation
\begin{equation}
\psi \mapsto \psi' \eqv \psi e^{\phi \isk}.
\label{G-d7}
\end{equation}
(Recall that multiplication of $|\psi\ra$ by the unit imaginary is
represented by right-sided multiplication of $\psi$ by $\isk$.)  The
transformation~\eqref{G-d7} is a special case of the more general
transformation
\begin{equation}
\psi \mapsto \psi R,
\label{G-d9I}
\end{equation}
where $R$ is a constant rotor.  Similarly, invariance of the
action~\eqref{G-d4} under spacetime rotations is described by
\begin{align}
\psi &\mapsto R \psi \\
\ho(a) &\mapsto R \ho(a) \Rrev.
\label{G-d9II}
\end{align}
In both cases, $\psi$ just picks up a single rotor.  From the previous
section we know that, when the rotor $R$ is position-dependent, the
quantity $\dmu R\Rrev$ is a bivector-valued linear function of
$e_\mu$.  Since
\begin{equation}
\dmu (R \psi) = R \dmu \psi + (\dmu R \Rrev) R\psi,
\end{equation}
with a similar result holding when the rotor acts from the right, we
need the following covariant derivatives for local internal and
external rotations:
\begin{align}
\mbox{Internal:} \quad & D_\mu^I \psi = \dmu \psi + \half \psi
\Om^I_\mu \\
\mbox{External:} \quad & D_\mu \psi = \dmu \psi + \half \Om_\mu
\psi. 
\end{align}
For the case of (internal) phase rotations, the rotations are
constrained to take place entirely in the $\isk$ plane.  It follows
that the internal connection takes the restricted form
\begin{equation} 
\Om^I(a) = 2e\, a\dt A \, \isk
\end{equation}
where $A$ is the conventional electromagnetic vector
potential and $e$ is the coupling constant (the charge). The
full covariant derivative therefore has the form
\begin{equation}
\ho(e^\mu) \Bigl( \dmu \psi + \half \Om_\mu \psi + e \psi \isk e_\mu
\dt A \Bigr) 
\end{equation}
and the full invariant action integral is now (in coordinate-free form)
\begin{equation}
S= \int |d^4 x|(\det\hu)^{-1} \Bigl\la \ho(\da) \bigl(a \dt \grad +
\half \Om(a) \bigr) \psi I\gk \psirev -e \ho(A) \psi \go \psirev  - m
\psi \psirev \Bigr\ra. 
\label{G-d12}
\end{equation}
The action~\eqref{G-d12} is invariant under the symmetry transformations
listed in Table~\ref{G-dtab1}.

{\renewcommand{\arraystretch}{1.6}
\begin{table}
\begin{center}
\begin{tabular}{c|cccc}
\hline \hline
&  \multicolumn{4}{c}{Transformed Fields} \\
{\renewcommand{\arraystretch}{0.8}
\begin{tabular}[b]{c}
Local \\ 
symmetry 
\end{tabular}}
& $\psi'(x)$ & $\ho'(a;x)$ & $\Om'(a;x)$ & $eA'(x)$ \\
\hline 
Displacements & $\psi(x')$ & $\ho(\fo^{-1}(a);x')$ & $\Om(\fu(a);x')$ &
$e\fo(A(x'))$ \\
{\renewcommand{\arraystretch}{0.8}
\begin{tabular}{c}
Spacetime \\ 
rotations
\end{tabular}}
& $R\psi$ & $R\ho(a)\Rrev$ & $R\Om(a)\Rrev - 2a\dt\grad R\Rrev$ &
$eA$ \\
{\renewcommand{\arraystretch}{0.8}
\begin{tabular}{c}
Phase \\ 
rotations
\end{tabular}}
& $\psi \et{\phi\isk}$ & $\ho(a)$ & $\Om(a)$ & $eA-\grad\phi$ \\
\hline \hline
\end{tabular}
\end{center}
\caption[dummy1]{\it The symmetries of the action integral~\eqref{G-d12}.}
\label{G-dtab1}
\end{table}}

\subsection{The coupled Dirac equation}

Having arrived at the action~\eqref{G-d12} we now derive the coupled Dirac
equation by extremising with respect to $\psi$, treating all other
fields as external.  When applying the Euler--Lagrange equations to the
action~\eqref{G-d12} the $\psi$ and $\psirev$ fields are not treated as
independent, as they often are in quantum theory.  Instead, we just
apply the rules for the multivector derivative discussed in
Section~\ref{O-GC} and Appendix~\ref{app-results}.  The Euler--Lagrange
equations can be written in the form
\begin{equation}
\dpsi \cll = \dmu (\partial_{\psi_{,\mu}} \cll)
\label{G-d15}
\end{equation}
where for simplicity we assume that the $x^\mu$ are a set of fixed
orthonormal coordinates given by $x^\mu = \gamum \dt x$ (see
Appendix~\ref{app-results}).  Applied to the action~\eqref{G-d12},
equation~\eqref{G-d15} yields
\begin{multline}
(\ho(\grad) \psi I\gk)^\tld + \half I\gk\psirev\ho(\da)\Om(a) + \half
(\ho(\da)\Om(a)\psi I\gk)^\tld - e\go \psirev \ho(A) \\
 - (e\ho(A) \psi \go)^\tld -2 m \psirev = \dmu \bigl( \dhi I\gk
\psirev \ho(\gamum) \bigr) \dh. 
\end{multline}
Reversing this equation and simplifying gives
\begin{multline}
\qquad \ho(\da)\bigl( a\dt\grad + \half \Om(a) \bigr) \psi I\gk
-e\ho(A)\psi\go -m \psi \\
= -\half \dh \cld_\mu  \bigl( \ho(\gamum) \dhi \bigr) \psi I\gk, \qquad
\label{G-d17}
\end{multline} 
where we have employed the $\cld_\mu$ derivative defined in
equation~\eqref{G5}.  If we now introduce the notation
\begin{align}
D \psi &\eqv \ho(\da)\bigl( a \dt \grad + \half \Om(a) \bigr) \psi =
g^\mu ( \dmu + \half \Om_\mu) \psi \\
\cla &\eqv  \ho(A),
\end{align}
we can write equation~\eqref{G-d17} in the form
\begin{equation}
D \psi \isk -e\cla \psi = m \psi \go -\half \dh \cld_\mu \bigl(
\ho(\gamum) \dhi \bigr) \psi \isk. 
\label{G-d18}
\end{equation}
This equation is covariant under the symmetries listed in
Table~\ref{G-dtab1}, as must be the case since the equation was
derived from an invariant action integral.  But equation~\eqref{G-d18}
is not what we would have expected had we applied the gauging
arguments at the level of the Dirac equation, rather than the Dirac
action.  Instead, we would have been led to the simpler equation
\begin{equation}
D \psi \isk -e\cla \psi = m \psi \go .
\label{G-d19}
\end{equation}
Clearly, equation~\eqref{G-d18} reduces to equation~\eqref{G-d19} only if the
$\ho(a)$ and $\Om(a)$ fields satisfy the identity
\begin{equation}
\dh \cld_\mu \bigl( \ho(\gamum) \dhi \bigr) = 0.
\end{equation}
The restriction to an orthonormal coordinate frame can be removed by
writing this equation as
\begin{equation}
\dh \dot{\cld}_\mu \bigl( \dot{\ho}(e^\mu) \dhi \bigr) = 0.
\label{G-d20}
\end{equation}
(It is not hard to show that the left-hand side of
equation~\eqref{G-d20} is a covariant vector; later it will be
identified as a contraction of the `torsion' tensor).  There are good
reasons for expecting equation~\eqref{G-d20} to hold.  Otherwise, the
minimally-coupled Dirac action would not yield the minimally-coupled
equation, which would pose problems for our use of action principles
to derive the gauged field equations.  We will see shortly that the
demand that equation~\eqref{G-d20} holds places a strong restriction
on the form that the gravitational action can take.

Some further comments about the derivation of~\eqref{G-d18} are now in
order.  The derivation employed only the rules of vector and
multivector calculus applied to a `flat-space' action integral.  The
derivation is therefore a rigorous application of the variational
principle.  This same level of rigour is not always applied when
deriving field equations from action integrals involving spinors.
Instead, the derivations are often heuristic --- $|\psi\ra$ and
$\la\psibar|$ are treated as independent variables and the
$\la\psibar|$ is just `knocked off' the Lagrangian density to leave
the desired equation.  Furthermore, the action integral given by many
authors for the Dirac equation in a gravitational background has an
imaginary component~\cite{nak-geom,goc-diff}, in which case the status
of the variational principle is unclear.  To our knowledge, only
Hehl~\& Datta~\cite{heh76} have produced a derivation that in any way
matches the derivation produced here.  Hehl~\& Datta also found an
equation similar to~\eqref{G-d20}, but they were not working within a
gauge theory setup and so did not comment on the consistency (or
otherwise) of the minimal-coupling procedure.

\subsection{Observables and covariant derivatives}

As well as keeping everything within the real spacetime algebra ,
representing Dirac spinors by elements of the even subalgebra offers
many advantages when forming observables.  As described in
Appendix~\ref{app-dirac}, observables are formed by the double-sided
application of a Dirac spinor $\psi$ to some combination of the fixed
$\{\gamum\}$ frame vectors.  So, for example, the charge current is
given by $\clj\eqv\psi\go\psirev$ and the spin current by $s\eqv\psi
\gk\psirev$.  In general, an observable is of the form
\begin{equation}
M \eqv \psi \Gamma \psirev,
\label{G-o1}
\end{equation}
where $\Gamma$ is a constant multivector formed from the $\{\gamum\}$.
All observables are invariant under phase rotations, so $\Gamma$ must
be invariant under rotations in the $\isk$ plane.  Hence $\Gamma$ can
consist only of combinations of $\go$, $\gk$, $\isk$ and their duals
(formed by multiplying by $I$).  An important point is that, in
forming the observable $M$, the $\Gam$ multivector is completely
`shielded' from rotations.  This is why the appearance of the $\go$
and $\gk$ vectors on the right-hand side of the spinor $\psi$ in the
Dirac action~\eqref{G-d12} does not compromise Lorentz invariance, and
does not pick out a preferred direction in space~\cite{DGL93-states}.
All observables are unchanged by rotating the $\{\gamum\}$ frame
vectors to $R_0\gamdm\Rrev_0$ and transforming $\psi$ to
$\psi\Rrev_0$.  (In the matrix theory this corresponds to a change of
representation.)

Under translations and rotations the observables formed in the above
manner~\eqref{G-o1} inherit the transformation properties of the
spinor $\psi$.  Under translations the observable
$M=\psi\Gamma\psirev$ therefore transforms from $M(x)$ to $M(x')$, and
under rotations $M$ transforms to $R\psi\Gamma\psirev\Rrev=RM\Rrev$.
The observable $M$ is therefore covariant.  These Dirac observables
are the first examples of quantities that transform covariantly under
rotations, but do not inherit this transformation law from the
$\lih$-field.  In contrast, all covariant forms of classical fields,
such as $\clf$ or the covariant velocity along a worldline
$\hu^{-1}(\dot{x})$, transform under rotations in a manner that that
is dictated by their coupling to the $\lih$-field.  Classical general
relativity in fact removes any reference to the rotation gauge from
most aspects of the theory.  Quantum theory, however, demands that the
rotation gauge be kept in explicitly and, as we shall show in
Section~\ref{DE}, Dirac fields probe the structure of the
gravitational fields at a deeper level than classical fields.
Furthermore, it is only through consideration of the quantum theory
that one really discovers the need for the rotation-gauge field.

One might wonder why the observables are {\em invariant\/} under phase
rotations, but only {\em covariant\/} under spatial rotations.  In
fact, the $\lih$-field enables us to form quantities like $\hu(M)$,
which are invariant under spatial rotations.  This gives an
alternative insight into the role of the $\lih$-field.  We will find
that both covariant observables ($M$) and their rotationally-invariant
forms ($\hu(M)$ and $\ho^{-1}(M)$) play important roles in the theory
constructed here.

If we next consider the directional derivative of $M$, we find that it
can be written as
\begin{equation}
\dmu M = (\dmu \psi) \Gamma \psirev + \psi \Gamma (\dmu \psi)^\tld.
\end{equation}
This immediately tells us how to turn the directional derivative
$\dmu M$ into a covariant derivative: simply replace the spinor
directional derivatives by covariant derivatives.  Hence we form
\begin{align}
(D_\mu \psi) \Gamma \psirev + \psi \Gamma (D_\mu \psi)^\tld 
&= (\dmu \psi) \Gamma \psirev + \psi \Gamma (\dmu \psi)^\tld + \half
\Omega_\mu  \psi \Gamma \psirev -\half \psi \Gamma \psirev
\Omega_\mu \nn  \\
&= \dmu (\psi \Gamma \psirev) + \Omega_\mu \crs (\psi \Gamma
\psirev).
\end{align}
We therefore recover the covariant derivative for observables:
\begin{equation}
\cld_\mu M \eqv \dmu M + \Omega_\mu \crs M .
\label{G-d25}
\end{equation}
This derivation shows that many features of the `classical' derivation
of gravitational gauge fields can be viewed as arising from more basic
quantum transformation laws.

Throughout this section we have introduced a number of distinct
gravitational covariant derivatives.  We finish this section by
discussing some of their main features and summarising our
conventions.  The operator $\cld_\mu$ acts on any covariant multivector
and has the important property of being a {\em derivation}, that is it
acts as a scalar differential operator,
\begin{equation}
\cld_\mu (AB) =  (\cld_\mu A)B + A (\cld_\mu B).
\end{equation}
This follows from Leibniz' rule and the identity
\begin{equation}
\Omega_\mu \crs (AB) =  \Omega_\mu \crs A \, B + A \, \Omega_\mu \crs B.
\end{equation}
Neither $D_\mu$ or $\cld_\mu$ are fully covariant, however, since they
both contain the $\Om_\mu$ field, which picks up a term in $\fu$ under
displacements~\eqref{GOmtrf}.  It is important in the applications to
follow that we work with objects that are covariant under
displacements, and to this end we define
\begin{equation}
\om(a) \eqv \Om (\hu(a)) = a \dt g^\mu \, \Om_\mu.
\end{equation}
We also define the full covariant directional derivatives $a\dt D$ and
$a\dt\cld$ by
\begin{align}
a \dt D \psi &\eqv a \dt \ho(\grad) \psi + \half \om(a) \psi \\
a \dt \cld M &\eqv a \dt \ho(\grad) M + \om(a) \crs M.
\end{align}
Note that these conventions imply that $\cld_\mu = g_\mu \dt \cld$,
with a similar result holding for $D_\mu$.

For the $a\dt\cld$ operator we can further define the covariant vector
derivative 
\begin{equation}
\cld M \eqv  \da \, a \dt \cld M = g^\mu \cld_\mu M.
\end{equation}
The covariant vector derivative contains a grade-raising and a
grade-lowering component, so that 
\begin{equation}
\cld A = \cld \dt A + \cld \wdg A,
\end{equation}
where
\begin{align}
\cld \dt A &\eqv \da \dt (a \dt \cld A) = g^\mu \dt (\cld_\mu A) \\
\cld \wdg A &\eqv \da \wdg (a \dt \cld A) = g^\mu \wdg (\cld_\mu A).
\end{align}
As with the vector derivative, $\cld$ inherits the algebraic
properties of a vector.

\begin{table}
\renewcommand{\arraystretch}{1.2}
\begin{center}
\begin{tabular}{lll}
\hline \hline
\\
& Gauge fields & 
\begin{minipage}[c]{7cm}
\( 
\begin{array}{ll}
\mbox{Displacements:} & \ho(a) \\
\mbox{Rotations:} & \Om(a), \quad \om(a) = \Om(\hu(a))
\end{array}
\)
\end{minipage} \\
\\
& Spinor derivatives &
\begin{minipage}[c]{6cm}
\( \begin{array}{l}
D_\mu \psi = \dmu \psi + \half \Om_\mu \psi \\
a \dt D \psi = a\dt\ho(\grad) \psi + \half \om(a) \psi
\end{array} \)
\end{minipage} \\
\\
& Observables derivatives &
\begin{minipage}[c]{6cm}
\( 
\begin{array}{l}
{} \cld_\mu M = \dmu M + \Om_\mu \crs M \\
a \dt \cld M = a\dt\ho(\grad) M + \om(a) \crs M \\
\cld M = \da a\dt\cld M = \cld \dt M + \cld \wdg M
\end{array} \)
\end{minipage}
\\
& Vector derivative &
\begin{minipage}[c]{2cm}
\begin{equation*}  
\grad = e^\mu \deriv{}{x^\mu}  
\end{equation*}
\end{minipage} 
\\
& Multivector derivative &
\begin{minipage}[c]{6cm}
\begin{equation*} 
\dX =  \sum_{i < \cdots < j} e^i \wdg \cdots \wdg e^j \, (e_j \wdg
\cdots \wdg e_i) \scp \dX 
\end{equation*}  
\end{minipage} \\
\\
\hline \hline
\end{tabular}
\end{center}
\caption[dummy1]{\sl Definitions and conventions}
\label{tab-convs}
\end{table}

A summary of our definitions and conventions is contained in
Table~\ref{tab-convs}.  We have endeavoured to keep these conventions
as simple and natural as possible, but a word is in order on our
choices.  It should be clear that it is a good idea to use a separate
symbol for the spacetime vector derivative ($\grad$), as opposed to
writing it as $\dx$.  This maintains a clear distinction between
spacetime derivatives, and operations on linear functions such as
`contraction' ($\da\cdot$) and `protraction' ($\da\wedge$).  It is
also useful to distinguish between spinor and vector covariant
derivatives, which is why we have introduced separate $D$ and $\cld$
symbols.  We have avoided use of the $d$ symbol, which already has a
very specific meaning in the language of differential forms.  It is
necessary to distinguish between rotation-gauge derivatives
($\cld_\mu$) and the full covariant derivative with the $\lih$-field
included ($a\dt\cld$).  Using $\cld_\mu$ and $a\dt\cld$ for these
achieves this separation in the simplest possible manner.  Throughout
this paper we will switch between using coordinate frames (as in
$\cld_\mu$) and the frame-free notation (such as $\om(a)$).  Seeing
which notation is appropriate for a given problem is something of an
art, though we hope to convey some of the basic rules in what
follows.


\section{The field equations}
\label{Field}

Having introduced the $\lih$  and $\Om$-fields, we now look to
construct an invariant action integral that will provide a set of
gravitational field equations.  We start by defining the
field-strength via
\begin{equation} 
[D_\mu, D_\nu] \psi = \half \liR_{\mu\nu} \psi,
\end{equation} 
so that
\begin{equation} 
\liR_{\mu\nu} = \dmu \Om_\nu - \dnu \Om_\mu + \Om_\mu \crs \Om_\nu.
\label{F1}
\end{equation} 
It follows that we also have 
\begin{equation}
[\cld_\mu, \cld_\nu] M = \liR_{\mu\nu} \crs M. 
\end{equation}
A frame-free notation is introduced by writing
\begin{equation}
\liR_{\mu\nu} = \liR(e_\mu \wdg e_\nu).
\end{equation}
The field $\liR(a \wedge b)$ is a bivector-valued linear function of its
bivector argument $a \wedge b$.  Its action on bivectors extends by
linearity to the function $\liR(B)$, where $B$ is an arbitrary bivector
and therefore, in four dimensions, not necessarily a pure `blade'
$a\wedge b$.  Where required, the position dependence is made explicit
by writing $\liR(B;x)$.

Under an arbitrary rotation, the definition~\eqref{F1} ensures that $\liR(B)$
transforms as
\begin{equation}
\liR(B) \mapsto \liR'(B)=R \liR(B)  \Rrev.
\end{equation}
Under local displacements we find that
\begin{align}
\liR'(e_\mu \wdg e_\nu) 
&= \dmu \Om'(e_\nu) - \dnu \Om'(e_\mu) + \Om'(e_\nu) \crs \Om'(e_\mu)
\nn \\
&= \lif(e_\mu) \dt \dgrad_{x'} \dot{\Om}\bigl(\lif(e_\nu);x'\bigr)
 -  \lif(e_\nu) \dt \dgrad_{x'} \dot{\Om}\bigl(\lif(e_\mu);x'\bigr) 
+ \Om'(e_\mu) \crs \Om'(e_\nu) \nn \\
&\quad + \Om\bigl( \dmu \lif(e_\nu) - \dnu \lif(e_\mu); x'\bigr) \nn \\
&= \liR\bigl(\lif(e_\mu \wdg e_\nu );x'\bigr) 
+  \Om\bigl(  \dmu \lif(e_\nu) - \dnu \lif(e_\mu); x'\bigr).
\end{align}
But we know that 
\begin{equation} 
 \dmu \lif(e_\nu) - \dnu \lif(e_\mu) = \dmu \dnu f(x) - \dnu\dmu f(x)
 = 0, 
\end{equation} 
so the field strength has the simple displacement transformation law
\begin{equation} 
\liR(B) \mapsto \liR'(B) = \liR\bigl(\lif(B); x'\bigr).
\end{equation} 
A covariant quantity can therefore be constructed by defining
\begin{equation}
\clr(B) \eqv \liR(\hu(B)).
\end{equation}
Under arbitrary displacements and local rotations, $\clr(B)$ has the
following transformation laws:
\begin{equation}
\begin{array}{rcl}
\mbox{Translations:} & & \clr'(B;x)=\clr(B;x') \\
\mbox{Rotations:} & & \clr'(B)=R\, \clr(\Rrev BR) \, \Rrev. 
\end{array}
\label{F3}
\end{equation} 
We refer to any linear function with transformation laws of this type
as a covariant tensor.  $\clr(B)$ is our gauge theory analogue of the
Riemann tensor.  We have started to employ a notation that is very
helpful for the theory developed here.  Certain covariant quantities,
such as $\clr(B)$ and $\cld$, are written with calligraphic symbols.
This helps keep track of the covariant quantities and often enables a
simple check that a given equation is gauge covariant.  It is not
necessary to write all covariant objects with calligraphic symbols,
but it is helpful for common objects such as $\clr(B)$ and $\cld$.

From $\clr(B)$ we define the following contractions:
\begin{alignat}{2}
\mbox{Ricci Tensor:}& & \quad  \clr(b)&= \da \dt \clr(a \wdg b) \\
\mbox{Ricci Scalar:}& & \quad \clr &= \da \dt \clr(a) \\
\mbox{Einstein Tensor:}& &\quad \clg(a)&= \clr(a) - \half a \clr.
\end{alignat}
The argument of $\clr$ determines whether it represents the Riemann or
Ricci tensors or the Ricci scalar.  Furthermore, we never apply the
extension notation of equation~\eqref{1defouter} to $\clr(a)$, so
$\clr(a \wedge b)$ unambiguously denotes the Riemann tensor.  Both
$\clr(a)$ and $\clg(a)$ are also covariant tensors, since they inherit
the transformation properties of $\clr(B)$.

The Ricci scalar is invariant under rotations, making it our first
candidate for a Lagrangian for the gravitational gauge fields.  We
therefore suppose that the overall action integral is of the form
\begin{equation}
S = \int |d^4x| \, \dhi (\half \clr - \kappa \cll_m),
\label{F-act}
\end{equation}
where $\cll_m$ describes the matter content and $\kappa=8 \pi G$.  The
independent dynamical variables are $\ho(a)$ and $\Om(a)$, and in
terms of these
\begin{equation}
\clr = \big\la \ho(\db \wdg \da) \bigl(a \dt \grad \Om(b) - b \dt \grad \Om(a)
+ \Om(a) \crs \Om(b)\bigr) \big\ra.
\end{equation}
We also assume that $\cll_m$ contains no second-order derivatives, so
that $\ho(a)$ and $\Om(a)$ appear undifferentiated in the matter Lagrangian.

\subsection[The $\hu$-equation]{The  $\hu$-equation}

The $\lih$-field is undifferentiated in the entire action, so its
Euler--Lagrange equation is simply
\begin{equation}
\dhoa (\dhi (\clr/2 - \kap \cll_m)) = 0.
\label{F-i1} 
\end{equation}
Employing some results from Appendix~\ref{app-results}, we find
that 
\begin{equation}
\dhoa \dhi = - \dhi \hu^{-1}(a)
\end{equation}
and
\begin{align}
\dhoa \clr &= \dhoa \la \ho(\dc \wdg \db) \liR(b\wdg c) \ra \nn \\
&= 2 \ho(\db) \dt \liR(b \wdg a), 
\end{align}
so that 
\begin{equation}
\dhoa (\clr \dhi) = 2 \clg(\hu^{-1}(a)) \dhi.
\end{equation}
If we now define the covariant matter stress-energy tensor $\clt(a)$
by
\begin{equation}
\dh \dhoa (\cll_m \dhi) = \clt (\hu^{-1}(a)),
\label{F-i2}
\end{equation}
we arrive at the equation
\begin{equation}
\clg(a) = \kap \clt(a).
\label{einsteqn} 
\end{equation}
This is the gauge theory statement of Einstein's equation.  Note,
however, that as yet nothing should be assumed about the symmetry of
$\clg(a)$ or $\clt(a)$.  In this derivation only the gauge fields have
been varied, and not the properties of spacetime.  Therefore, despite
the formal similarity with the Einstein equations of general
relativity, there is no doubt that we are still working in a flat
spacetime.

\subsection[The $\Om$-equation]{The {\boldmath $\Om$}-equation}

The Euler--Lagrange field equation from $\Om(a)$ is, after multiplying
through by $\dh$,
\begin{equation}
\partial_{\Omega(a)} \clr - \dh \db\dt\grad \bigl( \partial_{\Omega(a)_{,b}}
\clr \dhi \bigr) = 2\kappa \partial_{\Omega(a)} \cll_m,
\label{F5}
\end{equation}   
where we have made use of the assumption that $\Om(a)$ does not
contain any coupling to matter through its derivatives.  The
derivatives $\partial_{\Omega(a)}$ and $\partial_{\Omega(a)_{,b}}$ are
defined in Appendix~\ref{app-results}.  The only properties required
for this derivation are the following:
\begin{align}
\partial_{\Omega(a)} \la \Omega(b)M \ra &= a \dt b \la M \ra_2 \\
\partial_{\Omega(b)_{,a}} \la c \dt \grad \Omega(d) M \ra 
&=  a \dt c \, b \dt d  \la M \ra_2.
\end{align}
From these we derive
\begin{align}
\partial_{\Omega(a)} \la \ho(\dd \wdg \dc) \Om(c) \crs \Om(d) \ra 
&= \Om(d) \crs \ho(\dd \wdg a) + \ho(a \wdg \dc) \crs \Om(c) \nn \\
&= 2 \Om(b) \crs \ho(\db \wdg a) 
\label{F6}
\end{align}
and
\begin{align} 
\partial_{\Omega(a)_{,b}} \big\la \ho(\dd \wdg \dc) \bigl( c\dt\grad \Om(d) -
d\dt\grad \Om(c) \bigr) \big\ra &= \ho(a \wdg b) - \ho( b \wdg
a) \nn \\ 
&= 2 \ho(a \wdg b).
\label{F7}
\end{align}  
The right-hand side of~\eqref{F5} defines the `spin' of the matter, 
\begin{equation}
S(a) \eqv \partial_{\Omega(a)} \cll_m,
\end{equation} 
where $S(a)$ is a bivector-valued linear function of $a$. 
Combining~\eqref{F5}, \eqref{F6} and \eqref{F7} yields
\begin{multline}
\qquad \ho(\grad) \wdg \ho(a) + \dh \db \dt \grad \bigl( \ho(b)
\dhi \bigr) \wdg \ho(a) \\ 
+ \Om(\db) \crs \ho(b \wdg a) = \kap S(a). \qquad
\label{F8}
\end{multline} 
Recall that in an expression such as this, the vectors $a$ and $b$ are
viewed as being independent of position.

To make further progress we contract equation~\eqref{F8} with
$\hu^{-1}(\da)$.  To simplify this we require the result
\begin{align}
\la b\dt\grad \ho(a) \hu^{-1}(\da) \ra &= \dhi \la  b\dt\grad \ho(\da)
\, \dhoa \dh \ra \nn \\
&= \dhi  b\dt\grad \dh.
\end{align} 
We then obtain
\begin{equation}
-2  \dh \db \dt \grad \bigl( \ho(b)
\dhi \bigr) -2 \Om(\db) \dt \ho(b) = \kap \hu^{-1}(\da) \dt S(a).
\end{equation}
To convert this into its manifestly covariant form we first define the
covariant spin tensor $\cls(a)$ by

\begin{equation}
\cls(a) \eqv S \bigl(\ho^{-1}(a) \bigr).
\end{equation}
We can now write
\begin{equation}
\dh \dot{\cld}_\mu \bigl( \dot{\ho}(e^\mu) \dhi \bigr) = - \half \kap
\da \dt \cls(a). 
\end{equation} 
In Section~\ref{G-dirac} we found that the minimally-coupled Dirac
action gave rise to the minimally-coupled Dirac equation only when the
term on the left of this equation was zero.  We now see that this
requirement amounts to the condition that the spin tensor has zero
contraction.  But, if we assume that the $\Om(a)$ field only couples
to a Dirac fermion field, then the coupled Dirac action~\eqref{G-d12}
gives
\begin{equation}
\cls(a) = \cls \dt a,
\end{equation}
where $\cls$ is the spin trivector
\begin{equation}
\cls= \half \psi I\gk \psirev.
\end{equation}
In this case the contraction of the spin tensor does vanish:
\begin{equation}
\da \dt (\cls \dt a) = (\da \wdg a) \dt \cls = 0.
\end{equation} 

There is a remarkable consistency loop at work here.  The Dirac action
gives rise to a spin tensor of just the right type to ensure that the
minimally-coupled action produces the minimally-coupled equation.
\textit{But this is only true if the gravitational action is given by
the Ricci scalar}!  No higher-order gravitational action is consistent
in this way.  So, if we demand that the minimally-coupled field
equations should be derivable from an action principle, we are led to
a highly constrained theory.  This rules out, for example, the type of
`$R+R^2$' Lagrangian often considered in the context of Poincar\'{e}
gauge theory~\cite{rau82,hec91,kho92}.  In addition, the spin sector
is also tightly constrained. Satisfyingly, these constraints force us
to a theory that is first-order in the derivatives of the fields,
keeping the theory on a similar footing to the Dirac and Maxwell
theories.

The only freedom in the action for the gravitational fields is the
possible inclusion of a cosmological constant $\Lambda$.  This just
enters the action integral~\eqref{F-act} as the term $-\Lambda\dhi$.  The
presence of such a term does not alter equation~\eqref{F8}, but
changes~\eqref{einsteqn} to
\begin{equation}
\clg(a) - \Lam a = \kap \clt(a).
\end{equation}
The presence of a cosmological constant cannot be ruled out on
theoretical grounds alone, and this constant will be included when we
consider applications to cosmology.

Given that the spin is entirely of Dirac type, equation~\eqref{F8} now
takes the form
\begin{equation}
\cld \wdg \ho(a) = \db \wdg (b \dt \cld \ho(a)) = \kap \cls \dt \ho(a).
\label{F10}
\end{equation} 
This is the second of our gravitational field equations.  Between
them, equations~\eqref{einsteqn} and~\eqref{F10} define a set of 40
scalar equations for the 40 unknowns in $\ho(a)$ and $\Om(a)$.  Both
equations are manifestly covariant.  In the spin-torsion extension of
general relativity (the Einstein--Cartan--Sciama--Kibble theory),
$\cld\wedge\ho(a)$ would be identified as the gravitational torsion, and
equation~\eqref{F10} would be viewed as identifying the torsion with
the matter spin density.  Of course, in GTG torsion is not a property
of the underlying spacetime, it simply represents a feature of the
gravitational gauge fields.  Equation~\eqref{F10} generalises to the
case of an arbitrary vector $a=a(x)$ as follows:
\begin{equation}
\cld \wdg \ho(a) = \ho(\grad \wdg a) + \kap \cls \dt \ho(a).
\label{wdgeqn} 
\end{equation} 
Of particular use is the case where $a$ is the coordinate frame vector
$e^\mu$, so that we can write
\begin{equation}
\cld \wdg g^\mu = \kap \cls \dt g^\mu.
\end{equation}

\subsection{Covariant forms of the field equations}

For all the applications considered in this paper the gravitational
fields are generated by matter fields with vanishing spin.  So, to
simplify matters, we henceforth set $\cls$ to zero and work with the
second of the field equations in the form
\begin{equation}
\cld \wdg \ho(a) = \ho(\grad \wdg a).
\label{wdgnotors}
\end{equation} 
It is not hard to make the necessary generalisations in the presence
of spin.  Indeed, even if the spin-torsion sector is significant, one
can introduce a new field~\cite{DGL98-spintor}
\begin{equation}
\om'(a) \eqv \om(a) - \half \kap a \dt \cls
\end{equation}
and then the modified covariant derivative with $\om(a)$ replaced by
$\om'(a)$ still satisfies equation~\eqref{wdgnotors}.

The approach we adopt in this paper is to concentrate on the
quantities that are covariant under displacements.  Since both
$\ho(\grad)$ and $\om(a)$ satisfy this requirement, these are the
quantities with which we would like to express the field equations.
To this end we define the operator
\begin{equation}
L_a \eqv a \dt \ho(\grad)
\end{equation} 
and, for the remainder of this section, the vectors $a$, $b$ \etc\ are
assumed to be arbitrary functions of position.  From
equation~\eqref{wdgnotors} we write
\begin{align}
\ho(\dgrad) \wdg \dho(c) &= -\dd \wdg \bigl(\om(d) \dt \ho(c) \bigr)
\nn \\ 
\implies \; \la b \wdg a \, \ho(\dgrad) \wdg \dho(c) \ra &= - \bigl\la
b \wdg a \, \dd \wdg \bigl(\om(d) \dt \ho(c)\bigr) \bigr\ra  \nn \\
\implies \; \bigl(\dL_a \dhu(b) - \dL_b \dhu(a)\bigr) \dt c &= \bigl(a
\dt \om(b) - b \dt \om(a)\bigr) \dt \ho(c)
\label{F-i4}
\end{align}
where, as usual, the overdots determine the scope of a differential
operator.  It follows that the commutator of $L_a$ and $L_b$ is
\begin{align}
[L_a, L_b] 
&= \bigl(L_a \hu(b) - L_b \hu(a)\bigr)\dt \grad \nn \\
&= \bigl(\dL_a \dhu(b) - \dL_b \dhu(a)\bigr) \dt \grad + (L_a b - L_b a) \dt
\ho(\grad) \nn \\
&= \bigl(a \dt \om(b) - b \dt \om(a) + L_a b - L_b a\bigr) \dt \ho(\grad).
\end{align}
We can therefore write
\begin{equation}
[L_a, L_b] = L_c,
\label{F-i55}
\end{equation}
where
\begin{equation}
c = a \dt \om(b) - b \dt \om(a) + L_a b - L_b a = a\dt\cld b -
b\dt\cld a.
\label{F-i6}
\end{equation} 
This `bracket' structure summarises the intrinsic content of~\eqref{F10}.

The general technique we use for studying the field equations is to
let $\om(a)$ contain a set of arbitrary functions, and then
use~\eqref{F-i6} to find relations between them.  Fundamental to this
approach is the construction of the Riemann tensor $\clr(B)$, which
contains a great deal of covariant information.  From the definition
of the Riemann tensor~\eqref{F1} we find that
\begin{align}
\clr(a \wdg b) 
&= \dL_a \dOm( \hu(b)) - \dL_b \dOm (\hu(a)) + \om(a) \crs \om(b) \nn \\
&= L_a \om(b) - L_b \om(a) + \om(a) \crs \om(b) - \Om(L_a \hu(b) -
L_b \hu(a)),
\end{align}
hence
\begin{equation}
\clr(a \wdg b) = L_a  \om(b) - L_b \om(a) + \om(a) \crs \om(b) -
\om(c),
\label{F-i8}
\end{equation} 
where $c$ is given by equation~\eqref{F-i6}.  Equation~\eqref{F-i8} now
enables $\clr(B)$ to be calculated in terms of position-gauge
covariant variables.

\subsubsection*{Solution of the `wedge' equation} 

Equation~\eqref{wdgnotors} can be solved to obtain $\om(a)$ as a
function of $\ho$ and its derivatives.  We define
\begin{equation}
H(a) \eqv \ho(\dgrad \wdg \dot{\ho}^{-1}(a)) = - \ho(\dgrad) \wdg
\dho\,\ho^{-1}(a), 
\label{F11.5}
\end{equation} 
so that equation~\eqref{wdgnotors} becomes
\begin{equation}
\db \wdg (\om(b) \dt a) = H(a).
\label{F12}
\end{equation}
We solve this by first `protracting' with $\da$ to give
\begin{equation}
\da \wdg \db \wdg (\om(b) \dt a) = 2 \db \wdg \om(b) = \db \wdg H(b).
\end{equation}
Now, taking the inner product with $a$ again, we obtain
\begin{equation}
\om(a) - \db \wdg (a \dt \om(b)) = \half a \dt (\db \wdg H(b)).
\end{equation}
Hence, using equation~\eqref{F12} again, we find that
\begin{equation}
\om(a) = -H(a) + \half a \dt (\db \wdg H(b)) .
\label{F15}
\end{equation}
In the presence of spin the term $\half\kap a\dt\cls$ is added to the
right-hand side.

\subsection{Point-particle trajectories}
\label{F-PP}

The dynamics of a fermion in a gravitational background are described
by the Dirac equation~\eqref{G-d19} together with the
quantum-mechanical rules for constructing observables.  For many
applications, however, it is useful to work with classical and
semi-classical approximations to the full quantum theory.  The full
derivation of the semi-classical approximation is described
in~\cite{gap}, but the essential idea is to specialise to motion along
a single streamline defined by the Dirac current $\psi\go\psirev$.
Thus the particle is described by a trajectory $x(\lam)$, together
with a spinor $\psi(\lam)$ which contains information about the
velocity and spin of the particle.  The covariant velocity is
$\hu^{-1}(x')$ where, for this and the following subsection, dashes
are used to denote the derivative with respect to $\lam$.  The
covariant velocity is identified with $\psi\go\psirev$ and the
Lagrange multiplier $p$ is included in the action integral to enforce
this identification.  Finally, an einbein $e$ is introduced to ensure
reparameterisation invariance.  The resultant action is
\begin{equation}
S = \int d\lam \, \la \psi' \isk\psirev + \half \Om(x')
\psi\isk\psirev + p(v - m e \psi\go\psirev) +m^2 e \ra,
\label{F-pp1}
\end{equation}
where
\begin{equation}
v \eqv \hu^{-1}(x').
\end{equation}
The equations of motion arising from~\eqref{F-pp1} are discussed
elsewhere~\cite{gap}.  (An effect worth noting is that, due to the
spin of the particle, the velocity $v$ and momentum $p$ are not
collinear.)

We can make a full classical approximation by neglecting the spin
(dropping all the terms containing $\psi$) and replacing
$\psi\go\psirev$ by $p/m$.  This process leads to the action
\begin{equation}
S = \int d\lam \, \bigl( p \dt \hu^{-1}(x') - \half e (p^2-m^2) \bigr) .
\label{F-pp2}
\end{equation}
The equations of motion derived from~\eqref{F-pp2} are
\begin{align}
v &= e p  \\
p^2 &= m^2 \\ 
\partial_\lam \ho^{-1}(p) &= \dgrad p \dt \dot{\hu}^{-1}(x').
\label{F-pp2.5}
\end{align}
The final  equation yields
\begin{align}
\partial_\lam p &= \ho(\dgrad) \, p \dt \dot{\hu}^{-1}(x') - 
x' \dt \dgrad \,  \ho\bigl( \dot{\ho}^{-1}(p) \bigr) \nn \\
&= \ho\bigl( (\dgrad \wdg \dot{\ho}^{-1}(p)) \dt \hu(v) \bigr) \nn \\
&= H(p) \dt v 
\end{align}
where $H(a)$ is defined by equation~\eqref{F11.5} and we have used
equation~\eqref{1fnadj}.  From equation~\eqref{F15} we see that
$a\dt\om(a)=-a\dt H(a)$, hence
\begin{equation}
e \partial_\lam (v/e) = -\om(v)\dt v.
\end{equation}
This is the classical equation for a point-particle trajectory.  It
takes its simplest form when $\lam$ is the proper time $\tau$ along
the trajectory.  In this case $e=1/m$ and the equation becomes
\begin{equation}
\deriv{v}{\tau} = -\om(v)\dt v,
\end{equation}
or, in manifestly covariant form,
\begin{equation}
v \dt \cld \, v = 0.
\label{PPeom}
\end{equation}
This equation applies equally for massive particles ($v^2=1$) and
photons ($v^2=0$).  Since equation~\eqref{PPeom} incorporates only
gravitational effects, any deviation of $v\dt\cld \, v$ from zero can be
viewed as the particle's acceleration and must result from additional
external forces.

Equation~\eqref{PPeom} is usually derived from the action
\begin{equation}
S = m \int d\lam \, \sqrt{\hu^{-1}(x')^2} = m \int d\lam \, \left(
g_{\mu\nu} \deriv{x^\mu}{\lam} \deriv{x^\nu}{\lam} \right)^{1/2}  ,
\end{equation}
which is obtained from~\eqref{F-pp2} by eliminating $p$ and $e$ with
their respective equations of motion.  A Hamiltonian form such
as~\eqref{F-pp2} is rarely seen in conventional general relativity,
since its analogue would require the introduction of a vierbein.
Despite this, the action~\eqref{F-pp2} has many useful features,
especially when it comes to extracting conservation laws.  For
example, contracting equation~\eqref{F-pp2.5} with a constant vector
$a$ yields
\begin{equation}
\partial_\lam \bigl( a \dt \ho^{-1}(p) \bigr) = a\dt \dgrad \,  p \dt
\dot{\hu}^{-1}(x' ). 
\label{PP-cons}
\end{equation}
It follows that, if the $\lih$-field is invariant under translations
in the direction $a$, then the quantity $a\cdot\ho^{-1}(p)$ is
conserved.  In Section~\ref{SYMMCONS} we show that this result extends
to the case where $\hu^{-1}(a)$ is a Killing vector.

\subsection{The equivalence principle and the Newtonian limit}

In the preceeding section we derived the equation $v\dt\cld\, v=0$ from
the classical limit of the Dirac action.  This equation is the GTG
analogue of the geodesic equation (see Appendix~\ref{app-tens}).
Arriving at such an equation shows that GTG embodies the weak
equivalence principle --- the motion of a test particle in a
gravitational field is independent of its mass.  The derivation also
shows the limitations of this result, which only applies in the
classical, spinless approximation to quantum theory.  The strong
equivalence principle, that the laws of physics in a freely-falling
frame are (locally) the same as those of special relativity, is also
embodied in GTG through the application of the minimal coupling
procedure.  Indeed, it is clear that both of these `principles' are
the result of the minimal coupling procedure.  Minimal coupling
ensures, through the Dirac equation, that point-particle trajectories
are independent of particle mass in the absence of other forces.  It
also tells us how the gravitational fields couple to any matter field.
As we have seen, this principle, coupled with the requirement of
consistency with an action principle, is sufficient to specify the
theory uniquely (up to an unspecified cosmological constant).

The relationship between the minimal coupling procedure (the gauge
principle) and the equivalence principle has been widely discussed
previously.  Feynman~\cite{fey-grav}, for example, argues that a more
exact version of the equivalence between linear acceleration and a
gravitational field requires an equation of the form
\begin{equation}
\mbox{gravity}' = \mbox{gravity} + \mbox{acceleration},
\end{equation}
which clearly resembles a gauge transformation.  What is not often
stressed is the viewpoint presented here, which is that if gravity is
constructed entirely as a gauge theory, there is no need to invoke the
equivalence principle; the physical effects embodied in the principle
are simply consequences of the gauge theory approach.  This further
illustrates the different conceptual foundations of GTG and general
relativity.  Similarly, there is no need for the principle of general
covariance in GTG, which is replaced by the requirement that all
physical predictions be gauge-invariant.  It is often argued that the
principle of general covariance is empty, because any physical theory
can be written in a covariant form.  This objection cannot be levelled
at the statement that all physical predictions must be gauge
invariant, which has clear mathematical and physical content.

The simplest, classical measurements in GTG are modelled by assuming
that observers can be treated as frames attached to a single
worldline.  If this worldline is written as $x(\lam)$, then the
covariant velocity is $v=\hu^{-1}(x')$, and the affine parameter
for the trajectory is that which ensures that $v^2=1$.  The affine
parameter models the clock time for an observer on this trajectory.
Of course, there are many hidden assumptions in adopting this as a
realistic model --- quantum effects are ignored, as is the physical
extent of the observer --- but is is certainly a good model in weak
fields.  In strong fields a more satisfactory model would involve
solving the Dirac equation to find the energy levels of an atom in the
gravitational background and use this to model an atomic clock.

Equation~\eqref{PPeom} enables us to make classical predictions for
freely-falling trajectories in GTG, and the photon case ($v^2=0$) can
be used to model signalling between observers.  As an example,
consider the formula for the redshift induced by the gravitational
fields.  Suppose that a source of radiation follows a worldline
$x_1(\tau_1)$, with covariant velocity $v_1=\hu^{-1}(x_1')$.  The
radiation emitted follows a null trajectory with covariant velocity
$u$ ($u^2=0$).  This radiation is received by an observer with a
worldline $x_2(\tau_2)$ and covariant velocity
$v_2=\hu^{-1}(x_2')$.  The spectral shift $z$ is determined by the
ratio of the frequency observed at the source, $u(x_1)\dt v_1$, and
the frequency observed at the receiver, $u(x_2)\dt v_2$, by
\begin{equation} 
1 + z \eqv \frac{u(x_1) \dt v_1}{u(x_2) \dt v_2}.
\label{PPredshift}
\end{equation}
This quantity is physically observable since the right-hand side is a
gauge-invariant quantity.  This is because each of the four vectors
appearing in~\eqref{PPredshift} is covariant, which eliminates any
dependence on position gauge, and taking the dot product between pairs
of vectors eliminates any dependence on the rotation gauge.

A final point to address regarding the foundations of GTG is the
recovery of the Newtonian limit.  Derivations of this are easily
produced by adapting the standard works in general relativity .
Furthermore, in Section~\ref{SS-STAT} we show that the description of
the gravitational fields outside a static, spherically symmetric star
is precisely the same as in general relativity.  The trajectories
defined by equation~\eqref{PPeom} are those predicted by general
relativity, so all of the predictions for planetary orbits (including
those for binary pulsars) are unchanged.  Similarly, the results for
the bending of light are the same as in general relativity.  As we
show in the applications, any differences between GTG and general
relativity emerge through the relationship with quantum theory, and
through the global nature of the gauge fields in GTG.  These
differences have no consequences for classical tests of general
relativity, though they are potentially testable through the
interaction with quantum spin, and are certainly significant for
discussing more fundamental aspects of gravitational physics.


\section{Symmetries, invariants and conservation laws}

Having determined the gravitational gauge fields and their field
equations, we now establish some general results which are applied in
the sections that follow.  Again, we restrict to the case of vanishing
torsion.  The approach we adopt in solving the field equations is to
let $\om(a)$ be an arbitrary function, and then work with a set of
abstract first-order equations for the terms that comprise $\om(a)$.
However, in letting $\om(a)$ be an arbitrary function, we lose some of
the information contained in the `wedge' equation~\eqref{wdgnotors}.  This
information is recovered by enforcing various properties that the
fields must satisfy, including the symmetry properties of $\clr(B)$
and the Bianchi identities.  In addition, it is often necessary to
enforce some gauge-fixing conditions.  For the rotation gauge these
conditions are applied by studying $\clr(B)$, so it is important to
analyse its general structure.

We start with the result that, for an arbitrary multivector $A(x)$,
\begin{equation}
\cld \wdg (\cld \wdg A) = \cld \wdg \bigl( \ho(\grad \wdg \ho^{-1}(A))
\bigr) = \ho(\grad \wdg \grad \wdg \ho^{-1}(A)) = 0,
\end{equation} 
where we have made use of equation~\eqref{wdgnotors}.  It then follows
from the result
\begin{align}
g^\mu \wdg \bigl( \cld_\mu ( g^\nu \wdg (\cld_\nu A)) \bigr)
&= g^\mu \wdg g^\nu \wdg ( \cld_\mu \cld_\nu A ) \nn \\
&= \half \ho(e^\mu) \wdg
\ho(e^\nu) \wdg ( \liR(e_\mu \wdg e_\nu) \crs A),
\end{align}
that, for any multivector $A$,
\begin{equation}
\da \wdg \db \wdg (\clr(a \wdg b) \crs A) = 0.
\label{S-3}
\end{equation}
This derivation illustrates a useful point.  Many derivations can be
performed most efficiently by working with the $\cld_\mu$, since these
contain commuting partial derivatives.  However, the final expressions
take their most transparent form when the $\lih$-field is included so
that only fully covariant quantities are employed.

If we now set $A$ in equation~\eqref{S-3} equal to a vector $c$, and
protract with $\dc$, we find that
\begin{equation}
\dc \wdg \da \wdg \db \wdg (\clr(a \wdg b) \crs c) = -2 \da \wdg \db
\wdg \clr(a \wdg b) = 0.
\end{equation}
Taking the inner product of the term on the right-hand side with $c$
we obtain
\begin{equation}
c \dt \bigl( \da \wdg \db \wdg \clr(a \wdg b) \bigr) 
= \db \wdg \clr(c \wdg b) - \da \wdg \clr(a \wdg c) - 
\da \wdg \db \wdg (\clr(a \wdg b) \crs c) ,
\end{equation}
in which both the left-hand side and the final term on the right-hand
side vanish.  We are therefore left with the simple expression
\begin{equation}
\da \wdg \clr(a \wdg b) = 0,
\label{S-Rsym}
\end{equation}
which summarises all the symmetries of $\clr(B)$.  This equation says
that the trivector $\da\wedge\clr(a\wedge b)$ vanishes for all values
of the vector $b$, so gives a set of $4 \times 4=16$ equations.  These
reduce the number of independent degrees of freedom in $\clr(B)$ from
36 to the expected 20.  It should be clear from the ease with which
the degrees of freedom are calculated that the present geometric
algebra formulation has many advantages over traditional tensor
calculus.

\subsection{The Weyl tensor}
\label{symms-Weyl}

A good example of the power of the present approach is provided by an
analysis of the Riemann and Weyl tensors.  To illustrate this point a
number of examples of $\clr(B)$ for physical systems are included in
this section.  (These are stated here without derivation.)  The first
application of geometric algebra to the analysis of the Riemann tensor
in classical differential geometry was given by Hestenes~\&
Sobczyk~\cite{hes-gc,sob81}.  Here the formalism is developed and
extended for applications relevant to our gauge theory of gravity.

Six of the degrees of freedom in $\clr(B)$ can be removed by arbitrary
gauge rotations.  It follows that $\clr(B)$ can contain only 14
physical degrees of freedom.  To see how these are encoded in
$\clr(B)$ we decompose it into Weyl and `matter' terms.  Since the
contraction of $\clr(a\wedge b)$ results in the Ricci tensor
$\clr(a)$, we expect that $\clr(a\wdg b)$ will contain a term in
$\clr(a) \wdg b$.  This must be matched with a term in $a \wdg
\clr(b)$, since it is only the sum of these that is a function of
$a\wdg b$.  Contracting this sum we obtain
\begin{align}
\da \dt ( \clr(a) \wdg b + a \wdg \clr(b)) 
&= b \clr - \clr(b) + 4 \clr(b) - \clr(b) \nn \\
&= 2 \clr(b) + b \clr,
\end{align}
and it follows that
\begin{equation}
\da \dt \bigl( \half \left( \clr(a) \wdg b + a \wdg \clr(b) \right) -
{\textstyle \frac{1}{6}} a \wdg b \clr \bigr) = \clr(b).
\end{equation}
We can therefore write
\begin{equation}
\clr(a \wdg b) = \clw(a \wdg b) + \half \bigl(\clr(a) \wdg b + a \wdg
\clr(b) \bigr) - {\textstyle \frac{1}{6}} a \wdg b \, \clr,
\label{S-RWplusM}
\end{equation}
where $\clw(B)$ is the Weyl tensor, and must satisfy
\begin{equation}
\da \dt \clw(a \wdg b) = 0.
\label{S-W1}
\end{equation}

Returning to equation~\eqref{S-Rsym} and contracting, we obtain
\begin{equation}
\db \dt \bigl( \da \wdg \clr(a \wdg b) \bigr) = \da \wdg \clr(a) = 0, 
\end{equation}
which shows that the Ricci tensor $\clr(a)$ is symmetric.  It follows
that
\begin{equation}
\da \wdg \bigl( \half ( \clr(a) \wdg b + a \wdg \clr(b)) -
{\textstyle \frac{1}{6}} a \wdg b \, \clr  \bigr) = 0
\end{equation}
and hence that
\begin{equation}
\da \wdg \clw(a \wdg b) = 0.
\label{S-W2}
\end{equation}
Equations~\eqref{S-W1} and~\eqref{S-W2} combine to give the single equation
\begin{equation}
\da \clw(a \wdg b) = 0.
\label{S-Wsym}
\end{equation}
Since the $\da\cdot$ operation is called the `contraction', and
$\da\wedge$ the `protraction', Hestenes \& Sobczyk~\cite{hes-gc} have
suggested that the sum of these be termed the `traction'.
Equation~\eqref{S-RWplusM} thus decomposes $\clr(B)$ into a `tractionless'
term $\clw(B)$ and a term specified solely by the matter stress-energy
tensor (which determines $\clr(a)$ through the Einstein tensor
$\clg(a)$).  There is no generally accepted name for the part of
$\clr(B)$ that is not given by the Weyl tensor so, as it is entirely
determined by the matter stress-energy tensor, we refer to it as the
matter term.

\subsection{Duality of the Weyl tensor}

To study the consequences of equation~\eqref{S-Wsym} it is useful to
employ the fixed $\{\gamdm\}$ frame, so that equation~\eqref{S-Wsym}
produces the four equations
\begin{align}
\si\clw(\si) + \sj\clw(\sj) + \sk\clw(\sk) &= 0 \label{S-5} \\ 
\si\clw(\si) - \isj\clw(\isj) - \isk\clw(\isk) &= 0 \\ 
-\isi\clw(\isi) + \sj\clw(\sj) - \isk\clw(\isk) &= 0 \\ 
-\isi\clw(\isi) - \isj\clw(\isj) + \sk\clw(\sk) &= 0. \label{S-6}
\end{align}
Summing the final three equations, and using the first, produces
\begin{equation}
I\sigk \clw(I\sigk) = 0
\end{equation}
and substituting this into each of the final three equations produces
\begin{equation}
\clw(I\sigk) = I \clw(\sigk).
\end{equation}
It follows that the Weyl tensor satisfies
\begin{equation}
\clw(IB) = I \clw(B),
\label{S-Wdual}
\end{equation}
and so is `self-dual'.  This use of the term `self-dual' differs
slightly from its use in the 2-spinor formalism of Penrose~\&
Rindler~\cite{pen-I}.  However, the spacetime algebra pseudoscalar $I$
a similar role to the Hodge star operation (the duality
transformation) in differential form theory, so `self-duality' is
clearly an appropriate name for the relation expressed by
equation~\eqref{S-Wdual}.

The fact that tractionless linear functions mapping bivectors to
bivectors in spacetime satisfy equation~\eqref{S-Wdual} was first noted
in~\cite{hes-gc}.  Equation~\eqref{S-Wdual} means that $\clw(B)$ can be
analysed as a linear function on a three-dimensional complex space
rather as a function on a real six-dimensional space.  This is why
complex formalisms, such as the Newman--Penrose formalism, are so
successful for studying vacuum solutions.  The unit imaginary employed
in the Newman--Penrose formalism is a disguised version of the
spacetime pseudoscalar~\cite{DGL-polspin}.  Geometric algebra reveals
the geometric origin of this `imaginary' unit, and enables us to
employ results from complex analysis without the need for formal
complexification.  Furthermore, the complex structure only arises in
situations where it is geometrically significant, instead of being
formally present in all calculations.

Given the self-duality of the Weyl tensor, the remaining content of
equations \eqref{S-5}--\eqref{S-6} is summarised by the relation
\begin{equation}
\sigk \clw(\sigk) = 0.
\end{equation}
This equation says that, viewed as a three-dimensional complex linear
function, $\clw(B)$ is symmetric and traceless.  This gives $\clw(B)$
five complex, or ten real degrees of freedom.  (Since we frequently
encounter combinations of the form scalar $+$ pseudoscalar, we refer
to these loosely as `complex' scalars.)  The gauge-invariant information
in $\clw(B)$ is held in its complex eigenvalues and, since the sum of
these is zero, only two are independent.  This leaves a set of four
real intrinsic scalar quantities.

Overall, $\clr(B)$ has 20 degrees of freedom, 6 of which are contained
in the freedom to perform arbitrary local rotations.  Of the remaining
14 physical degrees of freedom, four are contained in the two complex
eigenvalues of $\clw(B)$, and a further four in the real eigenvalues
of the matter stress-energy tensor.  The six remaining physical
degrees of freedom determine the rotation between the frame that
diagonalises $\clg(a)$ and the frame that diagonalises $\clw(B)$.
This identification of the physical degrees of freedom contained in
$\clr(B)$ appears to be new, and is potentially very significant.

\subsection{The Petrov classification}
 
The algebraic properties of the Weyl tensor are traditionally encoded
in its Petrov type.  Here we present geometric algebra expressions for
the main Petrov types (following the conventions of Kramer \textit{et
al.}~\cite{kra-exact}).  The Petrov classification is based on the
solutions of the eigenvalue equation
\begin{equation}
\clw(B) = \alp B,
\end{equation}
in which $B$ is a bivector (the `eigenbivector') and $\alp$ is a
complex scalar.  There are five Petrov types: \slI, \slII, \slIII,
\slD\ and \slN.  Type~\slI\ are the most general, with two independent
eigenvalues and three linearly-independent orthogonal eigenbivectors.
Such tensors have the general form
\begin{equation}
\clw(B) = \half \alp_1 (B + 3 F_1 B F_1) + \half \alp_2 (B + 3 F_2 B
F_2) 
\end{equation}
where $\alp_1$ and $\alp_2$ are complex scalars and $F_1$ and $F_2$
are orthogonal unit bivectors ($F_1^2=F_2^2=1$).  The eigenbivectors
are $F_1$, $F_2$ and $F_3\eqv F_1F_2$, and the corresponding eigenvalues
are $(2\alp_1-\alp_2)$, $(2\alp_2-\alp_1)$ and $-(\alp_1+\alp_2)$.

Type~\slD\ (degenerate) are a special case of type~\slI\ tensors in
which two of the eigenvalues are the same.  Physical examples are
provided by the Schwarzschild and Kerr solutions.  The region outside
a spherically-symmetric source of mass $M$ has
\begin{equation}
\clr(B) = \clw(B) = - \frac{M}{2r^3} (B + 3 \sig_r B \sig_r)
\label{S-Rswz}
\end{equation}
where $\sig_r$ is the unit radial bivector.  The eigenbivectors of
this function are $\sig_r$, with eigenvalue $-2M/r^3$, and any two
bivectors perpendicular to $\sig_r$, with eigenvalue $M/r^3$.
Similarly, $\clr(B)$ for a stationary axisymmetric source described by
the Kerr solution is~\cite{gap,dor-thesis}
\begin{equation}
\clr(B) = \clw(B) = - \frac{M}{2(r-I L\cos\!\theta)^3} (B + 3 \sig_r B
\sig_r) .
\label{S-Rkerr}
\end{equation}
This differs from the radially-symmetric case~\eqref{S-Rswz} only in that
its eigenvalues contain an imaginary term governed by the angular
momentum $L$.  Verifying that~\eqref{S-Rswz} and~\eqref{S-Rkerr} are
tractionless is simple, requiring only the result that, for an
arbitrary bivector $B$,
\begin{equation}
\da F a \wdg b = \da F (ab - a \dt b) = - b F,
\end{equation}
which employs equation~\eqref{O-gc5} from Section~\ref{O-GC}.

The fact that the Riemann tensor for the Kerr solution is obtained
from that for the Schwarzschild solution by replacing $r$ by
$r-IL\cos\!\theta$ is reminiscent of a `trick' used to derive the Kerr
solution in the null tetrad formalism~\cite{new65}.  This is
particularly suggestive given that the unit imaginary employed in the
Newman--Penrose formalism is a disguised version of the spacetime
pseudoscalar $I$.  The significance of these observations is discussed
further in~\cite{DL04-KS,gap}.

For tensors of Petrov type other than~\slI, null bivectors play a
significant role.  Type~\slII\ tensors have eigenvalues $\alp_1$,
$-\alp_1$ and 0 and two independent eigenbivectors, one timelike and
one null.  Type~\slIII\ and type~\slN\ have all three eigenvalues zero,
and satisfy 
\begin{eqnarray}
\mbox{type~\slIII:} & & \clw^3(B) = 0,  \hs{0.6} \clw^2(B) \neq 0 \\ 
\mbox{type~\slN:} & & \clw^2(B) = 0.
\end{eqnarray}
An example of a type~\slN\ tensor is provided by gravitational
radiation.   For a plane-polarised gravitational wave
travelling in the $\gk$ direction $\clr(B)$ is given by
\begin{equation}
\clr^+(B) = \clw^+(B) = \qrt f(t-z) \, \gam_+( \gi B \gi - \gj B \gj
)\gam_+ 
\end{equation}
for waves polarised in the direction of the $\gi$ and $\gj$ axes,
and
\begin{equation}
\clr^\times(B) = \clw^\times(B) = \qrt f(t-z) \, \gam_+ ( \gi B \gj +
\gj B \gi ) \gam_+
\end{equation}
for waves polarised at $45^\circ$ to the axes.  In both cases $f(t-z)$
is a scalar function, and $\gam_+$ is the null vector
\begin{equation}
\gam_+ \eqv \go + \gk.
\end{equation}
The direct appearance of the null vector $\gam_+$ in $\clw(B)$ shows
that $\clw^2(B)=0$, and is physically very suggestive.  Expressions of
the type $\gam_+ B \gam_+$ project the bivector $B$ down the null
vector $\gam_+$, and such a structure is exhibited in the radiation
field generated by an accelerating point charge~\cite{DGL93-notreal}.

These examples illustrate the uniquely compact forms for
the Riemann tensor afforded by geometric algebra.  In terms of both
clarity and physical insight these expressions are far superior to
any afforded by tensor algebra, the Newman--Penrose formalism or
differential forms.  Only Wahlquist and Estabrook's (3+1) dyadic
notation~\cite{wah65,wah92} achieves expressions of comparable
compactness, although their formalism is of limited applicability.

\subsection{The Bianchi identities}

Further information from the wedge equation~\eqref{wdgnotors} is contained
in the Bianchi identity.  One form of this follows from a simple
application of the Jacobi identity:
\begin{equation}
[\cld_\alp, [\cld_\bet,\cld_\gam ]] A + \mbox{cyclic permutations} = 0
\hs{2} 
\end{equation}
which implies that
\begin{equation}
\cld_\alp \liR(e_\bet \wdg e_\gam) + \mbox{cyclic permutations} = 0.
\label{S-Bi1}
\end{equation}
Again, use of the $\cld_\mu$ derivatives makes this identity
straightforward, but more work is required to achieve a fully
covariant relation.

We start by forming the adjoint relation to~\eqref{S-Bi1}, which we
can write as
\begin{equation}
\da \wdg \db \wdg \dc \la \bigl( a \dt \grad \liR(b\wdg c) + \Om(a)
\crs \liR(b \wdg c) \bigr) B \ra = 0,
\label{S-Bi0}
\end{equation}
where $B$ is a constant bivector.  We next need to establish the
result that
\begin{equation}
B_1 \dt \clr(B_2) = B_2 \dt \clr (B_1).
\label{S-Bi2}
\end{equation}
This follows by contracting equation~\eqref{S-Rsym} with an
arbitrary vector and an arbitrary bivector to obtain the equations
\begin{align}
\dc \wdg ( a \dt \clr(c \wdg b)) &= \clr (a \wdg b) \\
(B \dt \da) \dt \clr(a \wdg b) &= - \da B \dt \clr(a \wdg b).
\end{align}
Protracting the second of these equations with $\db$ and using the
first, we obtain
\begin{equation}
\db \wdg \bigl( (B\dt\da) \dt \clr( a \wdg b) \bigr) = -2 \clr(B) = 
- \db \wdg \da B \dt \clr(a \wdg b).
\end{equation}
Taking the scalar product with a second bivector now gives
equation~\eqref{S-Bi2}.

Using this result in equation~\eqref{S-Bi0}, we now obtain
\begin{equation}
\grad \wdg \bigl( \ho^{-1}(\clr(B)) \bigr) - \da \wdg
\ho^{-1}(\clr(\Om(a) \crs B)) =0. 
\end{equation}
Finally, acting on this equation with $\ho$ and using
equation~\eqref{wdgnotors}, we establish the covariant result
\begin{equation}
\cld \wdg \clr(B) - \da \wdg \clr(\om(a) \crs B) = 0.
\end{equation}
This result takes a more natural form when $B$ becomes an arbitrary
function of position, and we write the Bianchi identity as
\begin{equation}
\da \wdg \bigl( a \dt \cld \, \clr(B) - \clr(a \dt\cld \, B) \bigr) = 0.
\label{S-Bianchi}
\end{equation}
We can extend the overdot notation of Section~\eqref{O-GC} in the obvious
manner to write equation~\eqref{S-Bianchi} as
\begin{equation}
\cldd \wdg \dot{\clr} (B) = 0,
\end{equation}
which is very compact, but somewhat symbolic and hard to apply without
unwrapping into the form of equation~\eqref{S-Bianchi}.

The self-duality of the Weyl tensor implies that
\begin{equation}
\cldd \wdg \dot{\clw}(I\,B) = -I \cldd \dt \dot{\clw}(B),
\end{equation}
so, in situations where the matter vanishes and $\clw(B)$ is the only
contribution to $\clr(B)$, the Bianchi identities reduce to
\begin{equation}
\cldd \dot{\clw}(B) = 0.
\end{equation}
The properties of a first-order equation such as this are discussed in
more detail in Section~\ref{charac}.

The contracted Bianchi identities are obtained from
\begin{align}
(\da \wdg \db) \dt \bigl( \cldd \wdg \dot{\clr} (a \wdg b) \bigr)
&= \da \dt \bigl( \dot{\clr}(a \wdg \cldd) + \cldd  \dot{\clr}(a) \bigr) \nn \\
&= 2 \dot{\clr}(\cldd) - \cld \clr,
\end{align}
from which we can write
\begin{equation}
\dot{\clg}(\cldd) = 0.
\label{S-Bicont1}
\end{equation}
An alternative form of this equation is obtained by taking the scalar
product with an arbitrary vector, and using the symmetry of $\clg(a)$
to write
\begin{equation}
\cldd \dt \dot{\clg}(a) = 0.
\end{equation}
Written out in full, this equation takes the form
\begin{equation}
\da \dt \bigl(L_a \clg(b) - \clg(L_a b) + \om(a) \crs \clg(b) -
\clg(\om(a) \crs b) \bigr) = 0.
\label{S-Bicont}
\end{equation}

\subsection{Symmetries and conservation laws}
\label{SYMMCONS}

We end this section with some comments on symmetries and conservation
laws.  These comments are not all directly relevant to the
applications discussed in this paper, but concern the general
structure of GTG.

The first significant point is that the theory is founded on an action
principle in a `flat' vector space.  It follows that all the familiar
equations relating symmetries of the action to conserved quantities
hold without modification.  (The geometric algebra approach to
Lagrangian field theory is developed in~\cite{DGL93-lft,gap}.)  Any
symmetry transformation of the total action integral that is
parameterised by a continuous scalar will result in a vector that is
conserved with respect to the vector derivative $\grad$.  To every
such vector there corresponds a covariant equivalent, as is seen from
the simple rearrangement
\begin{align}
\cld \dt \clj &= I \cld \wdg (I \clj) \nn \\
&= \dh I \grad \wdg (\ho^{-1}(I \clj))  \nn \\
&= \dh \grad \dt (\hu(\clj) \dhi).
\end{align}
So, if $J$ satisfies $\grad\dt J=0$, then the covariant equivalent
\begin{equation}
\clj = \hu^{-1}(J) \dh 
\end{equation}
satisfies the covariant equation $\cld\dt\clj=0$.  (This explains the
definition of $\clj$ in Section~\ref{FRGF}.)  Note that if we attempt
to form the canonical stress-energy tensor conjugate to translations
we obtain the quantity $\clg(a)-\kap\clt(a)$, which the field
equations set to zero.  The overall stress-energy tensor is therefore
clearly conserved, but this does not yield any new information.

A second feature of our use of flat background spacetime is that all
differential equations can be recast in integral form.  The integral
equation form is not always useful, since it often forces one to deal
with non-covariant quantities.  But integral equations are
particularly well-suited to handling singularities in the
gravitational fields. Just as Gauss' theorem in electromagnetism can
be used to determine the structure of an electric field source, so
integral equations can be used to uncover the structure of the matter
sources of gravitational fields.  An example of this is provided in
Section~\ref{SS-BH} where the Schwarzschild solution is shown to arise
from a matter stress-energy tensor containing a single $\del$-function
source of strength $M$.  A less obvious example is contained
in~\cite{DL04-KS,DGL96-erice}, where it is shown that the matter
generating the Kerr solution takes the form of a ring rotating at the
speed of light, supported by a disk of tension.  Such notions are
quite different from classical general relativity.

Killing vectors have played a significant role in the analysis of
symmetries and conserved quantities in general relativity, and their
properties are largely unchanged in GTG.  The simplest
covariant form of Killing's equation for a Killing vector $K$ is that 
\begin{equation}
a \dt (b \dt \cld K) + b \dt (a \dt \cld K) = 0
\label{S-Kill}
\end{equation}
for any two vector fields $a$ and $b$.  Contracting with $\da\cdot\db$
immediately yields the result that $K$ is divergenceless:
\begin{equation}
\cld \dt K = 0.
\end{equation}
Killing vectors are frequently obtained when, in some coordinate
system, the metric is independent of one of the coordinates.  Suppose
that $g_{\mu\nu} = g_\mu \cdot g_\nu$ is independent of the $x^0$
coordinate.  It follows that
\begin{equation}
\deriv{}{x^0} g_{\mu\nu} = g_\mu \dt (g_0 \dt\cld g_\nu) + g_\nu \dt
(g_0 \dt \cld g_\mu) = 0.
\label{S-K1}
\end{equation}
But, for a coordinate frame, 
\begin{equation}
g_\mu \dt \cld g_\nu - g_\nu \dt \cld g_\mu = \hu^{-1} (\partial_\mu
e_\nu - \partial_\nu e_\mu) = 0
\end{equation}
and using this in equation~\eqref{S-K1} we find that
\begin{equation}
g_\mu \dt (g_\nu \dt \cld K) + g_\nu \dt ( g_\mu \dt \cld K) = 0,
\label{S-K2}
\end{equation}
where $K=g_0$.  Equation~\eqref{S-K2} is entirely equivalent to the
frame-free equation~\eqref{S-Kill}.

A further consequence of equation~\eqref{S-Kill} is that, for any vector
$a$, 
\begin{equation}
a \dt (a \dt \cld K) = 0.  
\end{equation}
So, for a particle satisfying the geodesic equation $v\dt \cld \,
v=0$~\eqref{PPeom}, we see that
\begin{equation}
\partial_\tau (v\dt K) = v \dt \cld (v \dt K) = K \dt (v \dt \cld v) +
v \dt (v \dt \cld K) = 0.
\end{equation}
It follows that the quantity $v\cdot K$ is conserved along the worldline
of a freely-falling particle.


\newpage

\vspace*{5cm} 

\begin{center}
{\LARGE Part II --- Applications}
\end{center}

\newpage

\section{Spherically-symmetric systems}
\label{SS}

The first full application of our formalism is to time-dependent
spherically-symmetric fields.  For simplicity, we consider only the
case where the matter is described by a perfect fluid.  The equations
derived here are applicable to static and collapsing stars,
radially-symmetric black holes and many aspects of cosmology,
including inflation.  Furthermore, in a suitable gauge the relevant
equations are essentially Newtonian in form, making their physical
interpretation quite transparent.  Applications discussed here include
an analytic solution to the equations governing collapsing dust, and
the new understanding of horizons forced by our gauge theory.  This
section includes an extended version of the work presented
in~\cite{DGL-erice}.

\subsection{The `intrinsic' method}

The traditional approach to solving the gravitational field equations
is to start with the metric $g_{\mu \nu}$, which is usually encoded as
a line element
\begin{equation}
ds^2 = g_{\mu\nu} \, dx^\mu \, dx^{\nu}.
\end{equation} 
The analogous quantity in GTG is derived from the $\lih$-function via
\begin{equation}
g_{\mu\nu} = \hu^{-1}(e_\mu) \dt \hu^{-1}(e_\nu),
\label{SS-int1}
\end{equation} 
where the $\{e_{\mu}\}$ comprise a coordinate frame (see also
Appendix~C).  For a given matter stress-energy tensor, the field
equations then yield a set of non-linear, second-order differential
equations for the terms in $g_{\mu\nu}$.  These equations are
notoriously hard to solve.  On the other hand, \textit{any} metric is
potentially a solution to the Einstein equations --- one where the
matter stress-energy tensor is determined by the corresponding
Einstein tensor.  This approach, in which the tail wags the dog, has
recently probably been more popular!  Here we present a new approach
to solving the gravitational field equations.  The method is closely
tied to our gauge-theoretic understanding of gravity, but can always
be used to generate a metric which solves the Einstein equations.  So,
even if one rejects our gauge-theory description of gravity, the
techniques developed below can still be viewed as providing a new
method for studying the Einstein equations.

Under a local rotation the vector $\hu^{-1}(a)$ transforms as
\begin{equation}
\hu^{-1}(a) \mapsto R \hu^{-1}(a) \Rrev.
\end{equation}
It follows that the metric $g_{\mu\nu}$~\eqref{SS-int1} is invariant
under rotation-gauge transformations.  This is in keeping with our
earlier observation that, at the classical level, it is possible to
work with a set of equations that are invariant under rotation-gauge
transformations, and this is precisely what general relativity does.
This approach has the advantage of removing a number of degrees of
freedom from the theory, but one pays a heavy price for this: the
equations become second-order and one has to deal with complicated
non-linear terms.  The approach we develop here is quite different.
We keep the rotation-gauge field explicit, and work entirely with
quantities that transform covariantly under position-gauge
transformations.  Such quantities include $\ho(\grad)$, $\om(a)$ and
$\clr(B)$.  We therefore work with directional derivatives of the form
$L_a=a\dt\ho(\grad)$, and treat $\om(a)$ as an arbitrary field.  The
relationship between $\ho(a)$ and $\om(a)$ is then encoded in the
commutation relations of the $L_a$.  This setup is achieved by
initially making a suitably general ansatz for the $\lih$-function.
This trial form is then substituted into equation~\eqref{F15} to find
the general form of $\om(a)$.  An arbitrary $\om(a)$ field consistent
with this general form is then introduced, resulting in a set of
equations relating commutators of the $L_a$ derivatives to the
variables in $\om(a)$.

Next, the Riemann tensor $\clr(B)$ is constructed in terms of abstract
first-order derivatives of the $\om(a)$ and additional quadratic
terms.  The rotation-gauge freedom is then removed by specifying the
precise form that $\clr(B)$ takes.  For example, one can arrange that
$\clw(B)$ is diagonal in a suitable frame.  This gauge fixing is
crucial in order to arrive at a set of equations that are not
under-constrained.  With $\clr(B)$ suitably fixed, one arrives at a
set of relations between first-order abstract derivatives of the
$\om(a)$, quadratic terms in $\om(a)$, and matter terms.  The final
step is to impose the Bianchi identities, which ensure overall
consistency of the equations with the bracket structure.  Once all
this is achieved, one arrives at a fully `intrinsic' set of equations.
Solving these equations usually involves searching for natural
`integrating factors'.  These integrating factors provide `intrinsic'
coordinates, and many of the fields can be expressed as functions of
these coordinates alone.  The final step is to `coordinatise' the
solution by making an explicit (gauge) choice of the $\lih$-function.
The natural way to do this is to ensure that the coordinates used in
parameterising $\ho(a)$ match the intrinsic coordinates defined by the
integrating factors.

The method outlined above is quite general and can be applied to a
wide range of problems~\cite{gap,DLkerr03,DGL96-cylin}.  Here we
employ it in the analysis of time-dependent spherically-symmetric
systems.

\subsection{The intrinsic field equations}
\label{SSINT}

We start by introducing a set of spherical polar coordinates.  In
terms of the fixed $\{\gamdm\}$ frame we define:
\begin{equation}
\begin{alignedat}{3}
t &\eqv x \dt \go & \qquad \cos\! \theta &\eqv x\dt\gam^3 / r \\ 
r &\eqv \sqrt{(x \wdg \go)^2} & \qquad \tan\! \phi &\eqv
(x\dt\gam^2) / (x\dt\gam^1). 
\end{alignedat}
\end{equation}
The associated coordinate frame is
\begin{equation}
\begin{aligned}
e_t &\eqv \go \\
e_r &\eqv x \wdg \go \, \go /r \\
e_\theta  &\eqv r \cos\! \theta (\cos\!\phi \, \gi + \sin\!\phi
\, \gj) -r \sin\!\theta \, \gk \\
e_\phi &\eqv r \sin\!\theta (-\sin\!\phi \, \gi + \cos\!\phi \, \gj)
\end{aligned}
\end{equation}
and the dual-frame vectors are denoted by $\{e^t, e^r, e^\theta,
e^\phi\}$.  We will also frequently employ the unit vectors $\thht$
and $\phht$ defined by
\begin{equation}
\thht \eqv e_\theta /r, \qquad \phht \eqv e_\phi /(r
\sin\!\theta).
\end{equation}
Associated with these unit vectors are the unit timelike bivectors
\begin{equation}
\sigr \eqv e_r e_t, \qquad \sigth \eqv \thht e_t, \qquad \sigph
\eqv \phht e_t,
\end{equation}
which satisfy
\begin{equation}
\sigr \sigth \sigph = e_t e_r \thht \phht = I.
\end{equation}
The dual spatial bivectors are given by 
\begin{equation}
I \sigr = -\thht \phht, \qquad I\sigth = e_r \phht, \qquad I\sigph = - 
e_r \thht. 
\end{equation}
Throughout we use the abbreviations
\begin{equation}
\dr = \deriv{}{r}, \qquad \dift = \deriv{}{t}.
\end{equation}

\subsubsection*{The {\boldmath $\hu$}-function}

Our first step towards a solution is to decide on a general form of
the $\lih$-function that is consistent with spherical symmetry.
Suppose that $B$ is a constant spatial bivector ($e_t\dt B=0$), and
define
\begin{align}
R &= e^{B/2} \\
x' &= \Rrev x R.
\end{align}
Then, in analogy with electromagnetism, the gravitational fields will
be spherically symmetric if rotating $\ho(a)$ to $R \ho(a) \Rrev$ and
displacing it to the back-rotated position $x'$ leaves $\ho(a)$
unchanged.  Hence rotational symmetry is enforced through the
requirement that 
\begin{equation}
R \ho_{x'}(\Rrev a R) \Rrev = \ho(a).
\end{equation}
This symmetry immediately implies that the $\{e^r,e^t\}$ and
$\{e^\theta, e^\phi\}$ pairs decouple from each other, and the action
of $\ho(a)$ on the $\thht$ and $\phht$ vectors is further restricted
to the form
\begin{equation}
\begin{aligned}
\ho(\thht) &= \alp \thht + \beta \phht \\
\ho(\phht) &= \alp \phht - \beta \thht.
\end{aligned}
\end{equation}
However, the skew-symmetric term parameterised by $\beta$ can always
be removed by a rotation in the $I\sigr$ plane, so we can assume that
$\ho(a)$ is diagonal on $\{e^\theta, e^\phi\}$.  No such assumption
can be made for the $\{e^r,e^t\}$ vectors, so we take $\ho(a)$ as
having the general form
\begin{equation}
\begin{aligned}
\ho(e^t) &= f_1 e^t + f_2 e^r \\
\ho(e^r) &= g_1 e^r + g_2 e^t \\
\ho(e^\theta) &= \alp e^\theta \\
\ho(e^\phi) &= \alp e^\phi,
\label{SS-ho}
\end{aligned}
\end{equation}
where $f_1$, $f_2$, $g_1$, $g_2$ and $\alp$ are all functions of $t$
and $r$ only.  We retain the gauge freedom to perform a boost in the
$\sigr$ direction, and this freedom is employed later to simplify the
equations.  Our remaining position-gauge freedom lies in the freedom
to reparameterise $t$ and $r$, which does not affect the general form
of~\eqref{SS-ho}.  A natural parameterisation will emerge once the
`intrinsic' variables have been identified.

\subsubsection*{The {\boldmath $\om$}-function}

To find a general form $\om(a)$ consistent with~\eqref{SS-ho} we
substitute equation~\eqref{SS-ho} into equation~\eqref{F15} for
$\om(a)$ as a function of $\ho(a)$.  Where the coefficients contain
derivatives of terms from $\ho(a)$ new symbols are introduced.
Undifferentiated terms from $\ho(a)$ appearing in $\om(a)$ are left in
explicitly.  These arise from frame derivatives and the algebra is
usually simpler if they are included directly.  This procedure results
in the following form for $\om(a)$:
\begin{equation}
\begin{aligned}
\om(e_t) &= G e_r e_t \\
\om(e_r) &= F e_r e_t \\
\om(\thht) &= S \thht e_t + (T-\alp /r)e_r \thht \\
\om(\phht) &= S \phht e_t + (T-\alp /r)e_r \phht,
\label{SS-om}
\end{aligned}
\end{equation}
where $G$, $F$, $S$ and $T$ are functions of $t$ and $r$ only.  The
important feature of these functions is that they are position-gauge
covariant.

Substituting this definition for $\om(a)$ into equations~\eqref{F-i55}
and~\eqref{F-i6} we find that the bracket relations are as follows:
\begin{equation}
\begin{alignedat}{2}
[L_t, L_r] &= G L_t - F L_r &\qquad [L_r, L_\thht] &= -T L_\thht  \\
{} [L_t, L_\thht] &= -S L_\thht &\hs{0.8} [L_r, L_\phht] &= -T
L_\phht  \\
{} [L_t, L_\phht] &= -S L_\phht  &\hs{0.8} [L_\theta, L_\phi]
&= 0,  
\end{alignedat}
\label{SS-brckts}
\end{equation}
where
\begin{equation}
\barr{lcl}
L_t \eqv e_t \dt \ho(\grad) & \qquad & L_{\thht} \eqv \thht \dt
\ho(\grad) \\
L_r \eqv e_r\dt \ho(\grad) & \qquad & L_{\phht} \eqv \phht \dt
\ho(\grad). 
\earr
\end{equation}
The use of unit vectors in these derivatives eliminates the need to
calculate irrelevant coordinate derivatives.  A set of bracket
relations such as~\eqref{SS-brckts} is precisely what one aims to achieve
--- all reference to the $\lih$-function is removed and one deals
entirely with position-gauge covariant quantities.

\begin{table}
\begin{center}
\begin{tabular}{r|ccc}
     		&  $\sigr$ 	& $\sigth$ 	& $\sigph$    \\ 
\hline
$e_t \dt\cld$	&  $0$ 		&$G\,I\sigph$  	&$-G I\sigth $ \\
$e_r \dt\cld$	&  $0$ 		&$F\,I\sigph$   &$-F I\sigth $ \\
$\thht\dt\cld$  & $T\,\sigth-S\,I\sigph$ & $-T\,\sigr$ & $S\, I\sigr$ \\
$\phht\dt\cld$  & $T\,\sigph+S\,I\sigth$ & $-S\, I\sigr$ & $-T\,\sigr$ 
\end{tabular}
\end{center}
\caption{\em Covariant derivatives of the polar-frame unit timelike
bivectors.} 
\label{SS-t1}
\end{table}

\subsubsection*{The Riemann tensor}

Having found a suitable form for $\om(a)$ we next use
equation~\eqref{F-i8} to calculate $\clr(B)$.  This derivation is
simplified by judicious use of the results listed in
Table~\ref{SS-t1}.  The only subtlety in the derivation is the removal
of terms involving derivatives of $\alp/r$ using the bracket
relations~\eqref{SS-brckts}.  Since $\alp/r=L_\thht \theta$ we have
\begin{equation}
L_t (\alp /r) = L_t L_\thht \theta = [L_t, L_\thht]
\theta = -S \alp /r
\label{SS-eq1}
\end{equation}
and
\begin{equation}
L_r (\alp /r) = L_r L_\thht \theta = [L_r, L_\thht]
\theta = -T \alp /r.
\label{SS-eq2}
\end{equation}
Application of equation~\eqref{F-i8} is now straightforward, and leads to
the Riemann tensor
\begin{equation}
\begin{aligned}
\clr(\sigr) &= (L_r G - L_t F +G^2 - F^2) \sigr \\
\clr(\sigth) &= (-L_t S +GT -S^2) \sigth + (L_t T +ST-SG) I\sigph \\
\clr(\sigph) &= (-L_t S +GT -S^2) \sigph - (L_t T +ST-SG) I\sigth \\
\clr(I\sigph) &= (L_r T +T^2-FS) I\sigph - (L_r S+ST-FT) \sigth \\
\clr(I\sigth) &= (L_r T +T^2-FS) I\sigth + (L_r S+ST-FT) \sigph \\
\clr(I \sigr) &= (-S^2 + T^2 -(\alp /r)^2) I \sigr.
\label{SS-riem1}
\end{aligned}
\end{equation}

\subsubsection*{The matter field and gauge fixing}

Now that we have found $\clr(B)$ in terms of `intrinsic' functions and
their first derivatives, we must next decide on the form of matter
stress-energy tensor that the gravitational fields couple to.  We
assume that the matter is modelled by an ideal fluid so we can write
\begin{equation}
\clt(a) = (\rho + p) a \dt v v - p a,
\end{equation}
where $\rho$ is the energy density, $p$ is the pressure and $v$ is the
covariant fluid velocity ($v^2=1$).  Radial symmetry means that $v$
can only lie in the $e_t$ and $e_r$ directions, so $v$ must take the
form
\begin{equation}
v = \cosh\!u\, e_t + \sinh\!u\, e_r .
\end{equation}
But, in restricting the $\lih$-function to the form of
equation~\eqref{SS-ho}, we retained the gauge freedom to perform arbitrary
radial boosts.  This freedom can now be employed to set $v=e_t$, so
that the matter stress-energy tensor becomes
\begin{equation}
\clt(a) = (\rho + p) a \dt e_t e_t - p a.
\label{SS-mat}
\end{equation}
There is no physical content in the choice $v=e_t$ as all physical
relations must be independent of gauge choices.  In particular,
setting $v=e_t$ does not mean that the fluid is `at rest', or that we
are `comoving with the fluid'.  An observer comoving with the fluid
will have a covariant velocity $e_t$, but this implies no special
relationship with the time coordinate $t$, since the observer's
trajectory would have $\partial_\lam x=\hu(e_t)$ and nothing has yet
been said about the specific form of $\hu(a)$.

In setting $v=e_t$ all rotation-gauge freedom has finally been
removed.  This is an essential step since all non-physical degrees of
freedom must be removed before one can achieve a complete set of
physical equations.  Note that the rotation gauge has been fixed by
imposing a suitable form for $\clr(B)$, rather than restricting the
form of $\ho(a)$.  The reason for working in this manner is obvious
--- $\clr(B)$ deals directly with physically-measurable quantities,
whereas the algebraic structure of the $\lih$-function is of little
direct physical relevance.

From equation~\eqref{S-RWplusM} the source term in $\clr(B)$ is given by
\begin{equation}
\clr(a \wdg b) - \clw(a \wdg b) = 4 \pi \bigl(a \wdg \clt(b) + \clt(a)
\wdg b - \tthird \clt \,  a \wdg b \bigr)
\end{equation}
where $\clt=\da\dt\clt(a)$ is the trace of the matter stress-energy
tensor.  With $\clt(a)$ given by equation~\eqref{SS-mat}, $\clr(B)$ is
restricted to the form
\begin{equation}
\clr(B) = \clw(B) + 4\pi \bigl( (\rho+p) B\dt e_t e_t -\tthird \rho B\bigr) .
\end{equation}
Comparing this with equation~\eqref{SS-riem1} we find that $\clw(B)$
has the general form
\begin{equation}
\barr{lcl}
\clw (\sigr) = \alp_1 \sr & & \clw(I\sigr) = \alp_4 I \sigr \\
\clw(\sigth) = \alp_2 \sigth + \beta_1 I\sigph & & 
\clw(I\sigth) = \alp_3 I\sigth + \beta_2 \sigph \\
\clw(\sigph) = \alp_2 \sigph - \beta_1 I\sigth & & 
\clw(I\sigph) = \alp_3 I\sigph - \beta_2 \sigth. \\
\earr
\end{equation}
But $\clw(B)$ must be self-dual, so $\alp_1=\alp_4$, $\alp_2 = \alp_3$
and $\bet_1=-\bet_2$.  It must also be symmetric, which implies that
$\bet_1=\bet_2$.  It follows that $\bet_1=\bet_2=0$.  Finally,
$\clw(B)$ must be traceless, which requires that $\alp_1 + 2\alp_2 =
0$.  Taken together, these conditions reduce $\clw(B)$ to the form
\begin{equation}
\clw(B) = \frac{\alp_1}{4} (B + 3 \sigr B \sigr),
\end{equation}
which is of Petrov type~\slD.  It follows from the form of
$\clr(I\sigr)$ that if we set 
\begin{equation}
A \eqv \qrt (-S^2 + T^2 -(\alp/r)^2)
\end{equation}
then the full Riemann tensor must take the form
\begin{equation}
\clr(B) = (A+ \tthird \pi \rho) (B + 3\sig_r B \sig_r) + 
4\pi \bigl((\rho+p) B\dt e_t e_t -\tthird \rho B \bigr) .
\end{equation}
Comparing this with equation~\eqref{SS-riem1} yields the following set of
equations:
\begin{align}
L_t S &= 2A +GT -S^2 -4\pi p \label{SS-eq3} \\
L_t T &= S(G-T) \label{SS-eq3a} \\
L_r S &= T(F-S) \label{SS-eq3b} \\
L_r T &= -2A + FS -T^2 -4\pi \rho \label{SS-eq3c} \\
L_r G - L_t F &= F^2 - G^2 +4A +4 \pi(\rho + p) . 
\label{SS-eq4}
\end{align}

\subsubsection*{The Bianchi identity}

We are now close to our goal of a complete set of intrinsic equations.
The remaining step is to enforce the Bianchi identities.  The
contracted Bianchi identity~\eqref{S-Bicont1} for a perfect fluid results
in the pair of equations
\begin{gather}
\cld \dt (\rho v) + p \cld \dt v = 0 \label{SS-Bi4a} \\
(\rho +p) (v \dt \cld \, v) \wdg v - (\cld p) \wdg v = 0. 
\label{SS-Bi4b}
\end{gather}
Since $(v\cdot\cld \, v)\wedge v$ is the acceleration bivector, the
second of these equations relates the acceleration to the pressure
gradient.  For the case of radially-symmetric fields,
equations~\eqref{SS-Bi4a} and~\eqref{SS-Bi4b} reduce to
\begin{align}
L_t \rho &= - (F + 2S) (\rho +p) \label{SS-eq5} \\
L_r p &= -G (\rho +p) \label{SS-eq6} ,
\end{align}
the latter of which identifies $G$ as the radial acceleration.  The
full Bianchi identities now turn out to be satisfied as a consequence
of the contracted identities and the bracket relation
\begin{equation}
[L_t, L_r] = G L_t - F L_r.
\label{SS-bra}
\end{equation}

Equations~\eqref{SS-eq1}, \eqref{SS-eq2}, \eqref{SS-eq3}--\eqref{SS-eq4}, the
contracted identities~\eqref{SS-eq5} and \eqref{SS-eq6}, and the bracket
condition~\eqref{SS-bra} now form the complete set of intrinsic equations.
The structure is closed, in that it is easily verified that the
bracket relation~\eqref{SS-bra} is consistent with the known derivatives.
The derivation of such a set of equations is the basic aim of our
`intrinsic method'.  The equations deal solely with objects that
transform covariantly under displacements, and many of these
quantities have direct physical significance.

\subsubsection*{Integrating factors}

To simplify our equations we start by forming the derivatives of $A$.
From equations~\eqref{SS-eq1}, \eqref{SS-eq2} and
\eqref{SS-eq3}--\eqref{SS-eq4} it follows that
\begin{align}
L_t A + 3SA &= 2\pi Sp \\
L_r A + 3TA &= - 2\pi T\rho.
\end{align}
These results, and equations~\eqref{SS-eq3a} and~\eqref{SS-eq3b}, suggest that
we should look for an integrating factor for the $L_t+S$ and $L_r+T$
operators.  Such a function, $X$ say, should have the properties that
\begin{equation}
L_t X = SX, \qquad  L_r X = TX. 
\label{SS-Xeq}
\end{equation}
A function with these properties can exist only if the derivatives are
consistent with the bracket relation~\eqref{SS-bra}.  This is checked by
forming
\begin{align}
[L_t, L_r] X 
&= L_t(TX) - L_r (SX) \nn \\
&= X (L_t T - L_r S) \nn \\
&= X (SG-FT) \nn \\
&= G L_t X - F L_r X,
\label{SS-Xbra}
\end{align}
which confirms that the properties of $X$ are consistent
with~\eqref{SS-bra}.  Establishing the existence of integrating
factors in this manner is a key step in our method, because the
integrating factors play the role of intrinsically defined
coordinates.  If the $\lih$-function is parameterised directly in
terms of these functions, the physical status of the quantities in it
becomes clearer.  In the present case, equations~\eqref{SS-eq1}
and~\eqref{SS-eq2} show that $r/\alp$ already has the properties
required of $X$, so it is $r/\alp$ that emerges as the intrinsic
distance scale.  It is therefore sensible that the position-gauge
freedom in the choice of $r$ should be absorbed by setting $\alp=1$.
This then sets the intrinsic distance scale equal to $r$, lifting $r$
from the status of an arbitrary coordinate to that of a
physically-measurable quantity.

Having fixed the radial scale with the position-gauge choice
\begin{equation}
r = X, \qquad \alp= 1,
\label{SS-Xeqr}
\end{equation}
we can make some further simplifications.  From the form of
$\ho(a)$~\eqref{SS-ho} and equations~\eqref{SS-Xeq} and~\eqref{SS-Xeqr} we see
that
\begin{align}
g_1 &= L_r r = Tr \label{SS-int5} \\
g_2 &= L_t r = Sr,
\end{align}
which gives two of the functions in $\ho(a)$.  We also define
\begin{equation}
M \eqv -2 r^3 A = \half r ({g_2}^2 - {g_1}^2 +1),
\label{SS-defM}
\end{equation}
which satisfies
\begin{equation}
L_t M = -4\pi r^2 g_2 p
\label{SS-LtM}
\end{equation}
and
\begin{equation}
L_r M = 4\pi r^2 g_1 \rho.
\label{SS-LrM}
\end{equation}
The latter shows that $M$ plays the role of an intrinsic mass.

\subsubsection*{The `Newtonian' gauge}

So far a natural distance scale has been identified, but no natural
time coordinate has emerged.  To complete the solution it is
necessary to make a choice for the $t$ coordinate, so we now look for
additional criteria to motivate this choice.  We are currently free to
perform an arbitrary $r$ and $t$-dependent displacement along the
$e_t$ direction.  This gives us complete freedom in the
choice of $f_2$ function.  An indication of how this choice should be
made is obtained from equations~\eqref{SS-LtM} and~\eqref{SS-LrM} for the
derivatives of $M$~\eqref{SS-LtM}, which invert to yield
\begin{align}
\deriv{M}{t} &= \frac{-4 \pi g_1 g_2 r^2(\rho + p)}{f_1 g_1 - f_2
g_2} \label{SS-class1} \\ 
\deriv{M}{r} &= \frac{4 \pi r^2 (f_1 g_1 \rho + f_2 g_2 p)}{f_1 g_1 - f_2
g_2}. 
\end{align}

The second equation reduces to a simple classical relation if
we choose $f_2 = 0$, as we then obtain
\begin{equation}
\dr M = 4 \pi r^2 \rho,
\end{equation}
which says that $M(r,t)$ is determined by the amount of mass-energy in
a sphere of radius~$r$.  There are other reasons for choosing the time
variable such that $f_2=0$.  For example, we can then use the bracket
structure to solve for $f_1$.  With $f_2=0$ we have
\begin{align}
L_t &= f_1 \partial_t + g_2 \dr \\
L_r &= g_1 \dr,
\end{align}
and the bracket relation~\eqref{SS-bra} implies that
\begin{equation}
L_r f_1 = - G f_1.
\end{equation}
It follows that
\begin{equation} 
f_1 = \eps(t) \exp \left( - \int^r G /g_1 \, dr \right).
\end{equation}
The function $\eps(t)$ can be absorbed by a further $t$-dependent
rescaling along $e_t$ (which does not change $f_2$), so with $f_2=0$
we can reduce to a system in which
\begin{equation}
f_1 = \exp \left( - \int^r G /g_1 \, dr \right).
\label{SS-eqf1}
\end{equation}
Another reason why $f_2=0$ is a natural gauge choice is seen when the
pressure is zero.  In this case equation~\eqref{SS-eq6} forces $G$ to be
zero, and equation~\eqref{SS-eqf1} then forces $f_1=1$.  A free-falling
particle with $v=e_t$ ({\em i.e.} comoving with the fluid) then has
\begin{equation}
\tdot e_t + \rdot e_r = e_t + g_2 \, e_r,
\label{SS-newvel}
\end{equation}
where henceforth the dots denote differentiation with respect to the
proper time.  Since $\tdot=1$ the time coordinate $t$ matches the
proper time of all observers comoving with the fluid.  So, in the
absence of pressure, we are able to recover a global `Newtonian' time
on which all observers can agree (provided all clocks are correlated
initially).  Furthermore it is also clear from~\eqref{SS-newvel} that
$g_2$ represents the velocity of the particle.  Hence
equation~\eqref{SS-class1}, which reduces to
\begin{equation}
\dift M = -4 \pi r^2 g_2 \rho
\end{equation}
in the absence of pressure, has a simple Newtonian interpretation ---
it equates the work with the rate of flow of energy density.
Equation~\eqref{SS-defM}, written in the form
\begin{equation}
\frac{{g_2}^2}{2} -\frac{M}{r} = \frac{{g_1}^2-1}{2},
\end{equation}
is also now familiar from Newtonian physics --- it is a Bernoulli
equation for zero pressure and total (non-relativistic) energy
$({g_1}^2-1)/2$.

\begin{table}[t!!]
\renewcommand{\arraystretch}{1.2}
\begin{center}
\begin{tabular}{lll}
\hline \hline
\\
& The $\lih$-function &
\begin{minipage}[c]{6cm}
\fcolorbox{black}{white}{
\( \begin{array}{l}
\ho(e^t) = f_1 e^t \\
\ho(e^r) = g_1 e^r + g_2 e^t \\
\ho(e^\theta) = e^\theta \\
\ho(e^\phi)  = e^\phi 
\end{array} \) }
\end{minipage} \\
\\
& The $\om$ function & 
\begin{minipage}[c]{6cm}
\fcolorbox{black}{white}{
\( \begin{array}{l}
\om(e_t) = G e_r e_t \\
\om(e_r) = F e_r e_t \\
\om(\thht) = g_2/r \,\thht e_t + (g_1-1) /r \, e_r \thht \\
\om(\phht) = g_2/r \,\phht e_t + (g_1-1) /r \, e_r \phht 
\end{array} \) }
\end{minipage} \\
\\
& Directional derivatives &
\begin{minipage}[c]{6cm}
\fcolorbox{black}{white}{
\( \begin{array}{l}
L_t = f_1 \partial_t + g_2 \dr \\
L_r = g_1 \dr
\end{array} \) }
\end{minipage} \\
\\
& $G$, $F$ and $f_1$ & 
\begin{minipage}[c]{6cm}
\fcolorbox{black}{white}{
\( \begin{array}{l}
L_t g_1 = G g_2, \qquad L_r g_2 = F g_1  \\
f_1 = \exp \bigl( \int^r - G/ g_1 \, dr \bigr)
\end{array} \) }
\end{minipage} \\
\\
& Definition of $M$ &
\fcolorbox{black}{white}{
\( M \eqv \half r ({g_2}^2 - {g_1}^2 +1) \) } \\
\\
& Remaining derivatives &
\begin{minipage}[c]{6cm}
\fcolorbox{black}{white}{
\( \begin{array}{l}
L_t g_2 = G g_1 - M/r^2 -4\pi r p \\
L_r g_1 = F g_2 + M/r^2 - 4\pi r \rho
\end{array} \) }
\end{minipage} \\
\\
& Matter derivatives &
\begin{minipage}[c]{7cm}
\fcolorbox{black}{white}{
\( \begin{array}{l}
L_t M = -4\pi r^2 g_2 p, \qquad
L_r M = 4\pi r^2 g_1 \rho \\
L_t \rho = -(2 g_2/r+F)(\rho+p) \\  
L_r p = -G(\rho+p)
\end{array} \) }
\end{minipage} \\
\\
& Riemann tensor &
\begin{minipage}[c]{7cm}
\fcolorbox{black}{white}{
\( \begin{array}{l}
\clr(B) = 4 \pi \bigl((\rho+p) B \dt e_t e_t -2\rho/3 \, B \bigr) \\
\qquad - \half (M/r^3 - 4 \pi \rho /3) (B + 3 \sig_r B \sig_r)
\end{array} \) }
\end{minipage} \\
\\
& Stress-energy tensor &
\fcolorbox{black}{white}{
\(\clt(a) = (\rho+p) a \dt e_t e_t -pa \) }\\
\\
\hline \hline
\end{tabular}
\caption[dummy1]{\sl Equations governing a radially-symmetric perfect
fluid.  An equation of state and initial data $\rho(r,t_0)$ and
$g_2(r,t_0)$ determine the future evolution of the system.}
\label{rad-tab1}
\end{center}
\end{table}

For these reasons we refer to $f_2=0$ as defining the `Newtonian'
gauge.  The applications discussed in the following sections vindicate
our claim that this is the natural gauge for radially-symmetric
systems.  The full set of equations in the Newtonian gauge are
summarised in Table~\ref{rad-tab1}.  They underlie a wide range of
phenomena in relativistic astrophysics and cosmology.  The closest
analogue of the Newtonian gauge description of a spherically-symmetric
system is provided by Gautreau's `curvature coordinates'~\cite{gaut84}
(see also~\cite{gaut95,mart01}).  This description employs a set of
geodesic clocks in radial freefall, comoving with the fluid.  However,
such a description can only be applied if the pressure is independent
of radius, whereas the Newtonian gauge description is quite general.

One aspect of the equations in Table~\ref{rad-tab1} is immediately
apparent.  Given an equation of state $p=p(\rho)$, and initial data in
the form of the density $\rho(r,t_0)$ and the velocity $g_2(r,t_0)$,
the future evolution of the system is fully determined.  This is
because $\rho$ determines $p$ and $M$ on a time slice, and the
definition of $M$ then determines $g_1$.  The equations for $L_r p$,
$L_r g_1$ and $L_r g_2$ then determine the remaining information on
the time slice.  Finally, the $L_t M$ and $L_t g_2$ equations can be
used to update the information to the next time slice, and the process
can then start again.  The equations can thus be implemented
numerically as a simple set of first-order update equations.  This
fact considerably simplifies the study of collapsing matter, and
should be particularly significant in current studies of the
critical phenomena associated with horizon and singularity
formation~\cite{cho93,abr93}.

\subsection{Static matter distributions}
\label{SS-STAT}

As a simple first application we consider a static, radially-symmetric
matter distribution.  In this case $\rho$ and $p$ are functions of
$r$ only.  Since $M(r,t)$ is now given by
\begin{equation}
M(r) = \int_0^r 4 \pi {r'}^2 \rho(r') \, dr'
\label{SS-M4stat}
\end{equation}
it follows that 
\begin{equation}
L_t M = 4 \pi r^2 g_2 \rho = - 4 \pi r^2 g_2 p.
\label{star-1}
\end{equation}
For any physical matter distribution $\rho$ and $p$ must both be
positive, in which case equation~\eqref{star-1} can be satisfied only if
\begin{equation}
g_2 = F = 0 .
\end{equation}
Since $g_2=0$ we see that $g_1$ is given simply in terms of $M(r)$ by
\begin{equation}
{g_1}^2 = 1- 2M(r)/r,
\label{star-4}
\end{equation}
which recovers contact with the standard line element for a static,
radially-symmetric field.  It is immediately clear that a solution
exists only if $2M(r)<r$ for all $r$.  This is equivalent to the
condition that a horizon has not formed.

The remaining equation of use is that for $L_t g_2$, which now gives
\begin{equation}
G g_1 = M(r)/r^2 + 4\pi r p.
\label{star-5}
\end{equation}
Equations~\eqref{star-4} and~\eqref{star-5} combine with that for $L_r p$
to give the famous Oppen\-heimer--Volkov equation
\begin{equation}
\deriv{p}{r} = - \frac{(\rho + p)(M(r) + 4\pi r^3 p)}{r(r-2M(r))}. 
\label{star-OV}
\end{equation}
At this point we have successfully recovered all the usual equations
governing a non-rotating star, and the description is therefore
unchanged from that of standard general relativity.  The work involved
in recovering these equations from the full time-dependent case is
minimal, and the final form of $\ho(a)$ is very simple (it is a
diagonal function).  Furthermore, the meaning of the $t$ and $r$
coordinates is clear, since they have been defined operationally.

The solution extends straightforwardly to the region outside the star.
We now have $M$ constant, and
\begin{equation}
f_1 = 1/g_1 = (1-2M/r)^{-1/2},
\end{equation}
which recovers the Schwarzschild line element.  It follows that all
predictions for the behaviour of matter in the star's gravitational
field, including those for the bending of light and the perihelion
precession of Mercury, are unchanged.

\subsection{Point source solutions --- black holes}
\label{SS-BH}

The next solution of interest is obtained when the matter is
concentrated at a single point ($r=0$).  For such a solution,
$\rho=p=0$ everywhere away from the source, and the matter equations
reduce to
\begin{equation}
\left. \barr{l}
L_t M = 0 \\ L_r M = 0
\earr \right\}  \; 
\implies M = \mbox{constant}.
\label{bh-Mconst}
\end{equation}
Retaining the symbol $M$ for this constant we find that the equations
reduce to
\begin{align}
L_t g_1 &= G g_2 \\ 
L_r g_2 &= F g_1 \\
\intertext{and}
{g_1}^2 - {g_2}^2 &= 1- 2M/r.
\end{align}
No further equations yield new information, so we have an
under-determined system of equations and some additional gauge fixing
is needed to choose an explicit form of $\ho(a)$.  The reason for this
is that in the vacuum region the Riemann tensor reduces to
\begin{equation}
\clr(B) = - \frac{M}{2 r^3} (B + 3 \sigr B \sigr).
\label{SS-swzR}
\end{equation}
This tensor is now invariant under boosts in the $\sigr$ plane,
whereas previously the presence of the fluid velocity in the Riemann
tensor vector broke this symmetry.  The appearance of this new
symmetry in the matter-free case manifests itself as a new freedom in
the choice of $\lih$-function.

Given this new freedom, we should look for a choice of $g_1$ and $g_2$
that simplifies the equations.  If we attempt to reproduce the
Schwarzschild solution we have to set $g_2=0$, but then we immediately
run into difficulties with $g_1$, which is not defined for $r<2M$.  We
must therefore look for an alternative gauge choice.  We show in the
following section that, when $p=0$, $g_1$ controls the energy of
infalling matter, with particles starting at rest at $r=\infty$
corresponding to $g_1=1$.  A sensible gauge choice is therefore to set
\begin{equation}
g_1 = 1
\end{equation}
so that
\begin{align}
g_2 &= -\sqrt{2M/r} \\ 
G &= 0 \\ 
F &= - M/(g_2 r^2) \\ 
\intertext{and}
f_1 &= 1.
\end{align}
In this gauge the $\lih$-function takes the remarkably simple form
\begin{equation}
\ho(a) = a - \sqrt{2M/r} \, a \dt e_r \, e_t,
\label{bh-soln}
\end{equation}
which only differs from the identity through a single term.  From the
results of Section~\ref{O-Linalg} the extension to the action of $\ho$
on an arbitrary multivector $A$ is straightforward:
\begin{equation}
\ho(A) = A - \sqrt{2M/r} (A \dt e_r) \wdg e_t.
\end{equation} 
It follows that $\dh=1$ and the inverse of the adjoint function, as
defined by~\eqref{1fninv}, is given by
\begin{equation}
\hu^{-1}(A) = A + \sqrt{2M/r} (A \dt e_t) \wdg e_r. 
\end{equation} 

\subsubsection*{Point-particle trajectories}

To study the properties of the solution~\eqref{bh-soln} we consider the
equation of motion for infalling matter.  For a particle following the
trajectory $x(\tau)$, with $\tau$ the proper time, we have
\begin{equation}
v = \tdot \, e_t + (\tdot \sqrt{2M/r} + \rdot) e_r + \thetadot
e_\theta + \phidot e_\phi,
\end{equation}
where dots denote the derivative with respect to $\tau$.  Since the
$\lih$-function is independent of $t$ we have, from
equation~\eqref{PP-cons},
\begin{equation}
\hu^{-1}(e_t) \dt v = (1-2M/r) \tdot - \rdot \sqrt{2M/r} =
\mbox{constant} ,
\end{equation} 
and, for particles moving forwards in time ($\tdot>0$ for
$r\rightarrow\infty$), we can write
\begin{equation}
(1-2M/r) \tdot = \alpha + \rdot \sqrt{2M/r},
\label{bh-tdot}
\end{equation}
where the constant $\alpha$ satisfies $\alpha >0$.  The $\rdot$
equation is found from the constraint that $v^2=1$, which gives
\begin{equation}
\rdot^2 = \alpha^2 - (1-2M/r)(1 + r^2 (\thetadot^2 +
\sin^2\!\theta \, \phidot^2)).
\label{bh-rdot}
\end{equation} 
The horizon lies at $r=2M$ since, for $r<2M$, the velocity $\rdot$
must be negative. It might appear that an attempt to integrate
equation~\eqref{bh-tdot} will run into difficulties with the pole at
horizon, but this not the case.  At $r=2M$ we find that
$\rdot=-\alpha$ and this cancels the pole.  All particles therefore
cross the horizon and reach the singularity in a finite coordinate
time.

\begin{figure}[t!]
\begin{center}
\begin{picture}(300,240)
\put(0,240){\epsfig{figure=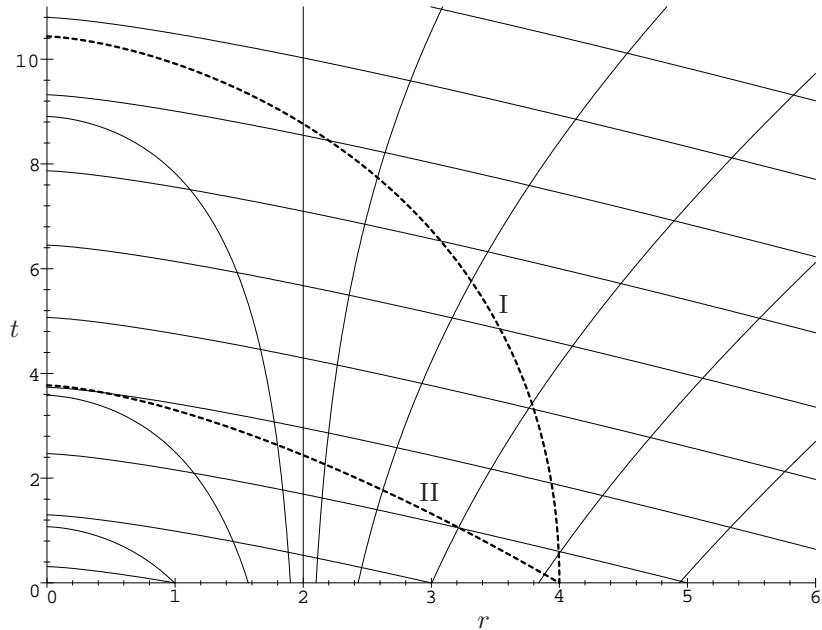,height=11cm,angle=-90}}
\put(177,0){$r$}
\put(0,110){$t$}
\put(185,119){I}
\put(155,48){II}
\end{picture}
\end{center}
\caption{\sl Matter and photon trajectories for radial motion in the
in the Newtonian gauge.  The solid lines are photon trajectories, and
the horizon lies at $r$=2.  The broken lines represent possible
trajectories for infalling matter.  Trajectory~I is for a particle
released from rest at $r=4$.  Trajectory~II is for a particle released
from rest at $r=\infty$.}
\label{bh-Fig1}
\end{figure}

Specialising to the case of radial infall, we see from
equation~\eqref{bh-rdot} that the constant $\alpha^2-1$ can be identified
with twice the particle's initial energy (for a unit mass particle).
Furthermore, equation~\eqref{bh-rdot} shows immediately that
$\ddot{r}$ satisfies
\begin{equation}
\ddot{r} = -M/r^2,
\end{equation}
a feature of motion in a spherically-symmetric gravitational field
that is rarely emphasised in general relativity texts.  Some possible
matter and photon trajectories are illustrated in
Figure~\ref{bh-Fig1}.  In the case where the particle is dropped from
rest at $r=\infty$ equations~\eqref{bh-tdot} and~\eqref{bh-rdot}
reduce to
\begin{equation}
\rdot = - \sqrt{2M/r}, \hs{0.5} \tdot = 1,
\label{bh-infall}
\end{equation}
and we recover an entirely Newtonian description of the motion.  The
properties of a black hole are so simple in the gauge defined
by~\eqref{bh-soln} that it is astonishing that this gauge is almost
never seen in the literature (see~\cite{gaut95a,mart01} for some
exceptions).  Presumably, this is because the line element associated
with~\eqref{bh-soln} does not look as natural as the $\lih$-function
itself and hides the underlying simplicity of the system.  Part of the
reason for this is that the line element is not diagonal, and
relativists usually prefer to find a coordinate system which
diagonalises $g_{\mu\nu}$.  Even when the freefall time coordinate $t$
is employed, a different radial coordinate is usually found to keep
the metric diagonal~\cite{steph-grav,lemait33}.

Since the gauge defined by $g_1=1$ and $g_2=-\sqrt{2M/r}$
extends our aim of keeping the equations in a simple Newtonian form,
we refer to this solution as defining the `Newtonian gauge' vacuum
solution.  We show in Section~\ref{COLLDST} that this gauge arises
naturally from the description of collapsing dust.  In the Newtonian
gauge one hardly needs to modify classical reasoning at all to
understand the processes involved --- all particles just cross the
horizon and fall into the singularity in a finite coordinate time.
And the horizon is located at $r=2M$ precisely because we can apply
Newtonian arguments!  The only departures from Newtonian physics lie
in relativistic corrections to the proper time taken for infall, and
in modifications to the equations for angular motion which lead to the
familiar results for orbital precession.

When extracting physical predictions in the Newtonian, or any other,
gauge, it is important to ensure that the problem is posed in a
gauge-invariant manner.  For example, one can envisage a simple
experiment with two observers initially at rest outside a black hole
at a distance $r_0$, where this distance is defined in terms of the
magnitude of $\clr(B)$.  One observer can then start free-falling, and
agree to emit photons of a chosen frequency at regular intervals.  If
one then computes what the remaining, stationary observer sees as a
function of their proper time, this is clearly something physically
meaningful.  It is not hard to show that the predictions for this are
gauge invariant, as they must be.  Furthermore, if everything takes
place outside the horizon, one can work in the `Schwarzschild' gauge
with $g_2=0$.  But, to see what happens as the free-falling observer
crosses the horizon, a global solution such as~\eqref{bh-soln} must
be used.  One still finds that the signal from the free-falling
observer becomes successively more red-shifted and less intense, as
predicted when working with the Schwarzschild metric, but the
free-falling observer crosses the horizon in a finite coordinate time.


\subsubsection*{Horizons and time-reversal asymmetry}

Our picture of the gravitational fields due to a radially-symmetric
point source is rather different from that of general relativity.
These differences are seen most clearly in the effects of time
reversal.  Time reversal is achieved by combining the displacement
\begin{equation}
f(x) = - e_t x e_t = x'
\label{SS-trevdis}
\end{equation}
with the reflection
\begin{equation}
\ho'(a) = - e_t \ho(a) e_t,
\label{SS-trevref}
\end{equation}
resulting in the the time-reversed solution
\begin{equation}
\ho^\ast(a) = e_t \ho_{x'}(e_t a e_t) e_t.
\label{SS-trev}
\end{equation}
As an example, the identity function $\ho(a)=a$ is time-reverse
symmetric --- as it should be.  The displacement~\eqref{SS-trevdis} is
a gauge transformation and cannot have any physical consequences.  The
reflection~\eqref{SS-trevref} is not a gauge transformation, however,
and can be used to transform between physically distinct gauge
sectors.  The reflection~\eqref{SS-trevref} is lost when the metric is
formed, so general relativity cannot handle time-reversal in the same
manner as GTG.

With the $\lih$-function described by equation~\eqref{SS-ho}, and with the
$\{f_i\}$ and $\{g_i\}$ functions of $r$ only, the effect
of~\eqref{SS-trev} is simply to change the sign of the off-diagonal
elements $f_2$ and $g_2$.  For example, applied to the
solution~\eqref{bh-soln}, the transformation~\eqref{SS-trev} produces the
time-reversed solution
\begin{equation}
\ho^\ast(a) = a + \sqrt{2M/r} \, a \dt e_r \, e_t.
\label{bh-solnT}
\end{equation}
The result is a solution in which particles inside the horizon are
swept out.  Once outside, the force on a particle is still attractive
but particles cannot re-enter through the horizon.

This lack of time-reversal symmetry is not a feature of the various
gauge choices made in arriving at~\eqref{bh-soln}; it is an inevitable
result of the presence of a horizon.  To see why, we return to the
equations in the form prior to the restriction to the Newtonian gauge.
The $\lih$-field is as defined by equation~\eqref{SS-ho} with $\alp=1$ and
the $\{f_i\}$ and $\{g_i\}$ functions of $r$ only.  The general set of
time-independent vacuum equations still have $M$ constant and
\begin{equation}
{g_1}^2 - {g_2}^2 = 1 - 2M/r
\label{SS-hor2}
\end{equation}
with
\begin{equation}
\dr g_1 = G, \qquad \dr g_2 = F.
\label{SS-int4}
\end{equation}
The bracket relation~\eqref{SS-bra} now gives
\begin{equation}
g_2 \dr f_2 - g_1 \dr f_1 = Gf_1 - F f_2 
\end{equation}
from which it follows that
\begin{equation}
\dr(f_1g_1 - f_2 g_2) = \dr \dh = 0.
\end{equation} 
Hence $\dh$ is a constant, with its value dependent on the choice of
position gauge.  Since $\clr(B)$ tends to zero at large $r$, we can
always choose the gauge such that $\ho(a)$ tends to the identity as
$r\rightarrow\infty$.  In this case $\dh$ must be one, so we can write
\begin{equation}
f_1g_1 - f_2 g_2 = 1.
\end{equation}

If we form the line element derived from our general
$\lih$-function we obtain
\begin{align}
ds^2 &= (1-2M/r) \, dt^2 +2 (f_1g_2 - f_2g_1) \, dt\, dr -({f_1}^2 -
{f_2}^2) \, dr^2 \nn \\
&\quad -r^2( d\theta^2 + \sin^2\! \theta \, d\phi^2).
\label{SS-gbh}
\end{align}
The off-diagonal term here is the one that breaks time-reversal
symmetry.  But we must have $g_1 = \pm g_2$ at the horizon, and we
know that $f_1g_1 - f_2g_2=1$ globally.  It follows that
\begin{equation}
f_1 g_2 - f_2 g_1 = \pm 1 \hs{0.5} \mbox{at} \,\, r = 2M,
\end{equation} 
so the line element~\eqref{SS-gbh} \textit{must} break time reversal
symmetry at the horizon~\cite{DGL-grav-bel}.  In fact, the asymmetry
is even more pronounced.  Once inside the horizon,
equation~\eqref{SS-hor2} forces a non-zero $g_2$, so the $\lih$-function
cannot be time-reverse symmetric anywhere inside the horizon.  This
link between the existence of a horizon and the onset of time-reversal
asymmetry is one of the most satisfying aspects of GTG.  Furthermore,
the requirement that a sign be chosen for $f_1g_2-f_2g_1$ at the
horizon shows that a black hole has more memory about its formation
than simply its mass $M$ --- it also remembers that it was formed in a
particular time direction.  We will see an example of this in the
following section.

At the level of the metric the discussion of time-reversal is much
less clear.  For example, inside the horizon a valid $\lih$-function is
obtained by setting $f_1$ and $g_1$ to zero.  Since $f_2$ and $g_2$
are non-zero, this $\lih$-function is manifestly not time-reverse
symmetric.  However, the line element generated by this $\lih$-function
is just the Schwarzschild line element
\begin{equation}
ds^2 = (1-2M/r) \,dt^2 - (1-2M/r)^{-1}\, dr^2 - r^2 (d\theta^2 +
\sin^2\theta \, d\phi^2),
\label{SS-swzsoln}
\end{equation}
which is usually thought of as being time-reverse symmetric.  Clearly,
our gauge theory probes questions related to time-reversal symmetry at
a deeper level than general relativity.  The consequences of our new
understanding of time-reversal will be met again in Section~\ref{DE},
where we study the Dirac equation in a black-hole background.

In general relativity one of the most important results for studying
radially-symmetric fields is Birkhoff's theorem.  This can be stated
in various ways, though the most usual statement is that the line
element outside a radially-symmetric body can always be brought to the
form of~\eqref{SS-swzsoln}.  As we have seen, this statement of
Birkhoff's theorem is correct in GTG only if no horizon is present.
However, the more general statement, that the fields outside a
spherically-symmetric source can always be made stationary, does
remain valid.

\subsubsection*{The Kruskal extension and geodesic completeness} 

In modern general relativity, the line element~\eqref{SS-swzsoln} is
not viewed as representing the final form of the metric for a
radially-symmetric black hole.  The full `maximal' solution was
obtained by Kruskal~\cite{kru60}, who employed a series of coordinate
transformations that mixed advanced and retarded Eddington--Finkelstein
coordinates.  The Kruskal extension describes a spacetime that
contains horizons and is time-reverse symmetric, so can have no
counterpart in GTG.  Furthermore, the Kruskal solution has two
distinct regions for each value of $r$~\cite{haw-large} and so is,
topologically, quite distinct from the solutions admitted in GTG.
This is because any solution of our equations must consist of $\ho(a)$
expressed as a function of position $x$.  The form of this position
dependence is arbitrary, but it must be present.  So, when the
coordinate $r$ is employed in defining the $\lih$-function, this
always represents a particular function of the vector $x$.  The point
$r=0$ is, by definition, a single point in space (an unbroken line in
spacetime).  No fields can alter this fact.  The Kruskal solution
contains two separate regions with the label $r=0$, so immediately
fails in GTG.  Instead of the full Kruskal extension with 4 sectors
(usually denoted I, II, $\mbox{I}'$ and
$\mbox{II}'$~\cite{haw-large}), GTG admits two distinct solutions, one
containing the sectors I and II, and the other containing $\mbox{I}'$
and $\mbox{II}'$.  These solutions are related by the discrete
operation of time reversal, which is not a gauge transformation.  This
splitting of a single time-reverse symmetric solution into two
asymmetric solutions is typical of the transition from a second-order
to a first-order theory.  Similar comments apply to the maximal
extensions of the Reissner--Nordstr\"{o}m and Kerr solutions.  The
infinite chain of `universes' general relativity admits as solutions
have no counterpart in our theory.  Instead, we use integral equations
to determine the nature of the matter singularities, precisely as one
would do in electromagnetism~\cite{DL04-KS}

In general relativity the Kruskal solution is the unique maximal
continuation of the Schwarzschild metric.  The fact that it has no
analogue in GTG means that our allowed solutions are not `maximal' and
forces us to address the issue of geodesic incompleteness.  For the
solution~\eqref{bh-soln} geodesics exist that cannot be extended into
the past for all values of their affine parameter.  But we have
already seen that the presence of a horizon commits us to a choice of
time direction, and in the following section we show how this choice
is fixed by the collapse process.  So, if we adopt the view that black
holes arise solely as the endpoint of a collapse process, then there
must have been a time before which the horizon did not exist.  All
geodesics from the past must therefore have come from a period before
the horizon formed, so there is no question of the geodesics being
incomplete.  We therefore arrive at a consistent picture in which
black holes represent the endpoint of a collapse process and the
formation of the horizon captures information about the direction of
time in which collapse occurred.  This picture is in stark contrast
with general relativity, which admits eternal, time-reverse symmetric
black-hole solutions.

\subsubsection*{Coordinate transformations and displacements}

The coordinate transformations employed in general relativity have two
distinct counterparts in GTG: as passive re-labellings of the
coordinates employed in a solution, such as changes of variables used
for solving differential equations; and as disguised forms of
position-gauge transformations.  An example of the latter is the
transformation between the Schwarzschild and advanced
Eddington--Finkelstein forms of the spherically-symmetric line element.
This is achieved with the coordinate transformations
\begin{align}
t' - r' &= t - (r+2M \ln(r-2M)) \\
r' &= r,
\end{align}
which can be viewed as the result of the displacement defined by
\begin{equation}
f(x) = x' = x - 2M \ln(r-2M) e_t.
\label{SStrf1}
\end{equation} 
This displacement is to be applied to the solution
\begin{equation}
\ho(a) = \Delta^{-1/2} a \dt e_t\, e^t + \Delta^{1/2} a \dt e_r \, e^r
+ a \wdg \sigr \, \sigr
\label{SStrf2}
\end{equation} 
where
\begin{equation}
\Delta = 1-2M/r.
\end{equation}
Clearly the gravitational fields are defined only outside the horizon,
and the aim is to achieve a form of $\ho(a)$ that is globally valid.

Differentiating the definition~\eqref{SStrf1} we find that
\begin{equation}
\fu(a) = a + \frac{2M}{r-2M} a \dt e_r \, e_t 
\end{equation}
and hence
\begin{equation}
\fo^{-1} (a) =  a - \frac{2M}{r-2M} a \dt e_t \, e_r.
\end{equation}
Now, the function~\eqref{SStrf2} is independent of $t$, so
$\ho(a;x')=\ho(a;x)$.  It follows that the transformed function
$\ho'(a)$ is given by
\begin{equation}
\ho'(a) = \ho\,\fo^{-1} (a) = \ho(a) - \frac{2M}{r} \Delta^{-1/2}  a \dt
e_t \, e_r. 
\end{equation} 
This new solution is not yet well-defined for all $r$, but if we now
apply the boost defined by the rotor
\begin{equation}
R = \exp(\sigr \chi/2),
\end{equation} 
where 
\begin{equation}
\sinh\!\chi = \half(\Delta^{-1/2} - \Delta^{1/2}),
\end{equation}
we obtain the solution
\begin{equation}
\ho''(a) = a + \frac{M}{r} a \dt e_- \, e_-,
\label{SSaef}
\end{equation}
where
\begin{equation}
e_- = e_t - e_r.
\end{equation}
The solution~\eqref{SSaef} is now globally defined.  It is the GTG
equivalent of the Kerr-Schild form of the Schwarzschild solution, and
has the property that infalling null geodesics are represented by
straight lines on a $t$--$r$ plot.  It is not hard to find a
transformation between~\eqref{SSaef} and the Newtonian gauge
solution~\eqref{bh-soln}.  This transformation consists of a displacement
and a rotation, both of which are globally well-defined.  On the other
hand, if one starts with the solution~\eqref{SSaef} and tries to recover a
version of the Schwarzschild solution by working in reverse, it is
clear that the process fails.  The boost needed is infinite at the
horizon and ill-defined for $r<2M$, as is the required displacement.
Such transformations fail to meet our requirement that gauge
transformations be well-defined over the whole region of physical
interest.

\subsubsection*{Integral equations and the singularity}

The Riemann tensor $\clr(B)$ contains derivatives of terms from
$\om(a)$ which fall off as $1/r^2$.  When differentiating such terms,
one must take account of the fact that 
\begin{equation}
\bgrad \dt (\bx/r^3) = 4\pi\del(\bx),
\end{equation}
where $\bx=x\wdg e_t$.  This fact will not affect the fields away from
the origin, but will show up in the results of integrals enclosing the
origin.  To see how, we again return to the setup before the Newtonian
gauge was chosen.  From equation~\eqref{SS-eq3c} we see that
\begin{equation}
e_t \dt \clg (e_t) =  8 \pi \rho = 2(- L_r T -T^2 + FS) -4A
\end{equation}
and, using equations~\eqref{SS-int5} and~\eqref{SS-int4}, this gives
\begin{align}
4 \pi \rho &= - (g_1 \dr g_1 - g_2 \dr g_2)/r + M/r^3 \nn \\
&= \dr(M/r) /r + M/r^3 \nn \\
&= (\sigr /r) \dt \bgrad (M/r) + (M/r) \bgrad \dt (\sigr /r) \nn \\
&= \bgrad \dt (M \bx /r^3).
\end{align}
It follows that $\rho=M \del(\bx)$, so the singularity generating the
radially-symmetric fields is a simple $\del$-function, of precisely
the same kind as the source of the Coulomb field in electrostatics.

The presence of the $\del$-function source at the origin is most
easily seen when the solution is analysed in the gauge defined
by~\eqref{SSaef}.  Solutions of this type are analysed
in~\cite{DL04-KS}, and we restrict ourselves here to a few basic
observations.  For the solution~\eqref{SSaef}, $\clr(B)$ is given
by
\begin{equation}
\ric(\ba + I\bb) = M \bigl( \ba \dt \bgrad (\bx/r^3) + 
I \bgrad \dt (\bb \wdg \bx/r^3) \bigr) 
\label{SS-EFadvR}
\end{equation} 
and it is simple to see that, away from the origin, \eqref{SS-EFadvR}
reduces to~\eqref{SS-swzR}.  In this section we adopt the convention
that, when all symbols are in bold face, the inner and outer products
drop down to their three-dimensional definitions.  The significance
of~\eqref{SS-EFadvR} is that it allows us to compute the integral of
the Riemann tensor over a region enclosing the origin simply by
converting the volume integral to a surface integral.  Taking the
region of integration to be a sphere of radius $r_0$ centred on the
origin, we find that
\begin{equation}
\int_{r \leq r_o} d^3\!x \, \ric(\ba) = M \int_{0}^{2 \pi} d\phi
\int_{0}^{\pi}d \theta \, \sin\! \theta \, \ba \dt \sigr \, \sigr =
\frac{4 \pi M }{3} \ba, 
\end{equation}
and
\begin{equation}
\int_{r \leq r_o} d^3\!x \, \ric(I \bb) = \int_{0}^{2 \pi} d\phi
\int_{0}^{\pi}  d \theta \, \sin\! \theta \, I \, \sigr \dt(\bb \wdg
\sigr) = -\frac{8 \pi M }{3} I \bb.
\end{equation} 
These results are independent of the radius of the spherical shell,
reflecting the spherical symmetry of the solution.  The above results
combine to give
\begin{equation}
\int_{r \leq r_o} d^3\!x \, \ric(B)  = \frac{4 \pi M }{3}B - 4 \pi M
B\wdg e_t \, e_t = -\frac{2 \pi M }{3}(B + 3 e_t B e_t) ,
\end{equation} 
which makes it clear what has happened.  The angular integral of the
Weyl component of $\clr(B)$ has vanished, because
\begin{equation}
\int_{0}^{2 \pi} d\phi \int_{0}^{\pi}d \theta \, \sin\! \theta (B + 3
\sig B \sigr) = 0,
\end{equation}
and what remains is the contribution from the stress-energy tensor,
which is entirely concentrated at the origin.  On contracting we find
that
\begin{align}
\int d^3\!x \, \ric(a) &= 4 \pi M e_t a e_t \nn \\
\int d^3\!x \, \ric &= -8 \pi M \label{EFadvint} \\
\int d^3\!x \, \clg(a) &= 8 \pi M a \dt e_t \, e_t \nn 
\end{align}
and, since $\ric(a) = 0$ everywhere except for the origin, the
integrals in~\eqref{EFadvint} can be taken over any region of space
enclosing the origin.  It is now apparent that the solution represents
a point source of matter, and we can therefore write
\begin{equation}
\clt(a) = M \delta(\bx) a \dt e_t \, e_t
\label{swzste}
\end{equation} 
for the matter stress-energy tensor.  This is consistent with the
definition of $M$ as the integral of the density for a static
system~\eqref{SS-M4stat}.

Analysing singularities in the gravitational fields by means of
integral equations turns out to be very powerful in GTG.  While the
above application does not contain any major surprises, we show
in~\cite{DL04-KS} that the same techniques applied to axisymmetric
fields reveal that the Kerr solution describes a ring of rotating
matter held together by a disk of isotropic tension --- a quite
different picture to that arrived at in general relativity.  This
clearly has implications for the ultimate fate of matter falling onto
the singularity, and could yield testable differences between GTG and
general relativity.

\subsection{Collapsing dust}
\label{COLLDST}

The equations in Table~\ref{rad-tab1} can be used to determine the
future evolution of a system given an equation of state and the
initial $\rho$ and $g_2$ distributions.  They are therefore
well-suited to the description of radial collapse and the formation of
horizons and singularities.  The simplest model, in which the pressure
is set to zero, describes collapsing dust.  This situation was first
studied by Oppenheimer~\& Snyder~\cite{opp39} and has been considered
since by many authors~\cite{gaut95,mis64,mis-grav,pan92}.  A feature
of these studies is the appearance of formulae which have a
suggestively Newtonian form.  This is usually dismissed as a
`coincidence'~\cite[Section 32.4]{mis-grav}.  Here we study the
collapse process in the Newtonian gauge and show that, far from being
coincidental, the Newtonian form of the results is a natural
consequence of the equations.  The distinguishing feature of the
Newtonian gauge approach is that the associated line element is not
diagonal.  This manifestly breaks time-reversal symmetry, as is
appropriate for the description of collapsing matter.  Working in this
gauge enables us to keep all fields globally defined, so the horizon
is easily dealt with and the matching onto an exterior vacuum region
is automatically incorporated.  This is quite different from previous
work~\cite{opp39,mis64,mis-grav}, which usually employs two distinct
diagonal metrics, one for the matter region and one for the vacuum.
Finding the correct matching conditions between these metrics is
awkward, and difficulties are encountered once the horizon has formed.

If $p=0$ it follows immediately that $G=0$ and hence $f_1=1$.  This
ensures that the global time coordinate $t$ agrees with the time
measured by observers comoving with the fluid.  Since $v\dt\cld\, v=0$
in the absence of pressure, such observers are also freely falling.
The function $g_2$ defines a velocity since, for a particle comoving
with the fluid, $g_2$ is the rate of change of $r$ (which is defined
by the Weyl tensor) with proper time $t$.  To emphasise its role as a
velocity we replace $g_2$ with the symbol $u$ for this section.  The
equations of Table~\ref{rad-tab1} now reduce to
\begin{align}
F &= \dr u \label{dst-F} \\ 
M(r,t) &= \int_0^r 4 \pi {r'}^2 \rho(r',t) \, dr', \label{dst-M}
\end{align}
which define $F$ and $M$ on a time slice, together with the update
equations
\begin{align}
\partial_t u + u \dr u &= - M/r^2 \label{dst-1} \\
\partial_t M + u \dr M &= 0. \label{dst-2}
\end{align}
Equations~\eqref{dst-1} and~\eqref{dst-2} afford an entirely Newtonian
description of the fluid.  Equation~\eqref{dst-1} is the Euler equation
with an inverse-square gravitational force, and~\eqref{dst-2} is the
equation for conservation of mass. The $L_t$ derivative plays the role
of the `matter' or `comoving' derivative for the fluid since, when
acting on a scalar, $v\dt\cld = L_t$.

The fact that $L_t M=0$ in the absence of pressure~\eqref{dst-2} is a
special case of a more general result.  Consider the integral
\begin{equation}
I(r,t) = \int_0^r 4 \pi s^2 \rho(s,t) f(s,t) \, ds 
\end{equation}
where $f(r,t)$ is some arbitrary function that is conserved along
fluid streamlines, that is, it obeys 
\begin{equation}
L_t f = 0.
\label{dst-Ltf}
\end{equation}
If we now construct $L_t I$ we find that
\begin{equation}
L_t I = u 4 \pi r^2 \rho f + \int_0^r 4\pi s^2 \bigl( \rho \dift f  +
f \dift \rho \bigr) \, ds.
\end{equation}
But, from
\begin{equation}
L_t \rho = -( 2 u /r + F ) \rho
\end{equation}
and equation~\eqref{dst-F}, we have
\begin{equation}
\dift(r^2 \rho) = - \dr(u r^2 \rho).
\end{equation}
Similarly, from equation~\eqref{dst-Ltf}, we see that
\begin{equation}
\dift f = - u \dr f,
\end{equation}
so it follows that
\begin{equation}
L_t I = u 4\pi r^2 \rho f - \int_0^r 4 \pi \bigl( u s^2 \rho
\dif{s} f + f \dif{s} (u s^2 \rho) \bigr) \, ds = 0.
\end{equation} 

Any integral of the type defined by $I$ leads to a quantity that is
conserved along the fluid streamlines.  The integral for
$M(r,t)$~\eqref{dst-M} is one such example, with $f$ set to 1.  It is
clear from its appearance in the Riemann tensor that $M$ represents
the `gravitating energy' of the region enclosed inside $r$.  

Since $G=0$, we have
\begin{equation}
L_t g_1 = 0
\end{equation}
and an alternative conserved quantity is therefore defined by
\begin{equation}
\mu(r,t) = \int_0^r 4 \pi s^2 \rho(s,t) \, \frac{ds}{g_1}.
\label{cos-restm}
\end{equation}
This is the covariant integral of the density, so is also a covariant
scalar quantity; it is simply the total rest-mass energy within $r$
(see Box 23.1 of `Gravitation'~\cite{mis-grav} for a discussion of
this point in the static case).  The relationship between the
rest-mass energy $\mu$ and the gravitating energy $M$ can be seen more
clearly by recalling that
\begin{equation}
g{_1}^2 = 1 - 2M/r + u^2.
\label{dst1.5}
\end{equation}
Since
\begin{equation}
M(r,t) - \mu(r,t) =  \int_0^r 4 \pi s^2 \rho(s,t) (g_1 - 1)\,
\frac{ds}{g_1} ,
\end{equation}	
the difference between the rest energy $\mu$ and the total energy $M$
is governed by $g_1-1$.  This is then multiplied by the term $4\pi
r^2\rho\, dr /g_1$, which is the rest mass of a shell of width $dr$.
For $|2M/r-u^2| \ll 1$ we can approximate~\eqref{dst1.5} to give
\begin{equation}
g_1 -1 \approx - M/r + \half u^2
\end{equation}
which explicitly shows the decomposition of the energy difference into
the sum of the Newtonian gravitational potential energy (always
negative) and the energy due to the bulk kinetic motion (always
positive).  It is clear that for a shell of material to escape it must
have $g_1-1 > 0$ so, with no approximation necessary, we recover the
Newtonian escape velocity $u^2 = 2M/r$.

As a further example of the insight provided by the Newtonian gauge,
consider the case where the interior of the shell is empty.  In this
case $M=0$, so
\begin{equation}
g_1 = (1+u^2)^{1/2},
\end{equation}
which shows that $g_1$ can be interpreted as a relativistic
$\gamma$-factor associated $u$.  This identification is justified if
we put $u=\sinh\alp$, which is reasonable since we know that $u$ can
be greater than 1.  It is the presence of this additional boost factor
in the formula for $M$ compared to $\mu$ that, in this case, makes
the total gravitating energy greater than the rest mass energy.  These
results should demonstrate that the physical picture of gravitational
collapse in the absence of pressure is really no different from that
afforded by Newtonian physics and special relativity.  It is therefore
no surprise that many of the results agree with those of Newtonian
physics.  Furthermore, abandoning a description in terms of
distorted volume elements and spacetime geometry has allowed us to
recover a clear physical picture of the processes involved.

\subsubsection*{Analytical solutions}

A useful property of the system of equations obtained when $p=0$ is
that it is easy to construct analytical solutions~\cite{tol34,bon47}.
To see this in the Newtonian gauge we write equation~\eqref{dst-2} in the
form
\begin{equation}
\left( \deriv{M}{t} \right)_{\!r} + u \left( \deriv{M}{r}
\right)_{\!t} = 0.
\end{equation} 
Since $M$ is a function of $r$ and $t$ only we can employ the
reciprocity relation
\begin{equation}
\left( \deriv{M}{t} \right)_{\!r}
\left( \deriv{t}{r} \right)_{\!M}
\left( \deriv{r}{M} \right)_{\!t} = -1
\end{equation}
to deduce that
\begin{equation}
\left( \deriv{t}{r} \right)_{\!M} = \frac{1}{u}.
\label{dst1.3}
\end{equation}
But we know that $u$ is determined by equation~\eqref{dst1.5}, and we also
know that both $M$ and $g_1$ are conserved along the fluid
streamlines.  We can therefore write $g_1=g_1(M)$, and
equation~\eqref{dst1.3} can be integrated straightforwardly to give $t$ as
a function of $r$ and $M$.

To perform the integration it is necessary to make a choice for the
sign of $u$.  For collapsing matter we clearly require $u<0$, while
for cosmology it turns out that $u>0$ is the appropriate
choice~\cite{DDGHL-I}.  For this section we can therefore write
\begin{equation}
\left( \deriv{t}{r} \right)_{\!M} = -\bigl( g_1(M)^2 - 1 + 2M/r
\bigr)^{-1/2}.
\label{dst1.7} 
\end{equation}
Finally, we need to choose a form for $g_1$.  This amounts to making
an initial choice of $u$, since $u$ and $g_1$ are related via
equation~\eqref{dst1.5}.  For this section we simplify to the case in
which the matter is initially at rest.  This might provide a
reasonable model for a star at the onset of a supernova, in which
there is a catastrophic loss of pressure due to vast amounts of
neutrino production, and the central core is suddenly left with no
supporting pressure.  With $u(r,0)=0$ we can write
\begin{equation}
{g_1}^2 = 1 - 2M(r_0)/r_0
\end{equation} 
where $r_0$ labels the initial $r$ coordinate.  We can view the value
of $r_0$ as carried along the streamline defined by it at $t=0$, so
can write $r_0=r_0(t,r)$ and treat $M$ and $g_1$ as functions of $r_0$
only.  Equation~\eqref{dst1.7} now becomes
\begin{equation}
\left( \deriv{t}{r} \right)_{\!r_0} = -\left( \frac{2M}{r} -
\frac{2M}{r_0} \right)^{-1/2}
\end{equation}
which integrates to give
\begin{equation}
t = \left( \frac{{r_0}^3}{2M} \right)^{1/2} \bigl( \pi/2 -
\sin^{-1}(r/r_0)^{1/2} + (r/r_0)^{1/2} (1-r/r_0)^{1/2} \bigr)
\label{dst-sol}
\end{equation}
where we have chosen the initial conditions to correspond to $t=0$.

Equation~\eqref{dst-sol} determines a streamline for each initial value
$r_0$, and can therefore be treated as implicitly determining the
function $r_0(r,t)$.  Since $M(r_0)$ and $g_1(r_0)$ are known, the
future evolution of the system is completely determined.  Furthermore,
quantities such as $\rho$ or $\dr u$ can be found directly once
$r_0(r,t)$ is known.  The above approach is easily extended to deal
with initial conditions other than particles starting from rest
since, once $M(r_0)$ and $g_1(r_0)$ are known, all one has to do is
integrate equation~\eqref{dst1.7}.  The ability to give a global
description of the physics in a single gauge allows for simple
simulations of a wide range of phenomena~\cite{DDGHL-I}.

An important restriction on the solution~\eqref{dst-sol} is that the
streamlines should not cross.  Crossed streamlines would imply the
formation of shock fronts, and in such situations our physical
assumption that $p=0$ will fail.  Streamline crossing is avoided if
the initial density distribution $\rho(r_0)$ is chosen to be either
constant or a monotonic-decreasing function of $r_0$.  This is
physically reasonable and leads to sensible simulations for a
collapsing star.

\subsubsection*{Singularity formation}

An immediate consequence of equation~\eqref{dst-sol} is that the time
taken before a given streamline reaches the origin is given by
\begin{equation}
t_1 = \pi  \bigl(r_0^3/8M \bigr)^{1/2}.
\label{dst-collt}
\end{equation}
Since the global time $t$ agrees with the proper time for observers
comoving with the fluid, equation~\eqref{dst-collt} is also the lapse of
proper time from the onset of collapse to termination at the
singularity as measured by any particle moving with the dust.  As
pointed out in Section~32.4 of~\cite{mis-grav}, the
formula~\eqref{dst-collt} agrees with the Newtonian result.  For the
reasons given above, this should no longer be a surprise.

Since $g_1$ and $M$ are conserved along a streamline,
equation~\eqref{dst1.5} shows that $u=g_2$ must become singular as
$r\rightarrow 0$.  Thus the central singularity forms when the first
streamline reaches the origin.  Near $r_0=0$ the initial density
distribution must behave as
\begin{equation}
\rho(r_0) \approx \rho_0 -\mbox{O}({r_0}^2)
\end{equation}
so the mass function $M(r_0)$ is
\begin{equation}
M(r_0) = \frac{4\pi}{3} {r_0}^3 - \mbox{O}({r_0}^5).
\end{equation}
It follows that 
\begin{equation}
\lim_{r_0\rightarrow 0} \, \pi  \bigl(r_0^3/8M \bigr)^{1/2} = \bigl( 3
\pi / 32\rho_0 \bigr)^{1/2} 
\end{equation}
so the central singularity forms after a time 
\begin{equation}
t_f = \left( \frac{3\pi}{32G\rho_0} \right)^{1/2} 
\label{cos-singform}
\end{equation}
where $\rho_0$ is the initial density at the origin and the
gravitational constant $G$ has been included.

\begin{figure}[t!]
\begin{center}
\epsfig{figure=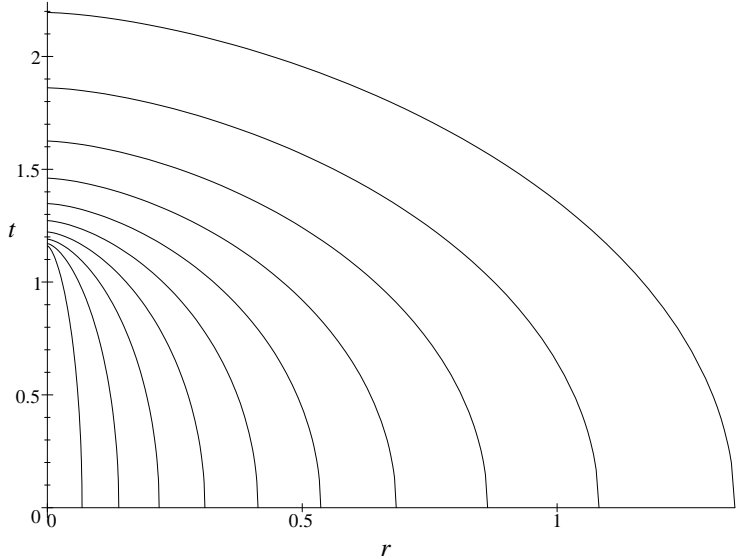,height=10cm,angle=-90}
\end{center}
\caption[dummy1]{\sl Fluid streamlines for dust collapsing from rest.
The initial density is given by equation~\eqref{dst-dens0} with
$\rho_0=0.22$ and $a=1$.  The singularity first forms at $t_f=1.16$.} 
\label{dst-fig1}
\end{figure}

A simulation of this process is shown in Figure~\ref{dst-fig1}, which
plots the fluid streamlines for the initial density function
\begin{equation}
\rho = \frac{\rho_0}{(1+(r/a)^2)^2}.
\label{dst-dens0}
\end{equation}
The streamlines all arrive at the singularity after a finite time,
and the bunching at $t_f$ can be seen clearly.  Solutions can be
extended beyond the time at which the singularity first forms by
including an appropriate $\delta$-function at the origin.

A further point revealed by such simulations is that it is possible to
have quite large differences between the total rest mass
$\mu_{\infty}\eqv\mu(r=\infty)$ and the final mass of the
singularity, $M$.  For example, for the case plotted in
Figure~\ref{dst-fig1} we find that $\mu_{\infty}=6.26$, whereas
$M=2.17$.  In this case nearly 3 times as many baryons have gone into
forming the black hole than is apparent from its mass $M$.  The
possibility of large differences between $M$ and $\mu$ is usually
ignored in discussions of the thermodynamics of black holes (see
footnote~14 of~\cite{bek74}).

\subsubsection*{Horizon formation}

\begin{figure}[t!]
\begin{center}
\begin{picture}(250,250)
\put(-50,250){\hbox{\epsfig{figure=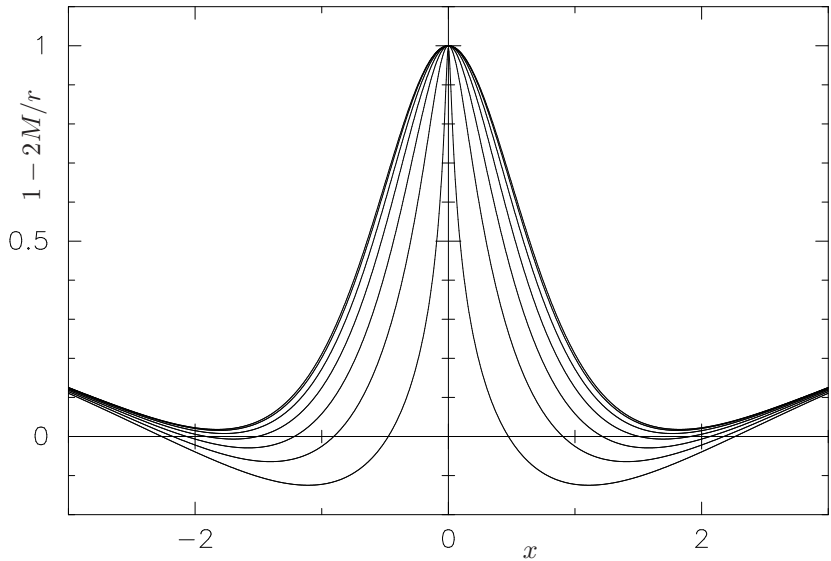,height=13cm,angle=-90}}}
\put(-27,135){\rotatebox{90}{$1-2M/r$}}
\put(163,0){$x$}
\end{picture}
\end{center}
\caption[dummy1]{\sl Simulation of collapsing dust in the Newtonian gauge.
Successive time slices for the horizon function $(1-2M(r,t)/r)$ {\em
versus\/} $x$ are shown, with the top curve corresponding to $t=0$ and
lower curves to successively later times. The initial velocity is
zero, and the initial density profile is given by
equation~\eqref{dst-dens0} with $\rho_0=0.22$ and $a=1$.  There is no
horizon present initially, but a trapped region quickly forms, since
in regions where $1-2M/r<0$ photons can only move inwards.}
\label{dst-Fig2}
\end{figure}

Any particle on a radial path has a covariant velocity of the form
\begin{equation}
v = \cosh\! u\, e_t + \sinh\! u \, e_r.
\end{equation} 
The underlying trajectory has $\xdot=\hu(v)$, so the radial motion is
determined by
\begin{equation}
\rdot = \cosh\! u\, g_2 + \sinh\! u \, g_1.
\end{equation} 
Since $g_2$ is negative for collapsing matter, the particle can only
achieve an outward velocity if ${g_1}^2 > {g_2}^2$.  A horizon
therefore forms at the point where
\begin{equation}
2 M(r,t)/r = 1.
\end{equation} 
To illustrate the formation of a horizon, we again consider the
initial density profile of equation~\eqref{dst-dens0}.  By
inverting~\eqref{dst-sol} at fixed $t$, $r_0$ is found as a function of
$r$.  From this, $(1-2M(r,t)/r)$ is calculated straightforwardly, and
this quantity is plotted on Figure~\ref{dst-Fig2} at different time
slices.  The plots show clearly that the horizon forms at a finite
distance from the origin.  It is conventional to extend the horizon
back in time along the past light-cone to the origin ($r=0$), since
any particle inside this surface could not have reached the point at
which $1-2M/r$ first drops to zero, and hence is also
trapped~\cite{nov-bh}.  The ease with which horizon formation is
treated again illustrates the advantages of working in a non-diagonal
gauge.  Such considerations will clearly be important when performing
numerical studies of more realistic collapse scenarios.

A final point is that, since $u$ is negative, it follows that
$f_1g_2-f_2g_1=u$ must also be negative.  This tells us that the
fields that remain after the collapse process has finished are in the
class defined by $f_1g_2-f_2g_1=-1$ at the horizon.  This time
direction is then frozen into the fields, as discussed in
Section~\ref{SS-BH}.

\subsection{Cosmology}
\label{SS-COS}

The equations of Table~\ref{rad-tab1} are sufficiently general to deal
with cosmology as well as astrophysics.  In recent years, however, it
has once more become fashionable to include a cosmological constant in
the field equations.  The derivation of Section~\ref{SSINT} is largely
unaffected by the inclusion of the cosmological term, and only a few
modifications to Table~\ref{rad-tab1} are required.  The full set of
equations with a cosmological constant incorporated are summarised in
Table~\ref{cos-tab1}.

\begin{table}[t!!]
\renewcommand{\arraystretch}{1.2}
\begin{center}
\begin{tabular}{lll}
\hline \hline 
\\
& The $\lih$-function &
\begin{minipage}[c]{6cm}
\fcolorbox{black}{white}{
\( \begin{array}{l}
\ho(e^t) = f_1 e^t \\
\ho(e^r) = g_1 e^r + g_2 e^t \\
\ho(e^\theta) = e^\theta \\
\ho(e^\phi)  = e^\phi 
\end{array} \) }
\end{minipage} \\
\\
& The $\om$ function & 
\begin{minipage}[c]{6cm}
\fcolorbox{black}{white}{
\( \begin{array}{l}
\om(e_t) = G e_r e_t \\
\om(e_r) = F e_r e_t \\
\om(\thht) = g_2/r \,\thht e_t + (g_1-1) /r \, e_r \thht \\
\om(\phht) = g_2/r \,\phht e_t + (g_1-1) /r \, e_r \phht 
\end{array} \) }
\end{minipage} \\
\\
& Directional derivatives &
\begin{minipage}[c]{6cm}
\fcolorbox{black}{white}{ \( \begin{array}{l}
L_t = f_1 \partial_t + g_2 \dr \\
L_r = g_1 \dr 
\end{array} \) }
\end{minipage} \\
\\
& $G$, $F$ and $f_1$ & 
\begin{minipage}[c]{6cm}
\fcolorbox{black}{white}{
\( \begin{array}{l}
L_t g_1 = G g_2, \qquad 
L_r g_2 = F g_1  \\
f_1 = \exp \bigl( \int^r - G/ g_1 \, dr \bigr)
\end{array} \) }
\end{minipage} \\
\\
& Definition of $M$ &
\fcolorbox{black}{light}{ \( M \eqv \half r ({g_2}^2 -
{g_1}^2 +1 - \Lambda r^2/3) \)} \\ 
\\
& Remaining derivatives &
\begin{minipage}[c]{7cm}
\fcolorbox{black}{light}{
\( \begin{array}{l}
L_t g_2 = G g_1 - M/r^2 + r\Lambda/3 - 4\pi r p \\
L_r g_1 = F g_2 + M/r^2 - r\Lambda/3 - 4\pi r \rho 
\end{array} \) }
\end{minipage} \\
\\
& Matter derivatives &
\begin{minipage}[c]{7cm}
\fcolorbox{black}{white}{
\( \begin{array}{l}
L_t M = -4\pi g_2 r^2 p, \qquad L_r M = 4\pi g_1 r^2 \rho \\
L_t \rho = -(2 g_2/r+F)(\rho+p) \\  
L_r p = -G(\rho+p)
\end{array} \) }
\end{minipage} \\
\\
& Riemann tensor & 
\begin{minipage}[c]{7cm}
\fcolorbox{black}{light}{
\( \begin{array}{l}
\clr(B) 4 \pi (\rho+p) B \dt e_t e_t - {\textstyle \frac{1}{3}}
(8 \pi \rho + \Lambda) B \\
\qquad - \half (M/r^3 - 4 \pi \rho /3) (B + 3 \sigr B \sigr )
\end{array} \) }
\end{minipage} \\
\\
& Stress-energy tensor &
\fcolorbox{black}{white}{ 
\( \clt(a) = (\rho+p) a \dt e_t e_t -pa \) } \\
\\
\hline \hline
\end{tabular}
\caption[dummy1]{\it Equations governing a radially-symmetric perfect
fluid with a non-zero cosmological constant $\Lambda$.  The
shaded equations differ from those of Table~\ref{rad-tab1}.} 
\label{cos-tab1}
\end{center}
\end{table}

In cosmology we are interested in homogeneous solutions to the
equations of Table~\ref{cos-tab1}.  Such solutions are found by
setting $\rho$ and $p$ to functions of $t$ only, and it follows
immediately from the $L_r p$ equation that
\begin{equation}
G=0, \quad \mbox{and} \quad  f_1 = 1.
\end{equation}
For homogeneous fields the Weyl component of the Riemann tensor must
vanish, since this contains directional information through the $e_r$
vector.  The vanishing of this term requires that
\begin{equation}
M(r,t) = \frac{4}{3} \pi r^3 \rho,
\label{cos-Mvsrho}
\end{equation}
which is consistent with the $L_r M$ equation.  The $L_tM$ and $L_t
\rho$ equations now reduce to
\begin{equation}
F = g_2 /r
\label{cos-F}
\end{equation}
and
\begin{equation}
\rhodot = -3 g_2 (\rho +p) /r.
\end{equation}
But we know that $L_r g_2 = F g_1$, which can only be consistent
with~\eqref{cos-F} if
\begin{equation}
F = H(t), \qquad g_2(r,t) = r H(t).
\end{equation}
The $L_t g_2$ equation now reduces to a simple equation for $\Hdt$,
\begin{equation}
\Hdt + H^2 - \Lambda /3 = - \frac{4 \pi}{3} (\rho + 3 p).
\end{equation}

Finally, we are left with the following pair of equations for $g_1$:
\begin{align}
L_t g_1 &= 0 \\
L_r g_1 &= ({g_1}^2 -1)/r.
\end{align}
The latter equation yields ${g_1}^2 = 1 +r^2 \phi(t)$ and the former
reduces to 
\begin{equation}
\dot{\phi} = -2H(t) \phi.
\end{equation}
Hence $g_1$ is given by 
\begin{equation}
{g_1}^2 = 1 - k r^2 \exp \left( -2 \int^t H(t') \, dt' \right),
\label{cos-g1}
\end{equation}
where $k$ is an arbitrary constant of integration.  It is
straightforward to check that~\eqref{cos-g1} is consistent with the
equations for $\Hdt$ and $\rhodot$.  The full set of equations
describing a homogeneous perfect fluid are summarised in
Table~\ref{cos-Tab2}.

\begin{table}[t!!]
\renewcommand{\arraystretch}{1.2}
\begin{center}
\begin{tabular}{lll}
\hline \hline 
\\
& The $\lih$-function &
\begin{minipage}[c]{7.5cm}
\fbox{
\( \begin{array}{l}
\ho(a) = a + a \dt e_r \bigl( (g_1-1) e^r + H(t) r e^t \bigr) \\
{g_1}^2 = 1 -k r^2 \exp \bigl( -2 \int^t H(t') \, dt' \bigr)
\end{array} \) } 
\end{minipage} \\
\\
& The $\om$ function & 
\fbox{ 
\( \om(a) = H(t) a \wdg e_t - (g_1-1)/r \, a \wdg (e_r e_t) e_t 
\) } \\
\\
& The density &
\begin{minipage}[c]{7.7cm}
\fbox{ \( 
8 \pi \rho /3  = H(t)^2 - \Lambda/3 +k \exp \bigl( -2 \int^t H(t')
\, dt' \bigr)  \) } 
\end{minipage} \\
\\
& \begin{tabular}{l}
Dynamical \\ equations \end{tabular} &
\begin{minipage}[c]{6.8cm}
\fbox{
\( \begin{array}{l}
\Hdt + H^2 - \Lambda/3 = - 4 \pi/3 \, (\rho + 3p)  \\
\rhodot = -3 H(t) (\rho +p) 
\end{array} \) }
\end{minipage} \\
\\
\hline \hline
\end{tabular}
\caption[dummy1]{\sl Equations governing a homogeneous perfect fluid.}
\label{cos-Tab2}
\end{center}
\end{table}

At first sight, the equations of Table~\ref{cos-Tab2} do not resemble
the usual Friedmann equations.  The Friedmann equations are recovered
straightforwardly, however, by setting
\begin{equation}
H(t) = \frac{\dot{S}(t)}{S(t)}.
\end{equation}
With this substitution we find that
\begin{equation}
{g_1}^2 = 1 - k r^2/ S^2
\end{equation}
and that the $\Hdt$ and density equations become
\begin{gather}
\frac{\ddot{S}}{S}-\frac{\Lambda}{3} = - \frac{4\pi}{3} (\rho +3p) \\
\intertext{and}
\frac{\dot{S}^2 +k}{S^2} - \frac{\Lambda}{3} = \frac{8\pi}{3} \rho, 
\end{gather}
recovering the Friedmann equations in their standard
form~\cite{nar-cosm}.  The intrinsic treatment has therefore led us to
work directly with the `Hubble velocity' $H(t)$, rather than the
`distance' scale $S(t)$.  There is a good reason for this.  Once the
Weyl tensor is set to zero, the Riemann tensor reduces to
\begin{equation}
\clr(B) = 4 \pi (\rho+p) B \dt e_t e_t - \third (8 \pi \rho +
\Lambda) B,
\end{equation}
and we have now lost contact with an intrinsically-defined distance
scale.  We can therefore rescale the radius variable $r$ with an
arbitrary function of $t$ (or $r$) without altering the Riemann
tensor.  The Hubble velocity, on the other hand, is intrinsic and it
is therefore not surprising that our treatment has led directly to
equations for this.

Among the class of radial rescalings a particularly useful one is to
rescale $r$ to $r' = S(t) r$.  This is achieved with the
transformation
\begin{equation}
f(x) = x \dt e_t \, e_t + S x \wdg e_t \, e_t,
\label{cos-trf1}
\end{equation}
so that, on applying equation~\eqref{G3}, the transformed $\lih$-function
is
\begin{equation}
\ho'(a) = a \dt e_t \, e_t + \frac{1}{S} \bigl( (1-kr^2)^{1/2} a \dt
e_r \, e^r + a \wdg \sigr \, \sigr \bigr).
\label{cos-stat}
\end{equation}
The function~\eqref{cos-stat} reproduces the standard line element used in
cosmology.  We can therefore use the transformation~\eqref{cos-trf1} to
move between the `Newtonian' gauge developed here and the gauge
of~\eqref{cos-stat}.  This is useful for later sections, where the Maxwell
and Dirac equations are solved in a cosmological background described
by~\eqref{cos-stat}.  The differences between these gauges can be
understood by considering geodesic motion.  A particle at rest with
respect to the cosmological frame (defined by the cosmic microwave
background) has $v=e_t$.  In the gauge of~\eqref{cos-stat} such a particle
is not moving in the flatspace background (the distance variable $r$
is equated with the comoving coordinate).  In the Newtonian
gauge, on the other hand, comoving particles are moving outwards
radially at a velocity $\rdot= H(t) r$, though this expansion centre
is not an intrinsic feature.  Of course, attempting to distinguish
these pictures is a pointless exercise, since all observables must be
gauge-invariant.  All that is of physical relevance is that, if two
particles are at rest with respect to the cosmological frame (defined
by the cosmic microwave background), then the light-travel time
between these particles is an increasing function of time and light is
redshifted as it travels between them.

\subsubsection*{Dust models}

The utility of the Newtonian gauge in cosmology has been independently
discovered by other authors~\cite{gaut84,ellis93}.  Here we illustrate
its advantages for dust models ($p=0$).  Setting $p$ to zero implies
that
\begin{equation}
H(t) = - \rhodot / 3\rho,
\end{equation}
so 
\begin{equation}
{g_1}^2 = 1-kr^2 \rho^{2/3}
\end{equation}
and
\begin{equation}
H(t) = \left( \frac{8 \pi}{3} \rho -k \rho^{2/3} + \frac{\Lambda}{3}
\right)^{1/2}.
\end{equation}
We are therefore left with a single first-order differential equation
for $\rho$.  Explicit solutions of this equations are often not needed,
as we can usually parameterise time by the density $\rho(t)$.

If we now look at the trajectories defined by the fluid, these have 
$v = e_t$ which implies that
\begin{equation}
\xdot = e_t + r H(t) e_r.
\end{equation}
It follows that
\begin{equation}
\rdot /r = H(t) = -\rhodot/3\rho
\end{equation}
and hence that
\begin{equation}
r/r_0 = ( \rho/\rho_0)^{-1/3}.
\end{equation}
The fluid streamlines form a family of spacetime curves spreading out
from the origin at the initial singularity (when $\rho=\infty$).  The
Newtonian gauge therefore describes an expanding universe in a very
simple, almost naive manner.  Since all points in a homogeneous
cosmology are equivalent, we can consider ourselves to be located at
$r=0$.  The Newtonian gauge then pictures our observable universe as a
ball of dust expanding outwards radially from us.

\begin{figure}[t!]
\begin{center}
\includegraphics[width=8cm,angle=-90]{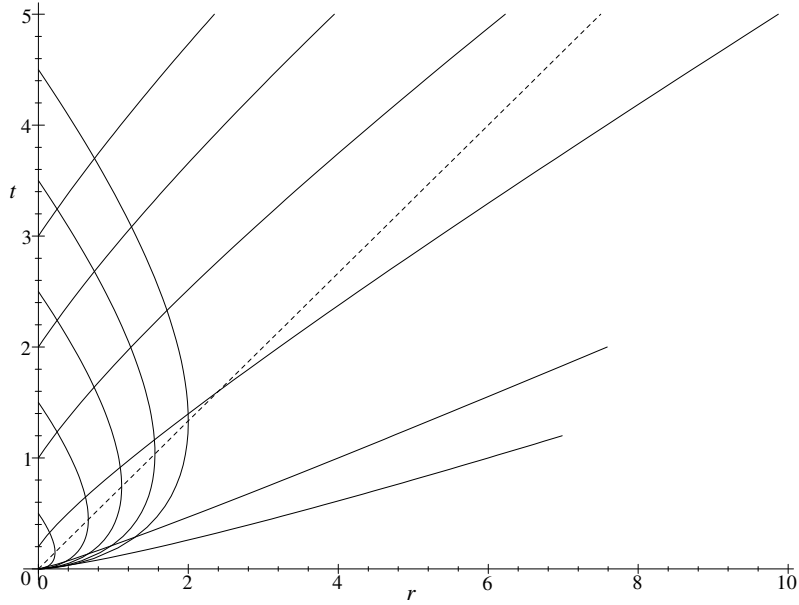}
\end{center}
\caption[dummy1]{\sl  Null geodesics for a dust-filled $k=\Lambda=0$
universe in the Newtonian gauge.  All geodesics start from the origin
at the initial singularity (here set at $t=0$).  The dashed line gives
the critical distance where photons `turn round' in this gauge.}
\label{cos-Fig1}
\end{figure}

While the picture provided by the Newtonian gauge has no physical
reality of its own, it does have some heuristic merit and can provide
a useful aid to one's physical intuition.  For example, consider the
familiar relationship between angular size and redshift.  An initially
surprising feature of this relationship is that, beyond a certain
redshift, angular sizes stop decreasing and start increasing.  This
result is easily understood in the Newtonian gauge.  Consider the
photon paths shown in Figure~\ref{cos-Fig1}. (These paths are for a
$k=0$ and $\Lambda=0$ universe, though the comments are applicable
more generally.)  Suppose that at some finite time $t_0$ we receive a
photon from the distant past.  This photon must have followed part of
a path which begins on the origin at $t=0$.  Before a certain time in
the past, therefore, photons received by us must have initially
travelled outwards before turning round and reaching us.  The angular
size of an observed object is then that appropriate to the actual $r$
coordinate of the object when it emitted the photons.  Since the value
of $r$ decreases before a certain time, angular sizes appear larger
for objects that emitted photons before this time.

A radial null geodesic has a trajectory
\begin{equation}
x(\tau) = t(\tau) e_t + r(\tau) e_r.
\end{equation}
For these trajectories
\begin{equation}
v = \hu^{-1}(\xdot) = \tdot \bigl( e_t - \frac{r H(t)}{g_1} \, e_r
\bigr) + \frac{\rdot}{g_1} \, e_r
\end{equation}
and the condition that $v^2=0$ reduces to
\begin{align}
\rdot /g_1 &= \tdot (r H(t) /g_1 \pm 1) \\
\implies \quad \frac{dr}{dt} &= r H(t) \pm g_1. 
\end{align}
Since $H(t)$ is positive in a cooling universe, the distance at which
photons `turn round' is defined by
\begin{equation}
r_c = \frac{g_1}{H(t)} = \left( \frac{8 \pi \rho}{3} + \frac{\Lambda}{3}
\right)^{-1/2}.
\end{equation} 
(In~\cite{DGL-erice} we mistakenly referred to this distance as a
particle horizon.)  For $k=\Lambda=0$ it is simple to show that this
distance corresponds to a redshift of $1.25$.  This result relates
physically measurable quantities, so is gauge invariant.

\subsubsection*{Closed universe models}

The Newtonian gauge presents a particularly simple picture for a
closed universe ($k>0$).  For $k>0$ the requirement that
$1-kr^2\rho^{2/3}$ is positive means that
\begin{equation}
r^3 \rho > k^{-3/2}.
\end{equation} 
This places a limit on the speed with which the dust can expand, so
the solution describes a finite ball of dust expanding into a vacuum.
This ball expands out to some fixed radius before turning round and
contracting back to the origin.  The turning point is achieved where $
H(t) = 0$ so, for $\Lambda=0$, this occurs when
\begin{equation}
\rho = (3k/8\pi)^3.
\end{equation}
The maximum radius is therefore
\begin{equation}
r_{\mbox{\small max}} = \frac{8\pi}{3k^{3/2}}.
\end{equation} 
The time taken to reach the future singularity is given
by~\eqref{cos-singform}, since this cosmological model is a special case
of spherical collapse in which the density is uniform.  This picture
of a model for a closed universe is both simple and appealing.  It
allows us to apply Newtonian reasoning while ensuring consistency
with the full relativistic theory.

The finite ball model for a $k>0$ cosmology is clearly useful when
considering experiments with particles carried out near the origin,
but globally one must consider the boundary properties of the ball.  A
crucial question is whether the particle horizon (the largest region
of the universe with which an observer at the origin is in causal
contact) extends past the edge of the ball or not.  It can be shown
that this horizon always lies inside the radius at which $g_1$ becomes
imaginary, except at the turnaround point (the point at which the ball
reaches its maximum radius), where the two radii coincide.  A suitable
choice of cutoff radius is therefore available in either the expanding
or contracting phase separately, but what happens at the turnaround
point is potentially ambiguous.  When discussing field theory,
however, the finite ball model is inadequate.  One must instead use a
global gauge so, in Section~\ref{MAX-COS}, we introduce the
`stereographic projection gauge'.  This provides a global solution
which can be shown to be spatially closed.  This solution is used in
the study of electromagnetism (Section~\ref{MAX-COS}) and the Dirac
equation (Section~\ref{DE-COS}) in a cosmological background.  It is
possible to treat the stereographic projection gauge solution in a
form of the Newtonian gauge, though this possibility is not explored
here.


\section[Electromagnetism in a gravitational background]{Electromagnetism in a gravitational \newline background} 
\label{MAX}

In Section~\ref{Field} we derived field equations for the
gravitational and Dirac fields.  We now turn to the derivation of the
Maxwell equations in a gravitational background.  A number of
applications of these equations are then discussed, including a simple
derivation of the characteristic surfaces for both the Maxwell and
Dirac equations.

The basic dynamical variable is the electromagnetic vector potential
$A$, for which the coupling to spinor fields was derived in
Section~\ref{G-dirac}.  Under phase rotations of the spinor field, $A$
transforms as
\begin{equation}
eA \mapsto eA - \grad \phi.
\label{em1}
\end{equation}
It follows that, under a displacement, $A$ must transform in the same
way as $\grad\phi$, that is,
\begin{equation}
A(x) \mapsto \fo(A(x')).
\end{equation}
The covariant form of the vector potential is therefore 
\begin{equation}
\cla=\ho(A),
\end{equation}
which is the term that appeared in the Dirac equation~\eqref{G-d19}.

From $A$, the Faraday bivector $F$ is defined by
\begin{equation}
F \eqv \grad \wdg A.
\end{equation}
This definition implies that, under displacements, $F(x)$ is transformed
to $F'(x)$, where
\begin{align}
F'(x) &= \grad \wdg \fo A(x') \nn \\
&= \fo(\grad_{x'} \wdg A(x')) \nn \\
&= \fo F(x').
\end{align}
It follows that the covariant analogue of $F$ is defined by
\begin{equation}
\clf = \ho(F),
\end{equation}
which is covariant under position and rotation-gauge transformations,
and is invariant under phase changes.

The same covariant quantity $\clf$ is obtained if one follows the
route used for the construction of $\liR(a\wedge b)$ at~\eqref{F1}.  In
particular, the contracted commutator of two covariant derivatives
gives (in the absence of torsion)
\begin{align}
\ho(e^\nu) \wdg \ho(e^\mu) [D_\mu, D_\nu] \psi 
&= \half \ho(\db)\ho(\da) \liR(a \wdg b) \psi \nn \\
&= \half \clr \psi.
\end{align}
The analogous construction for the `internal' covariant derivative
\begin{equation}
D_a^I \psi = a \dt \grad \psi - e a \dt A \, \psi \isk
\end{equation}
gives
\begin{align}
\ho(e^\nu) \wdg \ho(e^\mu) [D_\mu^I, D_\nu^I] \psi 
&= e \ho(\db) \wdg \ho(\da) (a\wdg b) \dt F \psi \isk \nn \\
&= 2 e \clf \psi \isk,
\label{em4}
\end{align}
which clearly identifies $\clf$ as a covariant quantity.  Unlike
$\clr$, however, $\clf$ is a bivector and equation~\eqref{em4} exhibits a
curious interaction between this bivector on the left of $\psi$, and
the fixed bivector $\isk$ on the right.

Having defined the covariant bivector $\clf$, it is clear that the
appropriate generalisation of the electromagnetic action to include
gravitational interactions is
\begin{equation}
S = \hint (\half \clf \dt \clf - \cla \dt \clj ),
\label{F-em2.5}
\end{equation}
where here $\clj$ is the covariant charge current.  Unlike the Dirac
action, the rotation-gauge field $\Om(a)$ does not appear in this
action.  It follows that the electromagnetic field does not act as a
source of spin.  The action~\eqref{F-em2.5} is varied with respect to $A$,
with $\ho(a)$ and $\clj$ treated as external fields.  The result of
this is the equation
\begin{equation}
\grad \dt G = J,
\label{F-em3}
\end{equation}
where
\begin{align}
G &\eqv \hu\, \ho(F) \dhi \\
\intertext{and}
J &\eqv \hu(\clj) \dhi.
\end{align}
Equation~\eqref{F-em3} combines with the identity 
\begin{equation}
\grad\wdg F=0
\label{F-em4}
\end{equation}
to form the full set of Maxwell equations in a gravitational
background.  We again see that the classical field equations can be
expressed in a form from which all reference to the rotation gauge has
been removed.

Some insight into the equations~\eqref{F-em3} and~\eqref{F-em4} is obtained by
performing a space-time split (see Section~\ref{O-STA}) and writing 
\begin{align}
F &= \bE + I\bB \\
G &= \bD +I\bH \\
J\go &= \rho + \bJ.
\end{align}
In terms of these variables Maxwell's equations in a gravitational
background take the familiar form
\begin{equation}
\begin{alignedat}{2}
\bgrad \dt \bld{B} &= 0 &\qquad \bgrad \dt \bld{D} &= \rho \\
\bgrad \btimes \bE &= - \deriv{\bB}{t}  &\qquad 
\bgrad \btimes \bH &= \bJ +\deriv{\bD}{t}
\end{alignedat}
\label{F-diel}
\end{equation} 
where $\bgrad=\go\wedge\grad=\sig_i\partial_{x^i}$ is the 3D vector
derivative, and the bold cross $\btimes$ is the traditional
vector cross product:
\begin{equation}
\ba \btimes \bb = -I \, \ba \crs \bb.
\end{equation}
Equation~\eqref{F-diel} shows that the $\lih$-field defines the dielectric
properties of the space through which the electromagnetic fields
propagate, with $\dhi\hu\,\ho$ determining the generalized
permittivity/permeability tensor.  Many phenomena, including the
bending of light, can be understood easily in terms of the properties
of the dielectric defined by the $\lih$-field.

\subsubsection*{Covariant form of the Maxwell equations}

So far we have failed to achieve a manifestly covariant form of the
Maxwell equations.  We have, furthermore, failed to unite the separate
equations into a single equation.  In the absence of gravitational
effects the equations $\grad\dt F=J$ and $\grad\wdg F=0$ combine into
the single equation
\begin{equation}
\grad F = J.
\end{equation}
The significance of this equation is that the $\grad$ operator is
invertible, whereas the separate $\grad\cdot$ and $\grad\wedge$
operators are not~\cite{hes-gc,DGL93-paths}.  Clearly, we expect that
such a unification should remain possible after the gravitational
gauge fields have been introduced.  To find a covariant equation, we
first extend the `wedge' equation~\eqref{F10} to include higher-grade
terms.  To make the derivation general we include the spin term, in
which case we find that
\begin{align}
\cldd \wdg \dho(a \wdg b) &= (\cldd \wdg \dho(a)) \wdg \ho(b) - \ho(a)
\wdg (\cldd \wdg \dho(b)) \nn \\
&= \kappa (\ho(a) \dt \cls) \wdg \ho(b) - \kappa \ho(a) \wdg (\ho(b) \dt
\cls) \nn \\
&= \kappa \cls \crs \ho(a \wdg b)
\end{align}
and, more generally, we can write
\begin{equation}
\cld \wdg \ho(A_r) = \ho(\grad \wdg A_r) + \kappa \la \cls \ho(A_r)
\ra_{r+1}.
\end{equation} 
We can therefore replace equation~\eqref{F-em4} by
\begin{equation}
\cld \wdg \clf - \kappa \cls \crs \clf = 0.
\label{F-em6}
\end{equation}
We next use the rearrangement
\begin{align}
\grad \dt (\hu(\clf) \dhi) 
&= I\grad \wdg (I \hu(\clf) \dhi) \nn \\
&= I \grad \wdg (\ho^{-1}(I \clf)) \nn \\
&= I \ho^{-1} \bigl( \cld \wdg (I \clf) + \kappa (I \clf) \crs \cls \bigr)
\end{align}
to write equation~\eqref{F-em3} as
\begin{equation}
\cld \dt \clf - \kappa  \cls \dt \clf = I \ho(J I) = \clj.
\label{F-em7}
\end{equation}
Equations~\eqref{F-em6} and~\eqref{F-em7} now combine into the single equation
\begin{equation}
\cld \clf - \kappa \cls \clf = \clj,
\label{F-em8}
\end{equation}
which achieves our objective.  Equation~\eqref{F-em8} is manifestly
covariant and the appearance of the $\cld\clf$ term is precisely what
one might expect on `minimal-coupling' grounds.  The appearance of the
spin term is a surprise, however.  Gauge arguments alone would not
have discovered this term and it is only through the construction of a
gauge-invariant action integral that the term is found.
Equation~\eqref{F-em8} should be particularly useful when considering
electromagnetic effects in regions of high spin density, such as
neutron stars.

To complete the description of electromagnetism in a gravitational
background we need a formula for the free-field stress-energy tensor.
Applying the definition~\eqref{F-i2} we construct 
\begin{align}
\clt_{\mbox{\scriptsize em}}( \hu^{-1}(a))
&= \half \dh \, \dhoa \bigl\la \ho(F) \ho(F) \dhi \bigr\ra \nn \\
&=\ho(a \dt F) \dt \clf - \half \hu^{-1}(a) \, \clf \dt \clf .
\end{align}
Hence,
\begin{align}
\clt_{\mbox{\scriptsize em}}(a) 
&= \ho(\hu(a) \dt F) \dt \clf - \half a \, \clf \dt \clf \nn \\
&= (a \dt \clf) \dt \clf - \half a \, \clf \dt \clf \nn \\
&= -\half \clf a \clf,
\label{F-elmste}
\end{align}
which is the natural covariant extension of the gravitation-free form
$-FaF/2$.  The tensor~\eqref{F-elmste} is symmetric, as one expects for
fields with vanishing spin density.

\subsection{Characteristic surfaces}
\label{charac}

In their spacetime algebra form the Maxwell and Dirac equations are
both first-order differential equations involving the vector
derivative $\grad$.  For electromagnetism, this first-order form of
the equations offers many advantages over the equivalent second-order
theory~\cite{hes-sta,DGL93-paths}.  We have now seen that
gravitational interactions modify both these equations in such a way
that the vector derivative $\grad$ is replaced by the position-gauge
covariant derivative $\ho(\grad)$.  As an illustration of the utility
of first-order equations, both without and with gravitational effects,
we now give a simple derivation of the properties of characteristic
surfaces.

Consider, initially, a generic equation of the type
\begin{equation}
\grad \psi = f(\psi,x)
\label{cha1}
\end{equation}
where $\psi$ is some arbitrary field, and $f$ is some known function.
Suppose that an initial set of data is given over some
three-dimensional surface in spacetime, and we wish to propagate this
information off the surface into the adjoining region.  We pick three
vectors, $a$, $b$ and $c$, which are tangent to the surface.  From our
initial data we can construct $a\dt\grad \psi$, $b\dt\grad \psi$ and
$c\dt\grad \psi$.  Now define
\begin{equation}
n \eqv I \, a \wdg b \wdg c
\end{equation}
and use $n\grad=n\dt\grad + n\wdg \grad$ to decompose~\eqref{cha1} into 
\begin{equation}
n\dt\grad \psi = - n \wdg \grad \psi + n f(\psi).
\label{F-ch1}
\end{equation}
The right-hand side of~\eqref{F-ch1} contains the term
\begin{align}
n \wdg \grad \psi 
&= I (a \wdg b \wdg c) \dt \grad \psi \nn \\
&=  I (a \wdg b \, c\dt\grad \psi - a \wdg c \, b\dt\grad \psi + b
\wdg c \, a\dt\grad \psi), 
\end{align}
which is therefore known.  It follows that we know all the terms on
the right-hand side of equation~\eqref{F-ch1} and can therefore construct
$n\dt\grad\,\psi$.  This gives us the derivative required to propagate
off the surface.  The only situation for which propagation is
impossible is when $n$ remains in the surface.  This occurs when
\begin{align}
n \wdg (a \wdg b \wdg c) &= 0 \nn \\
\implies \hs{.5} n \wdg (n I) &= 0 \nn \\
\implies \hs{.5} n \dt n &= 0.
\label{F-ch3}
\end{align}
It follows that the characteristic surfaces for any first-order
equation of the type defined by~\eqref{cha1} are null surfaces.  These
considerations automatically include the Maxwell and Dirac equations.
It is notable how this result follows from purely algebraic
considerations.

The generalisation to a gravitational background is straightforward.
Equation \eqref{F-em8} is generalised to
\begin{equation} 
\ho(\grad) \psi = f(\psi),
\label{F-ch6}
\end{equation}
and we assume that a gauge choice has been made so that all the fields
(apart from $\psi$) are known functions of $x$.  Again, we assume that
the initial data consist of values for $\psi$ over some
three-dimensional surface, so we can still determine $a\dt\grad\psi$
{\em etc.}  Since
\begin{equation}
a\dt\grad = \hu^{-1}(a)\dt\ho(\grad)
\end{equation}
it follows that the vector of interest is now
\begin{equation}
I \hu^{-1}(a) \wdg \hu^{-1}(b) \wdg \hu^{-1}(c) = I \hu^{-1} (n I)
= \ho(n) \dhi. 
\end{equation}
This time we multiply equation~\eqref{F-ch6} by $\ho(n)$ and find that
$\ho(n)\dt\ho(\grad)\psi$ can be constructed entirely from known
quantities.  We can therefore propagate in the $\hu\,\ho(n)$
direction, so we now require that this vector does not lie in the
initial surface.  The analogue of~\eqref{F-ch3} is therefore
\begin{equation}
\hu\,\ho(n) \dt n = 0, \qquad \mbox{or} \qquad \ho(n)^2=0,
\end{equation}
and the characteristic surfaces are now those for which $\ho(n)$ is
null.  This is the obvious covariant extension of $n$ being a null
vector.

\subsection{Point charge in a black-hole background}

The problem of interest here is to find the fields generated by a
point source held at rest outside the horizon of a radially-symmetric
black hole.  The $\lih$-function in this case can be taken as that of
equation~\eqref{bh-soln}.  The solution to this problem can be found by
adapting the work of Copson~\cite{cop28} and Linet~\cite{lin76} to the
present gauge choices.  Assuming, for simplicity, that the charge is
placed on the $z$-axis a distance $a$ from the origin ($a>2M$), the
vector potential can be written in terms of a single scalar potential
$V(r,\theta)$ as
\begin{equation}
A = V(r,\theta) \left( e_t + \frac{\sqrt{2Mr}}{r-2M} e_r \right).
\end{equation}
It follows that
\begin{alignat}{2}
\bE &= - \bgrad V & \qquad \bB &= -  \frac{\sqrt{2Mr}}{r(r-2M)}
\deriv{V}{\theta} \, \sigph \\
\bH &= 0 & \qquad \bD &= - \deriv{V}{r} \sigr - \frac{1}{r-2M}
\deriv{V}{\theta} \, \sigth 
\end{alignat}
and
\begin{equation}
\clf = -\deriv{V}{r} \, \sigr - \frac{1}{r-2M} \deriv{V}{\theta} \,
(\sigth + \sqrt{2M/r} \, I\sigph  ). \label{max-F} 
\end{equation} 
The Maxwell equations now reduce to the single partial differential
equation
\begin{equation}
\frac{1}{r^2} \deriv{}{r} \left( r^2 \deriv{V}{r} \right) +
\frac{1}{r(r-2M)} \frac{1}{\sin\! \theta} \deriv{}{\theta} \left( \sin\!
\theta \deriv{V}{\theta} \right) = - \rho,
\end{equation}
where $\rho= q\delta(\bx-\ba)$ is a $\delta$-function at $z=a$.  This
was the problem originally tackled by Copson~\cite{cop28} who obtained
a solution that was valid locally in the vicinity of the charge, but
contained an additional pole at the origin.  Linet~\cite{lin76}
modified Copson's solution by removing the singularity at the origin
to produce a potential $V(r,\theta)$ whose only pole is on the
$z$-axis at $z=a$.  Linet's solution is
\begin{equation}
V(r,\theta) = \frac{q}{ar} \frac{(r-M)(a-M) - M^2 \cos^2\!\theta}{D} +
\frac{qM}{ar},
\label{max-V}
\end{equation}
where
\begin{equation}
D = \bigl( r(r-2M) + (a-M)^2 -2(r-M)(a-M) \cos\!\theta + M^2
\cos^2\!\theta \bigr)^{1/2} .
\end{equation}

\begin{figure}[t!]
\begin{center}
\includegraphics[width=7cm,angle=-90,clip=]{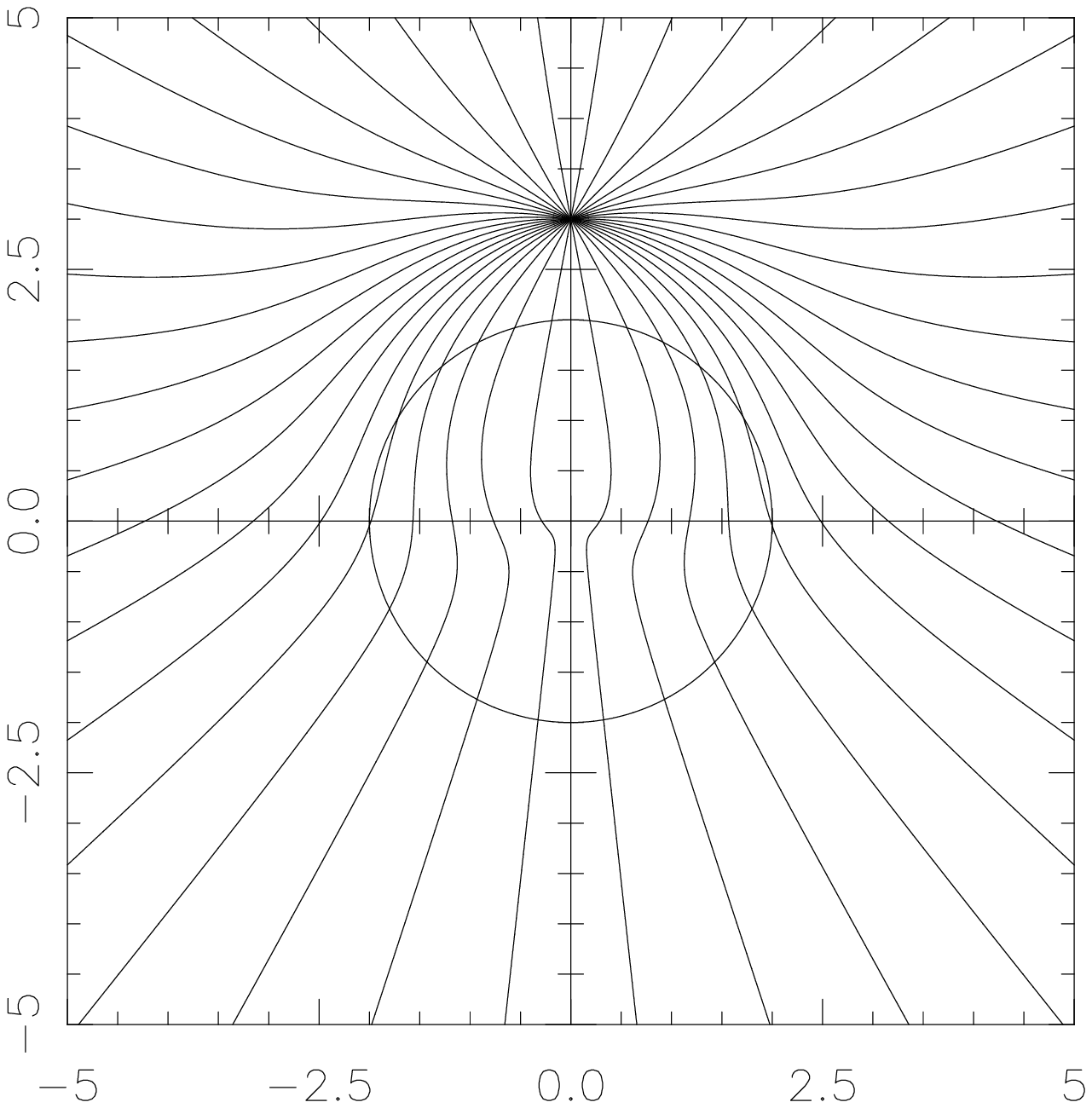}
\includegraphics[width=7cm,angle=-90,clip=]{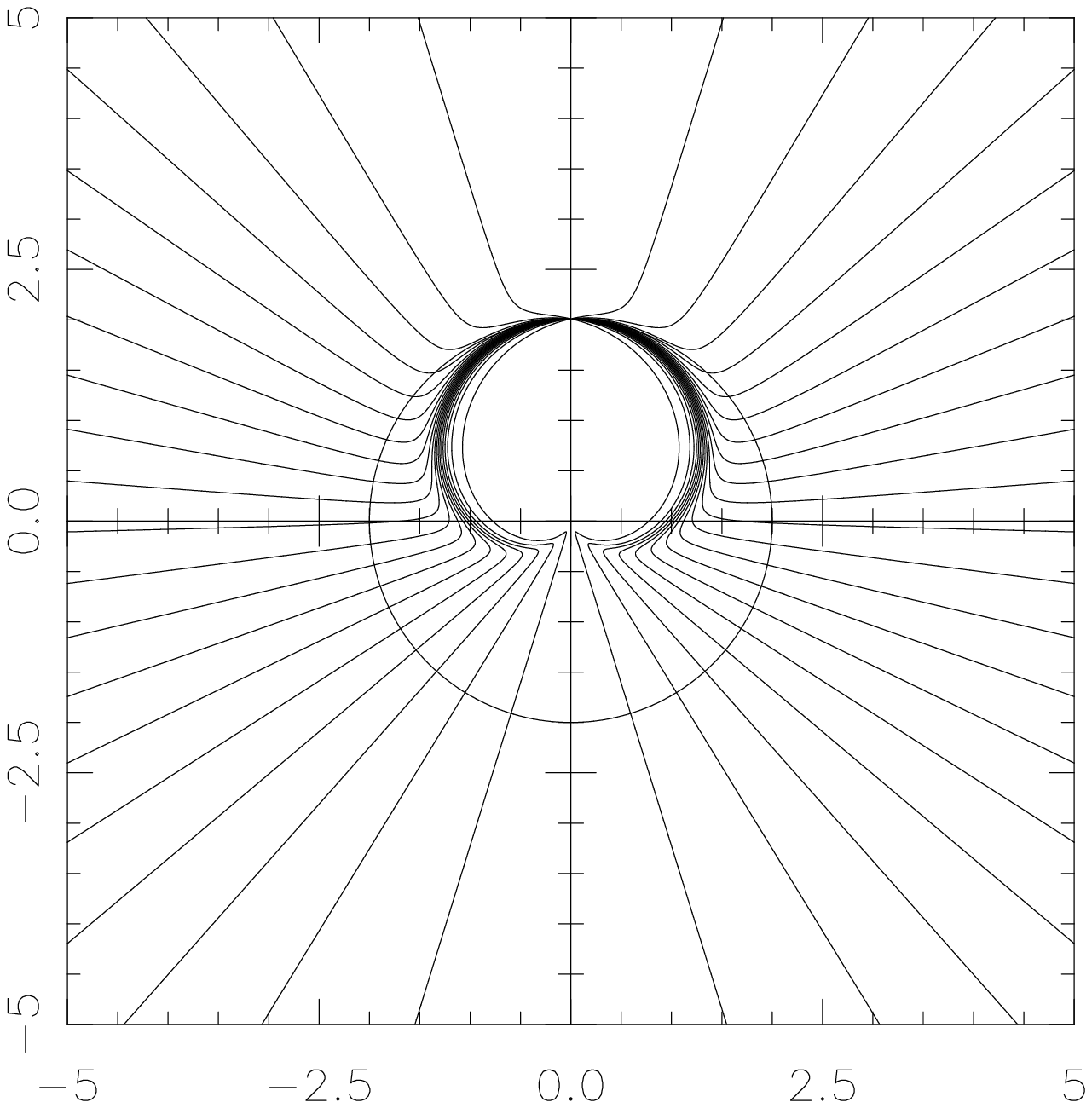}
\end{center}
\caption[dummy1]{\sl  Streamlines of the $\bD$ field.  The horizon
is at $r=2$ and the charge is placed on the $z$-axis.  The charge is
at $z=3$ and $z=2.01$ for the top and bottom diagrams respectively.
The streamlines are seeded so as to reflect the magnitude of $\bD$.
The streamlines are attracted towards the origin but never actually
meet it.  Note the appearance of a `cardiod of avoidance' as the
charge gets very close to the horizon.  The equation for this cardiod
is $r=M(1+\cos\!\theta)$, which is found by setting $D=0$ when
$a=2M$.}
\label{max-Fig1}
\end{figure}

An important feature of this solution is that once~\eqref{max-V} is
inserted back into~\eqref{max-F} the resultant $\clf$ is both finite and
continuous at the horizon.  Furthermore, since $\ho$ is well-defined
at the horizon, both $F$ and $G$ must also be finite and continuous
there.  Working in the Newtonian gauge has enabled us to construct a
global solution, and we can therefore study its global properties.
One simple way to illustrate the global properties of the solution is
to plot the streamlines of $\bD$ which, from equation~\eqref{F-diel}, is
divergenceless away from the source.  The streamlines should therefore
spread out from the charge and cover all space.  Since the distance
scale $r$ is fixed to the gravitationally-defined distance, the
streamlines of $\bD$ convey genuine intrinsic information.  Hence the
plots are completely unaffected by our choice for the $g_1$ or $g_2$
functions, or indeed our choice of $t$-coordinate.
Figure~\ref{max-Fig1} shows streamline plots for charges held at
different distances above the horizon.  Similar plots were first
obtained by Hanni~\& Ruffini~\cite{han73} although, as they worked
with the Schwarzschild metric, they were unable to extend their plots
through the horizon.  The plots reveal an effective contraction in the
radial direction.  It is not hard to show that the contraction is
precisely that of a particle moving with the free-fall velocity
$(2M/a)^{1/2}$ relative to a fixed observer.

The description presented here of the fields due to a point charge is
very different from that advocated by the `Membrane
Paradigm'~\cite{tho-mem}.  The membrane paradigm was an attempt to
develop the theory of black holes in a way that, as far as possible,
employed only familiar physical concepts.  In this way, gravitational
effects could be incorporated correctly without requiring an
understanding of the full theory.  The hope was that
astrophysicists would adopt this paradigm when modelling regions where
black-hole physics could be significant, such as at the heart of a
quasar.  The paradigm works by drawing a veil over the horizon (the
membrane) and concentrating on the physics outside the horizon as seen
by observers remaining at a fixed distance (fiducial observers).  Our
view, however, is that it is the Newtonian gauge which provides the
clearest understanding of the physics of black holes while requiring
minimal modification to Newtonian and special-relativistic ideas.
Furthermore, writing the Maxwell equations in the form~\eqref{F-diel}
removes any difficulties in applying conventional reasoning to the
study electromagnetism in a gravitational background.

There are other ways that our approach offers advantages over the
membrane paradigm.  When applying the membrane paradigm one has to
work with quantities which are singular at the horizon, and this is
hardly a recipe for applying traditional intuition!  As we have seen,
once formulated in the Newtonian gauge (or any other gauge admissible
in GTG) all physical quantities are finite.  The membrane paradigm
also warns physicists against producing plots such as
Figure~\ref{max-Fig1}, because such plots depend on the choice of
radial coordinate.  But our intrinsic approach makes it clear that
such plots \textit{are} meaningful, because $r$ is determined uniquely
by the Riemann tensor.  Presenting the plots in the form of
Figure~\ref{max-Fig1} enables direct physical information to be read
off.  In short, the simple physical picture provided by our intrinsic
method and Newtonian-gauge solution disposes of any need to adopt the
artificial ideas advocated by the membrane paradigm.

\subsection{Polarisation repulsion}

An interesting feature of the above solution~\eqref{max-V} is the
existence of a repulsive `polarisation' force~\cite{smi80}, one effect
of which is that a smaller force is needed to keep a charged particle
at rest outside a black hole than an uncharged one.  In their
derivation of this force, Smith~\& Will~\cite{smi80} employed a
complicated energy argument that involved renormalising various
divergent integrals.  Here we show that the same force can be derived
from a simple argument based on the polarisation effects of the
dielectric described by a black hole.  First, however, we must be
clear how force is defined.  In the presence of an electromagnetic
field the equation of motion for a point particle~\eqref{PPeom} is
modified to
\begin{equation}
m \vdot = (q \clf - m \om(v)) \dt v.
\label{force2}
\end{equation}
We therefore expect that any additional force should also be described
by a covariant bivector which couples to the velocity the same way
that $\clf$ does.  So, if we denote the externally applied force as
$W$, the equation of motion for a neutral test particle becomes
\begin{equation}
m\vdot = (W - m \om(v)) \dt v.
\label{eom1} 
\end{equation} 
Now, suppose that $W$ is chosen so that the particle remains at a
fixed distance $a$ outside the horizon of a black hole.  The equation
for the trajectory is
\begin{equation}
x (\tau) = t(\tau) e_t + a e_r(\theta_0, \phi_0),
\end{equation} 
where the constants $\theta_0$ and $\phi_0$ specify the angular
position of the particle.  The covariant velocity is
\begin{equation}
v = \tdot \hu^{-1}(e_t),
\end{equation}
and the condition that $v^2=1$ forces
\begin{equation}
\tdot^2 (1- 2M/a) = 1,
\end{equation} 
so that $\tdot$, and hence $v$, are constant.  Equation~\eqref{eom1} now
reduces to
\begin{gather}
W - m\om(v) = 0 \\
\implies \quad W = m \tdot \Om(e_t) = \frac{Mm}{a^2}
\left(1-\frac{2M}{a} \right)^{-1/2} \sig_a,
\end{gather}
where $\sig_a$ is the unit outward spatial vector from the source to
the particle.  The magnitude of the force is therefore 
\begin{equation}
|W| =  \frac{Mm}{a^2} \left(1-\frac{2M}{a} \right)^{-1/2}
\label{force1}
\end{equation} 
and it is not hard to check that this result is gauge-invariant.
Equation~\eqref{force1} reduces to the Newtonian formula at large
distances and becomes singular as the horizon is approached, where an
infinite force is required to remain at rest.

We now want to see how this expression for the force is modified if
the particle is charged and feels a force due to its own polarisation
field.  From~\eqref{force2} the extra term in the force is simply $\clf$,
and only the radial term contributes.  From equation~\eqref{max-F} this is
just $-\dr V \,\sigr$.  Since the charge lies on the $z$ axis, we need
only look at $V$ along this axis, for which
\begin{equation}
V(z) =  \frac{q}{|z-a|} - \frac{qM}{|z-a|} \left( \frac{1}{a} +
\frac{1}{z} \right) + \frac{qM}{az}.
\end{equation} 
The singular terms must be due to the particle's own Coulomb field,
and so do not generate a polarisation force.  The only term that
generates a force is therefore the final one, which is precisely the
term that Linet added to Copson's original formula!  This term
produces an outward-directed force on the charge, of magnitude $q^2
M/a^3$.  The applied force is therefore now
\begin{equation}
W = \left( \frac{Mm}{a^2} \left(1-\frac{2M}{a} \right)^{-1/2} -
\frac{Mq^2}{a^3} \right) \sig_a,
\end{equation}  
so a smaller force is needed to keep the particle at rest outside the
horizon.  This result agrees with that found in~\cite{smi80}, though
our derivation avoids the need for infinite mass renormalisation and
is considerably simpler.  This result is a good example of the
importance of finding global solutions.  The polarisation force is
felt outside the horizon, yet the correction term that led to it was
motivated by the properties of the field at the origin.

\subsection[Point charge in a $k>0$ cosmology]{Point charge in a
{\boldmath $k>0$} cosmology} 
\label{MAX-COS}

We saw in Section~\ref{SS-COS} that one form of the $\lih$-function for
a homogeneous cosmology is defined by~\eqref{cos-stat}
\begin{equation}
\ho(a) = a \dt e_t \, e_t + \alp \bigl( (1-kr^2)^{1/2} a \dt e_r \, e^r +
a \wdg \sigr \, \sigr \bigr),
\label{EM-cos}
\end{equation}
where $\alp=1/S$.  When $k$ is positive, however, the
function~\eqref{EM-cos} is undefined for $r>k^{-1/2}$ and so fails to
define a global solution.  A globally-valid solution is obtained with
the displacement
\begin{equation}
f(x) = x\dt e_t \, e_t + \frac{r}{1+k r^2 /4} \, e_r,
\label{EM-cosdisp}
\end{equation}
which results in the particularly simple solution
\begin{equation}
\ho'(a) = a \dt e_t \, e_t +  \alp (1+ kr^2 /4) a \wdg e_t \, e_t .
\label{EM-cossol}
\end{equation}
This solution is well-defined for all $x$ and generates an
`isotropic' line element.  (The solution can also be viewed as
resulting from a stereographic projection of a 3-sphere.)

We want to find the fields due to a point charge in the background
defined by~\eqref{EM-cossol}.  Since the $\lih$-function is diagonal we
start with the obvious ans\"{a}tz
\begin{equation}
A = \alp V(\bx) \, e_t,
\end{equation}
so that
\begin{equation}
\bE = - \alp \bgrad V
\end{equation}
and
\begin{equation}
\bD = - ( 1+ k r^2/4 )^{-1} \bgrad V.
\end{equation}
It follows that the equation we need to solve is simply
\begin{equation}
- \bgrad \dt \bigl(  ( 1+ k r^2/4 )^{-1}\bgrad V \bigr) = q
\del(\bx-\ba),
\label{EM-coseq}
\end{equation}
where the charge $q$ is located at $\bx=\ba$.  This equation can be
solved using the general technique described by Hadamard (see also
Copson~\cite{cop28} for a discussion of a similar problem.)  The
solution to~\eqref{EM-coseq} turns out to be
\begin{equation}
V = \frac{1+ ka^2/4 - ks/2}{(s(1+ ka^2/4 - ks/4))^{1/2}},
\label{EM-cos2}
\end{equation}
where
\begin{equation}
s= \frac{(\bx-\ba)^2}{1+kr^2/4}, \qquad a = \sqrt{\ba^2}.
\end{equation}
As a simple check, $V$ reduces to the usual Coulomb potential when
$k=0$.

\begin{figure}[t!]
\begin{center}
\begin{picture}(300,270)
\put(0,270){\hbox{\epsfig{figure=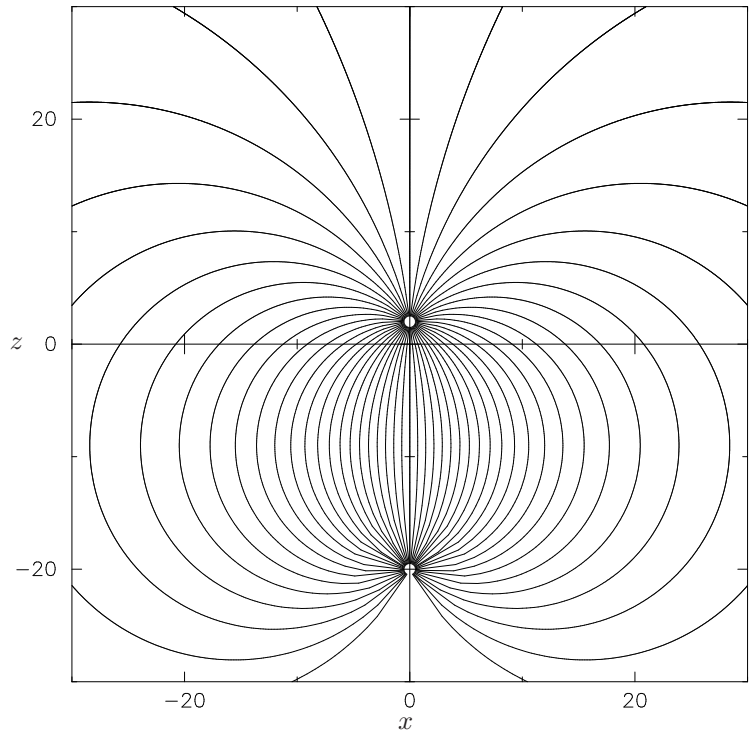,height=10cm,angle=-90}}}
\put(147,-10){$x$}
\put(0,134){$z$}
\end{picture}
\end{center}
\caption[dummy1]{\sl Fieldlines of the $\bD$-field for a point charge
in a $k >0$ universe.  The fieldlines follow null geodesics, which are
circles in this gauge.  The existence of an image charge is clear.}
\label{max-Fig3}
\end{figure}

Fieldline plots of the $\bD$ field defined from~\eqref{EM-cos2} are shown
in Figure~\ref{max-Fig3}.  The fieldlines follow null geodesics and
clearly reveal the existence of an image charge.  The reason for this
is can be seen in the denominator of $V$.  Not only is $V$ singular at
$s=0$, it is also singular at
\begin{equation}
1+ ka^2/4 - ks/4 = 0.
\end{equation}
So, if the charge lies on the $z$ axis, the position of its
image is found by solving
\begin{equation}
\frac{k(z-a)^2}{4+kz^2} = 1 +  \frac{ka^2}{4}
\end{equation}
from which we find that the image charge is located at
\begin{equation}
z = -4/(ka).
\end{equation}
(This is the stereographic projection of the opposite point to the
charge on a 3-sphere.)  However, if we try to remove this image charge
by adding a point source at its position, we find that the fields
vanish everywhere, since the new source has its own image which cancels
the original charge.  The image charge is therefore an unavoidable
feature of $k>1$ cosmologies.  We comment on this further in
Section~\ref{Cos}.


\section{The Dirac equation in a gravitational background} 
\label{DE} 

In this paper we began by introducing the gravitational gauge fields
and minimally coupling these to the Dirac action.  For our final major
application we return to the Dirac equation in a gravitational
background.  In the absence of an $\cla$ field, the minimally-coupled
equation~\eqref{G-d19} is
\begin{equation}
D \psi \isk = m \psi \go .
\label{D-eq}
\end{equation}
Here we consider two applications: the case of a black-hole
background; and cosmological models in which the universe is not at
critical density.

\subsection{Black hole background}

We have already demonstrated that the Newtonian gauge solution
dramatically simplifies the study of black-hole physics, so this is
the natural gauge in which to study the Dirac equation.  We therefore
start with an analysis in this gauge, and then consider the
gauge-invariance of the predictions made.  With the gravitational
fields as described in Section~\ref{SS-BH}, the Dirac
equation~\eqref{D-eq} becomes
\begin{equation}
\grad \psi \isk - (2M/r)^{1/2} \go \bigl( \dr \psi + 3/(4r) \psi
\bigr) \isk = m \psi \go. 
\label{DBH1}
\end{equation}
If we pre-multiply by $\go$ and introduce the symbol $i$ to represent
right-sided multiplication by $\isk$, so that $i\psi\eqv\psi \isk$,
then equation~\eqref{DBH1} becomes
\begin{equation}
i \dift \psi = -i \bgrad \psi + i (2M/r)^{1/2} r^{-3/4} \dr \bigl(
r^{3/4} \psi \bigr) + m \psibar ,
\label{DBH2}
\end{equation}
where $\psibar\eqv\go\psi\go$.  One feature that emerges immediately
is that the Newtonian gauge has recovered a Hamiltonian form of the
Dirac equation (see~\cite{DGL95-elphys} for a discussion of operators,
Hamiltonians and Hermiticity in the spacetime algebra approach to
Dirac theory).  Since the Newtonian gauge involves the notion of a
global time, it might have been expected that this gauge would lend
itself naturally to a Hamiltonian formulation.  An open question of
some interest is whether it is always possible to make such a gauge
choice (see~\cite{D00-kerr,gap} for an extension to the Kerr
solution).

The Hamiltonian~\eqref{DBH2} contains a subtlety: it is Hermitian only
away from the origin.  To see why, consider the interaction term
\begin{equation}
H_I (\psi) = i  (2M/r)^{1/2} r^{-3/4} \dr \bigl(r^{3/4} \psi \bigr) .
\end{equation}
For this we find that
\begin{align}
\int d^3 x \, \la \phidag H_I(\psi) \ra_S
&= \sqrt{2M} \int d\Om \, \int_0^\infty r^2 \,dr\, r^{-5/4} \la
\phidag \dr (r^{3/4} \psi) \isk \ra_S \nn \\
&= \sqrt{2M} \int d\Om \, \int_0^\infty dr\, \la r^{3/4} \phidag \dr
(r^{3/4} \psi) \isk \ra_S \nn \\
&= \int  d^3 x \, \la (H_I(\phi)^\dagger \psi \ra_S \, + 
\sqrt{2M} \int d\Om \, \Bigl[ r^{3/2} \la \phidag \psi \isk \ra_S
\Bigr]_0^\infty ,
\label{DBH3}
\end{align} 
where $\la \, \ra_S$ denotes the projection onto the `complex' $1$ and
$\isk$ terms and $\phidag\eqv\go\phirev\go$ (see
Appendix~\ref{app-dirac} and~\cite{DGL95-elphys}).  For all normalised
states the final term in~\eqref{DBH3} tends to zero as
$r\rightarrow\infty$.  But it can be shown from~\eqref{DBH2} that
wavefunctions tend to the origin as $r^{-3/4}$, so the lower limit is
finite and $H_I$ is therefore not a Hermitian operator.  This
immediately rules out the existence of normalisable stationary states
with constant real energy~\cite{bstates}.

Equation~\eqref{DBH2} can be used to propagate a spinor defined over some
initial spatial surface, and numerical simulations based on this
equation give a good picture of the scattering induced by a black
hole.  Here, however, we wish to focus on an analytical approach.  A
result that follows immediately from the Hamiltonian form of the Dirac
equation is that~\eqref{DBH2} is manifestly separable, so that we can
write
\begin{equation}
\psi(x) =  \psi(\bx) \alp(t).
\end{equation}
As usual, the solution of the $t$-equation is
\begin{equation}
\alp(t) = \exp(-\isk \, Et),
\end{equation}
where $E$ is the separation constant.  The non-Hermiticity of $H_I$
means that $E$ cannot be purely real if $\psi$ is normalisable.  The
imaginary part of $E$ is determined by equation~\eqref{DBH3} and, for
suitably normalised states, we find that
\begin{equation}
\mbox{Im}(E) = - \lim_{r\rightarrow 0} \, 2 \pi \sqrt{2M} \la \psidag \psi
\ra r^{3/2}.
\end{equation}
This equation shows that the imaginary part of $E$ is necessarily
negative, so the wavefunction decays with time.  This is consistent
with the fact that the streamlines generated by the conserved current
$\psi\go\psirev$ are timelike curves and, once inside the horizon,
must ultimately terminate on the origin.

With the $t$-dependence separated out, equation~\eqref{DBH2} reduces to
\begin{equation}
\bgrad \psi -  (2M/r)^{1/2} r^{-3/4} \dr \bigl(r^{3/4} \psi \bigr) = i
E \psi - i m \psibar. 
\label{DBH4}
\end{equation}
To solve this equation we next separate out the angular dependence.
This is achieved using the spherical monogenics, which are the
spacetime algebra equivalent of the $\chi_{jlm}$ Pauli spinors.  These
are described in detail in~\cite{DGL95-elphys} and here we quote the
necessary formulae.  The unnormalised monogenic $\psi^m_l$ is defined
by
\begin{equation}
\psi_l^m = \bigl((l+m+1)P_l^m(\cos\!\theta) -
P_l^{m+1}(\cos\!\theta) I\sig_\phi \bigr) \et{m\phi\isk},
\end{equation} 
where $l\geq 0$, $-(l+1) \leq m \leq l$, and $P^m_l$ are the associated
Legendre polynomials.  The two properties of the $\psi^m_l$ relevant
here are
\begin{align}
\bgrad \psi^m_l &= - (l/r) \sig_r \psi_l^m \\
\intertext{and}
\bgrad \bigl(\sigr \psi^m_l \bigr) &= (l+2)/r \, \psi_l^m .
\end{align}
Now, the operator $K(\psi) = \go (1-\bx \wdg \bgrad ) \psi \go$
commutes with the Hamiltonian defined by~\eqref{DBH2}.  Constructing
eigenstates of this operator with eigenvalue $\kap$, leads to
solutions of the form 
\begin{equation}
\psi(\bx, \kap) = 
\begin{cases}
\psi^m_l u(r) + \sigr \psi^m_l v(r) \isk & \kap = l+1 \\
\sigr \psi^m_l u(r) \sk + \psi^m_l Iv(r) & \kap = -(l+1) ,
\end{cases}
\label{DBH5}
\end{equation}
where $\kap$ is a non-zero integer and $u(r)$ and $v(r)$ are complex
functions of $r$ (\textit{i.e.}  sums of a scalar and an $\isk$ term).
Substituting~\eqref{DBH5} into equation~\eqref{DBH4}, and using the properties
of the spherical monogenics, we arrive at the coupled radial
equations
\begin{equation}
\begin{pmatrix} 1 & - (2M/r)^{1/2}  \\ - (2M/r)^{1/2} & 1 
\end{pmatrix}
\begin{pmatrix} u_1' \\ u_2' \end{pmatrix} 
= \bA \begin{pmatrix} u_1 \\ u_2 \end{pmatrix}
\label{DBH7}
\end{equation}
where
\begin{equation}
\bA \eqv 
\begin{pmatrix} \kap/r & i(E+m) - (2M/r)^{1/2}(4r)^{-1} \\ 
i(E-m) - (2M/r)^{1/2}(4r)^{-1} & - \kap/r \end{pmatrix},
\end{equation} 
$u_1$ and $u_2$ are the reduced functions defined by
\begin{equation}
u_1 = ru \qquad u_2 = i r v,
\end{equation}
and the primes denote differentiation with respect to $r$.  (We
continue to employ the abbreviation $i$ for $\isk$.)

To analyse~\eqref{DBH7} we first rewrite it in the equivalent form
\begin{equation}
\bigl(1-2M/r \bigr)
\begin{pmatrix} u_1' \\ u_2' \end{pmatrix} 
= \begin{pmatrix} 1 & (2M/r)^{1/2}  \\ (2M/r)^{1/2} & 1 
\end{pmatrix} \bA \begin{pmatrix} u_1 \\ u_2 \end{pmatrix}.
\label{DBH8}
\end{equation} 
This makes it clear that the equations have regular singular points at
the origin and horizon ($r=2M$), as well as an irregular singular
point at $r=\infty$.  To our knowledge, the special function theory
required to deal with such equations has not been developed.  Without
it we either attempt a numerical solution, or look for power series
with a limited radius of convergence.  Here we consider the latter
approach, and look for power-series solutions around the horizon.  To
this end we introduce the series
\begin{equation}
u_1 = \eta^s \sum_{k=0}^{\infty} a_k \eta^k, 
\qquad u_2 = \eta^s \sum_{k=0}^{\infty} b_k \eta^k, 
\label{DBH9}
\end{equation}	
where $\eta = r-2M$.  The index $s$ controls the radial dependence of
$\psi$ at the horizon, so represents a physical quantity.  To
find the values that $s$ can take, we substitute equation~\eqref{DBH9}
into~\eqref{DBH8} and set $\eta=0$.  This results in the equation
\begin{equation}
\frac{s}{2M}
\begin{pmatrix} a_0 \\ b_0 \end{pmatrix} 
= 
\begin{pmatrix} 1 & 1  \\ 1 & 1 \end{pmatrix}
\begin{pmatrix} \kap/(2M) & i(E+m) - (8M)^{-1} \\
i(E-m) - (8M)^{-1} & - \kap/(2M) \end{pmatrix}
\begin{pmatrix} a_0 \\ b_0 \end{pmatrix}
\end{equation}  
Rewriting this in terms of a single matrix and setting its determinant
to zero yields the two indicial roots
\begin{equation}
s=0 \hs{0.5} \mbox{and} \hs{0.5} s = -\half + 4iME.
\end{equation} 
The $s=0$ solution is entirely sensible --- the power series is
analytic, and nothing peculiar happens at the horizon.  The existence
of this root agrees with our earlier observation that one can evolve
the time-dependent equations without encountering any difficulties at
the horizon.  The second root is more problematic, as it leads to
solutions that are ill-defined at the horizon.  Before assessing the
physical content of these roots, however, we must first check that
they are gauge invariant.

If, instead of working in the Newtonian gauge, we keep the gauge
unspecified then, after separating out the angular dependence, the
equations reduce to
\begin{equation}
\begin{pmatrix} L_r & L_t  \\ L_t & L_r \end{pmatrix} 
\begin{pmatrix} u_1 \\ u_2 \end{pmatrix} 
=
\begin{pmatrix} \kap/r-G/2 & im -F/2  \\ -im-F/2 &  -\kap/r-G/2
\end{pmatrix}  
\begin{pmatrix} u_1 \\ u_2 \end{pmatrix} .
\label{DBH12}
\end{equation}
We can still assume that the $t$-dependence is of the form
$\exp(-iEt)$, so that equation~\eqref{DBH12} becomes
\begin{equation}
\begin{pmatrix} g_1 & g_2  \\ g_2 & g_1 
\end{pmatrix}
\begin{pmatrix} u_1' \\ u_2' \end{pmatrix} 
= \bB
\begin{pmatrix} u_1 \\ u_2 \end{pmatrix}
\label{DBH15}
\end{equation} 
where
\begin{equation}
\bB \eqv \begin{pmatrix} 
\kap/r-G/2 +if_2 E & i(m+f_1 E) -F/2  \\ -i(m-f_1 E) -F/2 & 
-\kap/r-G/2 +if_2 E \end{pmatrix} . 
\end{equation}
Now, since ${g_1}^2-{g_2}^2 = 1-2M/r$ holds in all gauges, we obtain
\begin{equation}
\bigl(1-2M/r \bigr)
\begin{pmatrix} u_1' \\ u_2' \end{pmatrix} 
= 
\begin{pmatrix} g_1 & -g_2 \\ -g_2 & g_1 \end{pmatrix} 
\, \bB \, 
\begin{pmatrix} u_1 \\ u_2 \end{pmatrix},
\end{equation}
and substituting in the power series~\eqref{DBH9} and setting $\eta=0$
produces the indicial equation
\begin{equation}
\det \left[ \begin{pmatrix} g_1 & -g_2 \\ -g_2 & g_1 \end{pmatrix} \,
\bB - \frac {s}{r} I \right]_{r=2M} = 0,
\end{equation}
where $I$ is the identity matrix.  Employing the result that
\begin{equation}
g_1G - g_2 F = \half \dr ({g_1}^2-{g_2}^2) = M/r^2
\end{equation}
we find that the solutions to the indicial equation are now
\begin{equation}
s=0 \hs{0.5} \mbox{and} \hs{0.5} s = -\half + 4iME (g_1 f_2-g_2 f_1).
\end{equation}
But in Section~\ref{SS-BH} we established that $g_1 f_2-g_2 f_1=+1$ at
the horizon for all solutions with a forward time direction.  This
demonstrates that the indices are indeed intrinsic, with the sign of
the imaginary term for the singular root picking up information about
the time direction implicit in the presence of the horizon.

The fact that $s=0$ is always a solution of the indicial equation
means that solutions always exist that are analytic at the horizon.
Determining the split between ingoing and outgoing states of these
solutions enables one to calculate reflection coefficients and
scattering amplitudes~\cite{gap}.  The question we wish to consider is
whether the second, singular, root can be physically significant.  To
address this we look at the current.  The covariant current $\clj$ is
given by $\psi\go\psirev$, and satisfies $\cld\dt\clj=0$.  The
corresponding non-covariant quantity is therefore
\begin{equation}
J = \hu(\psi\go\psirev) \dhi
\end{equation}
which satisfies the flatspace conservation equation $\grad\dt J=0$.
It is the streamlines of $J$ that are plotted as functions of $x$ and
determine the flow of density.  The crucial terms in $J$ are the time
component and radial component, which (ignoring the overall
exponential decay term) are given by
\begin{align}
\go\dt J &= \frac{1}{r^2} \bigl( f_1 (u_1 \urev_1 + u_2 \urev_2) + f_2
(u_1 \urev_2 + \urev_1 u_2 ) \bigr) \psi_l^m \psirev_l^m  \\
\intertext{and}
e^r \dt J &= \frac{1}{r^2} \bigl( g_1 (u_1 \urev_2 + \urev_1 u_2 ) +
g_2  (u_1 \urev_1 + u_2 \urev_2) \bigr) \psi_l^m \psirev_l^m. 
\end{align}
The $\{f_i\}$ and $\{g_i\}$ are finite for all admissible solutions
so, for the $s=0$ solution, the components of $J$ are well-defined at
the horizon.  Furthermore, it is easily shown that for $s=0$ the
radial flux at the horizon always points inwards.  The $s=0$ root
therefore describes the case where the flux crosses the horizon and
continues onto the singularity.

For the singular root we must first decide on a branch for the
solution so that $\psi$ is fully specified on both sides of the
horizon.  We can then assess whether the discontinuity in $\psi$, and
the discontinuity in the current generated by it, are physically
acceptable.  To do this, we first write $\eta^s$ as
\begin{equation}
\eta^s = \exp \bigl( (-\half + 4iME) \ln(r-2M) \bigr).
\end{equation}
We can now write
\begin{equation}
\ln(r-2M) = \ln|r-2M| + i \arg(r-2M),
\end{equation}
and for the choice of argument we take
\begin{equation}
\arg(r-2M) =
\begin{cases}
0 & r> 2M \\
-\pi & r < 2M.
\label{DBH12a}
\end{cases}
\end{equation}
(The choice of sign will be discussed further below.)  If we now take
the limit $r\rightarrow 2M$ from above and below we find that the
$\go$ component of $J$ is given by
\begin{equation}
\go \dt J = A(\theta,\phi) \et{-2\eps t} |r-2M|^{\mbox{\footnotesize
$-1 + 8M\eps$}} \times 
\begin{cases}
1 & r > 2M \\
\exp (8 \pi M E_r) & r < 2M
\end{cases}
\label{DBH20}
\end{equation}
where $A(\theta,\phi)$ is a positive-definite, finite term and we have
split $E$ into real and imaginary parts as
\begin{equation}
E = E_r - i \eps.
\end{equation} 
Equation~\eqref{DBH20} is valid in all gauges for which $g_1 f_2-g_2
f_1=+1$ at the horizon.  While the density $\go\dt J$ is singular
at the horizon, the presence of the positive term $8M\eps$ ensures
that any integral over the horizon is finite and the solution is
therefore normalisable.  This link between the properties of $\psi$ at
the horizon and at the origin (where $\eps$ is determined) provides
another example of the importance of finding global solutions to the
field equations.  The radial current now turns out to be
\begin{equation}
e^r \dt J = \frac{A(\theta,\phi)}{4M} \et{-2\eps t}
|r-2M|^{\mbox{\footnotesize $8M\eps$}} \times  
\begin{cases}
1 & r > 2M \\
-\exp(8 \pi M E_r) & r < 2M
\end{cases}
\label{DBH21}
\end{equation}
and is therefore zero at the horizon, and inward-pointing everywhere
inside the horizon.  It appears that the existence of the imaginary
contribution to $E$ does ensure that that the singular solutions have
sensible physical properties, and the singularity in $\psi$ at the
horizon is no worse than that encountered in the ground state of the
hydrogen atom~\cite{bjo-rel1}.  What is less clear, however, is the
extent to which the properties of $\psi$ at the horizon are compatible
with the original equation~\eqref{D-eq}.  In particular, since $\psi$ is
both singular and non-differentiable at the horizon, it does not
appear that the singular root can be viewed as defining a solution
of~\eqref{D-eq} over all space.

\subsection{The Hawking temperature}

A number of authors have attempted to give derivations of the Hawking
temperature and particle flux due to a black hole from an analysis of
first-quantised theory, \textit{i.e.\/} from the properties of wave
equations alone~\cite{dam76,liu81,zha82}.  This work has generated
some controversy~\cite{mart77}, so it is interesting to assess how the
ideas stand up in GTG.  These derivations focus on the singular
solutions to the wave equation (either Klein--Gordon or Dirac), and
study the properties of these solutions under the assumption that the
energy is real.  If one ignores the problems that $\eps=0$ introduces
for the normalisability of $\psi$ and presses ahead, then
from~\eqref{DBH21} there is now a non-zero current at the horizon and,
furthermore, there is a net creation of flux there.  The ratio of the
outward flux to the total flux is simply
\begin{equation}
\frac{e^r\dt J_+}{e^r\dt J_+ - e^r\dt J_-} = \frac{1}{\et{8\pi ME} +1} 
\label{DBH30}
\end{equation}
which defines a Fermi--Dirac distribution with temperature
\begin{equation}
T = \frac{1}{8\pi M k_B}.
\end{equation}
Remarkably, this is the temperature found by Hawking~\cite{haw74}.
The fact that both the correct Fermi--Dirac statistics and Hawking
temperature can be derived in this manner is astonishing, since both
are thought to be the result of quantum field theory.  But what can we
really make of this derivation?  The first problem is that setting
$\eps$ to zero means that the density is no longer normalisable at the
horizon --- any integral of the density over the horizon region
diverges logarithmically, which is clearly unphysical.  A further
problem relates to the choice of branch~\eqref{DBH12a}.  Had the opposite
branch been chosen we would not have obtained~\eqref{DBH30} and, as
pointed out in~\cite{mart77}, there is no \textit{a priori\/}
justification for the choice made in~\eqref{DBH12a}.

For the above reasons, the `derivation' of~\eqref{DBH30} cannot be viewed
as being sound.  The remarkable thing is that the same techniques can
be used to `derive' the correct temperatures for the
Reissner--Nordstr\"{o}m and Rindler cases, as well as the Schwinger
production rate in a constant electric field.  This will be discussed
elsewhere.  These further analyses contain another surprise: the
temperature at the interior horizon of a Reissner--Nordstr\"{o}m black
hole is necessarily negative!  However, while these analyses are both
interesting and suggestive, it is only through a study of the full
quantum field theory in a black hole background that one can be sure
about particle production rates.

A final point in this section is that all our analyses have been based
on working with the correct time-asymmetric solutions admitted in GTG.
On attempting to force through the analysis in the `Schwarzschild'
gauge ($g_2=f_2=0$), one discovers that the indices are now given by
\begin{equation}
s = -\half \pm 4iME .
\end{equation}
In this case no analytic solution is possible, and even the presence
of an exponential damping factor does not produce a normalisable
current at the horizon.  This only serves to reinforce the importance
of working with global solutions, since there is no doubt that the
presence of non-singular, normalisable solutions is an intrinsic
feature of horizons.

\subsection{The Dirac equation in a cosmological background}
\label{DE-COS}

As a second application we consider the Dirac equation in a
cosmological background.  We have a choice of form of
$\lih$-function to use, of which the simplest is that defined
by~\eqref{cos-stat}, 
\begin{equation}
\ho'(a) = a \dt e_t e_t + \frac{1}{S} \bigl( (1-kr^2)^{1/2} a \dt e_r
\, e^r + a \wdg \sigr \, \sigr \bigr).
\end{equation}
The Dirac equation in this background takes the form
\begin{multline}
\qquad
\left( e^t \partial_t + \frac{1}{S} \bigl( (1-kr^2)^{1/2} e^r \dr + e^\theta
\partial_\theta + e^\phi \partial_\phi \bigr) \right) \psi \isk \\
+ \left( \frac{3}{2} H(t) e_t - \frac{1}{rS}\bigl((1-kr^2)^{1/2} -1 \bigr)
e_r \right) \psi \isk = m \psi \go, \qquad
\label{DCB1}
\end{multline}
where the various functions are as defined in Section~\ref{SS-COS}.
Our question is this: can we find solutions to~\eqref{DCB1} such that the
observables are homogeneous?  There is clearly no difficulty if $k=0$
since, with $p=0$, equation~\eqref{DCB1} is solved by 
\begin{equation}
\psi = \rho^{1/2} \et{-\isk mt}
\label{dirac-sol1}
\end{equation} 
and the observables are fixed vectors which scale as $\rho(t)$ in
magnitude~\cite{DGL-cosm-bel}.  But what happens when $k\neq 0$?  It
turns out that the solution~\eqref{dirac-sol1} must be modified
to~\cite{DGL-cosm-bel}
\begin{equation}
\psi = \frac{\rho^{1/2}}{1+ \sqrt{1-kr^2}} \et{-\isk mt}.
\label{dirac-sol2}
\end{equation} 
For the case of $k>0$ both $\ho(a)$ and $\psi$ are only defined for
$r<k^{-1/2}$.  This problem is overcome by using the
displacement~\eqref{EM-cosdisp} to transform to the global solution of
equation~\eqref{EM-cossol}.  In this case $\psi$ is given by
\begin{equation}
\psi = \half (1+ k r^2 /4) \rho^{1/2} \et{-\isk mt},
\end{equation} 
which diverges as $r\rightarrow\infty$.

In both the $k>0$ and $k<0$ cases, $\psi$ contains additional
$r$-dependence and so is not homogeneous.  Furthermore, the
observables obtained from $\psi$ are also inhomogeneous.  In principle
one could therefore determine the origin of this space from
measurements of the current density.  This clearly violates the
principle of homogeneity, though it is not necessarily inconsistent
with experiment.  The implications for cosmology of this fact are
discussed in the following section.  (Some consequences for
self-consistent solutions of the Einstein--Dirac equations are
discussed in~\cite{DGL-cosm-bel,DGL97-selfcons} and, in the context of
general relativity, in~\cite{ish74}.)

The fact that quantum fields see this `preferred' direction in
$k\neq0$ models, whereas classical phenomena do not, reflects the
gauge structure of the theory.  Dirac spinors are the only fields
whose action couples them directly to the $\om(a)$-function.  All
other matter fields couple to the gravitational field through the
$\lih$-field only.  Dirac spinors therefore probe the structure of the
gravitational fields directly at level of the $\om$-field, which is
inhomogeneous for $k\neq 0$ models.  On the other hand, classical
fields only interact via the covariant quantities obtained from the
gravitational fields, which are homogeneous for all values of $k$.
This conclusion is reinforced by the fact that the Klein--Gordon
equation, for which the action does not contain the $\om(a)$-field,
\textit{does\/} have homogenous solutions in a $k\neq 0$ universe.

 
\section{Implications for cosmology}
\label{Cos}

In Section~\ref{SS-COS} we discussed some aspects of cosmology as
examples of the general treatment of time-varying spherically-symmetric
systems.  There we drew attention to the utility of the Newtonian
gauge as a tool for tackling problems in cosmology.  In addition, in
Sections~\ref{MAX-COS} and~\ref{DE-COS} we studied the Maxwell and
Dirac equations in various cosmological backgrounds.  In this section
we draw together some of our conclusions from these
sections. Specifically, we discuss redshifts, difficulties with $k\neq
0$ models, and the definitions of mass and energy for cosmological
models.

\subsection{Cosmological redshifts}

As a final demonstration of the use of the Newtonian gauge, consider a
photon following a null path in the $\theta=\pi/2$ plane.  In this case
the photon's momentum can be written as
\begin{align}
P &= \Phi R (\go + \gi) \Rrev, \\
\intertext{where}
R &= \et{\alp/2 \, \isk}
\end{align}
and $\Phi$ is the frequency measured by observers comoving with the
fluid.  We restrict to the pressureless case, so $G=0$ and $f_1=1$,
but will allow $\rho$ to be $r$-dependent.  A simple application
of equation~\eqref{PPeom} produces
\begin{equation}
\partial_\tau \Phi = - \Phi^2 \bigl( \frac{g_2}{r} \sin^2\chi + \dr
g_2 \cos^2\chi \Bigr),
\end{equation}
where $\chi=\phi+\alp$.  But, since $f_1=1$, we find that $\partial_\tau
t =\Phi$, so
\begin{equation}
\frac{d\Phi}{dt} = - \Phi \Bigl( \frac{g_2}{r} \sin^2\chi + \dr
g_2 \cos^2\chi \Bigr),
\label{ifc1.1}
\end{equation} 
which holds in any spherically-symmetric pressureless fluid.

For the case of a cosmological background we have $g_2 = H(t) r$, so
the angular terms drop out of equation~\eqref{ifc1.1}, and we are left
with the simple equation
\begin{equation}
\frac{d\Phi}{dt} = - H(t) \Phi = \frac{\rhodot}{3\rho} \Phi,
\end{equation}
which integrates to give the familiar redshift versus density relation
\begin{equation}
1 + z = (\rho_1/\rho_0)^{1/3}.
\label{ifc1}
\end{equation}
Other standard cosmological relations, such as the luminosity distance
and angular diameter versus redshift formulae, can be easily derived
in this gauge (see also Section~\ref{SS-COS}).

In~\cite{DGL-cosm-bel} equation~\eqref{ifc1} was derived in the gauge of
equation~\eqref{cos-stat}, in which all particles comoving with the
cosmological fluid are at rest in the background spacetime.  In this
gauge the redshift can be attributed to a loss of energy to the
gravitational field, although this is a gauge-dependent viewpoint ---
the only physical statement that one can make is embodied in
equation~\eqref{ifc1}.  The explanation of cosmological redshifts in our
theory has nothing to do with `tired light', or spacetime playing a
dynamic role by expanding, or even anything to do with Doppler shifts.
The redshift is simply a consequence of the assumption of homogeneity.
Ultimately, all physical predictions are independent of the gauge in
which they are made, although certain gauges may have useful
computational or heuristic value.

\subsection[$k\neq 0$ cosmologies]{{\boldmath $k\neq 0$} cosmologies} 

At the level of classical (\textit{i.e.} non-quantum) physics, there
is no doubt that $k\neq 0$ cosmologies are homogeneous.  This is true
in both general relativity and GTG.  No prediction derived for
classical systems of point particles or for electromagnetic fields can
reveal a preferred spatial direction in these models.  However, we saw
in Section~\ref{DE-COS} that it is impossible to find homogeneous
solutions of the Dirac equation in a $k\neq 0$ universe.  The
consequences of this are, in principle, observable, since the local
density gradient will reveal a preferred radial direction.  It has
already been pointed out that it is impossible to find a
\textit{self-consistent} solution of the combined system of
Einstein--Dirac equations for any case other than a spatially flat
cosmology~\cite{DGL-cosm-bel,ish74}.  We believe that this is the
first time that it has been pointed out that even a {\em
non}-self-consistent Dirac field would be observably inhomogeneous.
This is a more damaging result for $k\neq 0$ cosmologies, since it
reveals inhomogeneity without assuming that spin-torsion effects have
anything to do with the dynamics of the universe.

While the properties of Dirac fields pose theoretical difficulties for
$k\neq 0$ models, there is no contradiction with present observations.
Furthermore, one could question the validity of inferences drawn in
extrapolating the Dirac equation to cosmological scales.  There is,
however, a purely classical effect that does lead one to question the
validity of $k>0$ models.  As we have seen, when looking at the
properties of fields in a $k>0$ background it is necessary to work in
a globally-defined gauge, such as that of equation~\eqref{EM-cossol}.  In
this case the Maxwell equations show that each point charge must have
an image charge present in a remote region of the universe.  This is a
consequence of a closed universe that we have not seen discussed,
although it has doubtless been pointed out before.  The necessity for
this image charge raises many problems in attempting to take such a
universe seriously.

\subsection{Mass and energy in cosmological models}

Setting aside the problems with $k\neq 0$ models, a further issue on
which our theory sheds some light is discussions of the total matter
and energy content of the universe.  In Section~\ref{COLLDST} we
discussed the distinction between the rest-mass energy and the total
gravitating energy inside a sphere of radius $r$.  Since cosmological
models are a special case of the general theory outlined in
Section~\ref{SS}, this same distinction should be significant in
cosmology.

In Section~\ref{COLLDST} we identified the total gravitating energy of
a sphere of radius $r$ with the function $M(r,t)$.  For all
cosmological models, this is given by~\eqref{cos-Mvsrho}
\begin{equation}
M(r,t) = \frac{4}{3} \pi r^3 \rho(t).
\label{ifc3}
\end{equation}
In strictly homogeneous models the Weyl tensor vanishes, and we lose
an intrin\-sically-defined distance scale.  But, if we consider
cosmological models as the limiting case of spherically-symmetric
systems, then there seems little doubt that~\eqref{ifc3} is still the
correct expression for the gravitating energy within a sphere of
radius $r$ surrounding the origin.  Moreover, attempts to discover the
gravitating content of a region rely on perturbations away from ideal
uniformity.  In these cases an intrinsic distance scale is well
defined, since a Weyl tensor is again present.  (Determining the
gravitating content of a region is important in, for example,
determinations of $\Om$ --- the ratio of the actual density of the
universe to the critical density.)  On the other hand, the total rest
mass energy within a sphere of radius $r$ centred on the origin must
still be given by~\eqref{cos-restm}
\begin{equation}
\mu(r,t) = \int_0^r 4 \pi s^2 \rho(t) \, \frac{ds}{g_1}.
\label{ifc5}
\end{equation}
This remains a covariant scalar quantity, and is just $\rho(t)$
multiplied by the covariant volume integral (this is the volume one
would measure locally using light paths or fixed rods).

We have now defined two covariant scalar quantities, $M(r,t)$ and
$\mu(r,t)$, both of which are conserved along fluid streamlines in the
absence of pressure.  If the identifications made in the
spherically-symmetric case remain valid in the homogeneous case, then
the difference between these should give the additional contribution
to the total energy beyond the rest-mass energy.  For spatially-flat
universes we have $g_1=1$ so there is no difference.  (In terms of the
Newtonian-gauge description of Section~\ref{COLLDST}, the
gravitational potential energy cancels the kinetic energy.)  But, for
$k\neq 0$ models, there is a difference because $\mu(r,t)$ is now
given by
\begin{equation}
\mu(r) = \int_0^r \frac{4 \pi s^2 \rho(s)}{(1-ks^2\rho^{2/3})^{1/2}}
\, ds.
\end{equation}
An interesting place to study the difference between $M$ and $\mu$ is
in a $k>0$ universe at its `turnaround' point, as described in
Section~\ref{SS-COS}.  There one finds that, to lowest order in $r$,
the difference is given by
\begin{equation}
M(r) - \mu(r) \approx -\frac{3 M(r)^2}{5r},
\end{equation}
which is precisely the Newtonian formula for the self-potential of a
uniform sphere of mass $M(r)$.  This is what we would have expected
since, at the turnaround point, the kinetic energy vanishes.  

The above should only be viewed as suggestive, but one idea that it
appears to rule out is the popular suggestion that the total energy
density of the universe should be zero~\cite{try73,try84}.  If the
above analysis is correct then there is no possibility of the total
energy density $M(r,t)$ ever being zero.  Furthermore, for
spatially-flat models --- which we consider the most likely --- the
total energy density resides entirely in the rest-mass energies of the
particles in the universe and cannot be cancelled by a negative
gravitational contribution.


\section{Conclusions}

In this paper we developed a theory of gravity consisting of gauge
fields defined in a flat background spacetime.  The theory is
conceptually simple, and the role of the gauge fields is clearly
understood --- they ensure invariance under arbitrary displacements
and rotations.  While it is possible to maintain a classical picture
of the rotation gauge group, a full understanding of its role is only
achieved once the Dirac action is considered.  The result is a theory
which offers a radically different interpretation of gravitational
interactions from that provided by general relativity.  Despite this,
the two theories agree in their predictions for a wide range of
phenomena.  Differences only begin to emerge over issues such as the
role of topology, our insistence on the use of global solutions, and
in the interaction with quantum theory.  Furthermore, the separation
of the gauge fields into one for displacements and one for local
rotations is suggestive of physical effects being separated into an
inertia field and a force field.  Indeed, there is good reason to
believe that mass should enter relativistic multiparticle wave
equations in the manner of the $\lih$-field~\cite{DGL95-elphys}.  It
is possible that, in the development of a multiparticle theory, the
$\hu$ and $\Om$ fields will be extended in quite distinct ways.
Such possibilities do not appear to be open to a theory such as
general relativity, with its reliance on the metric as the `foundation
of all'~\cite{mis-grav}.  Probably the closest approach to the theory
developed here is the spin-2 field theory discussed by many authors
(see Box 17.2 of~\cite{mis-grav} and~\cite{fey-grav}).  This theory is
usually viewed as reproducing general relativity exactly, albeit in a
somewhat ugly form due to the existence of a background spacetime and
the reliance on infinite series of the field variable.  By contrast,
we hope to have demonstrated that GTG has an internal attractiveness
of its own, as well as simplicity due to its first-order nature.

A crucial question to address is whether any experimental tests are
likely to distinguish between general relativity and GTG in the
immediate future.  The biggest differences between general relativity
and GTG to emerge to date lie in the treatment of black hole
singularities~\cite{DL04-KS,DGL96-erice}, but these are
unlikely to be testable for some considerable time!  A more promising
area is the link between gravity and quantum spin.  GTG makes a clear
prediction for the type and magnitude of this interaction, whereas it
is not uniquely picked out in general spin-torsion theories, or in
more general Poincar\'{e} gauge theory.  Any experiment measuring this
interaction would therefore provide a clear test of GTG.  A partial
exploration of the effects of spin interactions in GTG is contained
in~\cite{DGL98-spintor}.

The techniques developed here reveal some remarkable properties of
spher\-ically-symmetric systems.  It has been known since the
1930's~\cite{mil34} that, in the absence of pressure, the dynamical
equations of cosmology can be cast in a Newtonian form.  We have now
shown that a single, unified, Newtonian treatment can be given for all
spherically-symmetric pressureless fluids, whether homogeneous or not.
Furthermore, effects that have hitherto been viewed as the result of
spacetime curvature can now be understood in a simple alternative
fashion.  The result is a physical picture in which the background
spacetime has no effect on either dynamics or kinematics.  This, we
believe, is both new and potentially very useful.  For example,
simulations of black hole formation, and studies of the behaviour of
the universe as a whole, can be carried out in exactly the same
framework.  All previous studies have relied on cutting and pasting
various metrics together, with the result that no clear, global view
of the underlying physics can be achieved.  These advantages are
exploited in~\cite{DDGHL-I} to model the growth of a
spherically-symmetric perturbation in a homogeneous background
cosmology, and to study the effect of the perturbation on the cosmic
microwave background.

The intrinsic method described here, and used to study
spherically-symmet\-ric systems, is quite general and can be applied
to a wide range of problems.  In~\cite{DGL96-cylin} the method is
applied to a restricted class of cylindrically-symmetric systems, and
elsewhere we have presented treatments of axisymmetric
systems~\cite{DLkerr03,gap}.  In all cases studied to date, the
intrinsic method has brought considerable clarity to what would
otherwise be a largely mysterious mess of algebra.  This is achieved
by removing the dependence on an arbitrary coordinate system, and
instead working directly with physical quantities.  The same technique
also looks well suited to the study of cosmological perturbation
theory, about which there has been considerable recent
debate~\cite{ellis-erice94}.

The interaction between Dirac theory and the gauge theory developed
here revealed a number of surprises.  The first was that consistency
of the action principle with the minimal-coupling procedure restricted
us to a theory that is unique up to the possible inclusion of a
cosmological constant.  The second was that spatially flat cosmologies
are the only ones that are consistent with homogeneity at the level of
the single-particle Dirac equation.  The final surprise was provided
by a study of the Dirac equation in a black hole background, which
revealed a remarkable link with the Hawking temperature and quantum
field theory.  Much work remains to settle the issues raised by this
final point, however.

As a final remark, we also hope to have demonstrated the power of
geometric algebra in analysing many physical problems.  Many of the
derivations performed in this paper would have been far more
cumbersome in any other language, and none are capable of the compact
expressions provided by geometric algebra for, say, the Riemann
tensor.  In addition, use of geometric algebra enabled us to remove
all reference to coordinate frames from the fundamental equations.
This is a real aid to providing a clear physical understanding
of the mathematics involved.  We would encourage anyone interested in
studying the consequences of our theory to take time to master the
techniques of geometric algebra.

\section*{Acknowledgements}

We thank Anton Garrett for his careful reading of this manuscript and
his many suggestions for improvements.  We also thank David Hestenes
for many enjoyable and thought-provoking discussions, and George Ellis
for a number of helpful suggestions.

\appendix

\section{The Dirac operator algebra} 
\label{app-dirac}

In the Dirac--Pauli representation the $\gamma$-matrices are defined as
\begin{equation}
\hat{\gamma}_0 = \left( \begin{array}{cr}
        I & 0 \\
        0 & -I
      \end{array} \right) , \quad
\hat{\gamma}_k = \left( \begin{array}{cr}
        0 & -\hat{\sig}_k \\
        \hat{\sig}_k & 0
      \end{array} \right) ,
\end{equation}
where the $\hat{\sig}_k$ are the standard Pauli
matrices~\cite{bjo-rel1,itz-quant}.  The Dirac $\gam$-matrices act on
spinors, which are four-component complex column vectors.  A spinor
$|\psi \ra$ is placed in one-to-one correspondence with an 8-component
even element of the spacetime algebra
via~\cite{DGL93-states,gul-steps}
\begin{equation}
|\psi\ra = \left( \begin{array}{r} 
 a^0 + ia^3  \\
-a^2 + ia^1  \\
-b^3 + ib^0  \\
-b^1 - ib^2
\end{array} \right) \lra \psi = a^0 + a^k I\sig_k + I(b^0 + b^k I\sig_k).
\end{equation}
The action of the $\{\hgamdm\}$, $i$ and
$\hat{\gam}_5=-i\hgo\hgi\hgj\hgk$ operators maps to
\begin{equation}
\begin{array}{rcl}
\hgamdm |\psi\ra & \lra & \gamdm \psi \gamma_0 \\
i |\psi\ra & \lra & \psi \isk \\
\hat{\gamma}_{5}  |\psi\ra & \lra & \psi  \sk.
\end{array}
\end{equation}
The Dirac equation,
\begin{equation}
\hgamum(i \partial_\mu - e A_\mu) |\psi\ra
 = m |\psi\ra ,
\end{equation}
now takes the spacetime algebra form
\begin{equation}
\gamum(\dmu \psi \isk - eA_\mu \psi) \go = m \psi.
\end{equation}
Recombining to form the vectors $\grad=\gamum\dmu$ and $A=\gamum
A_\mu$, and postmultiplying by $\go$, we arrive at the Dirac equation
in the form
\begin{equation}
\grad \psi \isk - eA \psi = m \psi \go.
\end{equation}
Under Lorentz transformations the spinor $\psi$ transforms
single-sidedly to $R\psi$, hence the presence of the fixed $\go$ and
$\gk$ vectors on the right-hand side of $\psi$ does not break Lorentz
invariance.

The role of the Dirac adjoint is played by the geometric operation of
reversion, and the quantum inner product projects out the $\{1,\isk\}$
components from a general multivector.  So, for example, the real part
of the inner product $\la \psibar_1 |\psi_2\ra$ is given in the
spacetime algebra by $\la\psirev_1\psi_2\ra$ and the imaginary part by
$-\la\psirev_1\psi_2\isk\ra$.  The Dirac current $J^\mu=\la\psibar
|\hgamum |\psi\ra$ is now replaced by the set of components
\begin{equation}
\la \psirev \gamum \psi \go \ra = \gamum \dt (\psi \go \psirev).
\end{equation}
These are simply the components of the vector $\psi\go\psirev$,
decomposed in the $\{\gamum\}$ frame.  Reference to the frame is
removed from the vector by defining the current as
\begin{equation}
\clj = \psi \go \psirev.
\end{equation}
Similarly, the role of the spin current is played by the vector
\begin{equation}
s = \psi \gk \psirev,
\end{equation}
and the spin trivector is simply $Is$.  The Dirac Lagrangian has the
equivalent form
\begin{equation}
\la \psibar |(\hgamum(i \dmu -eA_\mu) - m)| \psi\ra
\lra \la \grad \psi I\gk \psirev - eA\psi \go \psirev 
- m \psi \psirev \ra ,
\end{equation}
which is the form used in the main text.  A more detailed discussion
of the spacetime algebra formulation of Dirac theory is contained
in~\cite{DGL95-elphys}.

\section{Some results in multivector  calculus}
\label{app-results}

We begin with a set of results for the derivative with respect to the
vector $a$ in an $n$-dimensional space~\cite{hes-gc}:
\begin{equation}
\begin{alignedat}{2}
\da a \dt b &= b &\qquad \da  a^2 &= 2a  \\
\da \dt a  &= n  &\qquad \da a \dt A_r  &= rA_r \\
\da \wdg a  &= 0  &\qquad \da a \wdg A_r  &= (n-r)A_r \\
\da a  &= n  &\qquad \pardot_a A_r \dot{a}  &= (-1)^r (n-2r)A_r. 
\end{alignedat}
\end{equation}
The results needed for the multivector derivative in this paper
are:
\begin{equation}
\begin{aligned} 
\dX \la X A \ra &= P_{X}(A) \\
\dX \la \tilde{X} A \ra &= P_{X}(\tilde{A}), 
\end{aligned}
\end{equation} 
where $ P_{X}(A)$ is the projection of $A$ onto the grades contained
in $X$.  These results are combined using Leibniz' rule; for
example,
\begin{equation}
\dpsi \la \psi \psirev \ra = \pardot_\psi \la \psidot \psirev \ra + 
\pardot_\psi \la \psi \dot{\psirev} \ra = 2 \psirev.
\end{equation}

For the action principle we also require results for the multivector
derivative with respect to the directional derivatives of a field
$\psi$.  The aim is again to refine the calculus so that it becomes
possible to work in a frame-free manner.  (The derivations presented
here supersede those given previously in~\cite{DGL93-lft}.)  We first
introduce a fixed orthonormal frame $\{e^j\}$, with reciprocal
$\{e_k\}$, so that $e^j \dt e_k = \delta_k^j$.  The partial derivative
of $\psi$ with respect to the coordinate $x^j=e^j \dt x$ is
abbreviated to $\psi_{,j}$ so that
\begin{equation}
\psi_{,j} \eqv e_j \dt \grad \psi.
\end{equation}
We can now define the frame-free derivative
\begin{equation}
\partial_{\psi_{,a}} \eqv a \dt e_j \, \partial_{\psi_{,j}}.
\end{equation}
The operator $\partial_{\psi_{,a}}$ is the multivector derivative with
respect to the $a$-derivative of $\psi$.  The fundamental property of
$\partial_{\psi_{,a}}$ is that
\begin{equation}
\partial_{\psi_{,a}} \la b \dt \grad \psi M \ra = a \dt b P_\psi(M).
\end{equation}
Again, more complicated results are built up by applying Leibniz'
rule.  The Euler--Lagrange equations for the Lagrangian density
$\cll=\cll(\psi, a\dt\grad\psi)$ can now be given in the form
\begin{equation}
\dpsi \cll = \da \dt \grad (\partial_{\psi_{,a}} \cll),
\end{equation}
which is the form applied in the main text of this paper.

We also need a formalism for the derivative with respect to a linear
function.  Given the linear function $\hu(a)$ and the fixed frame
$\{e_i\}$, we define the scalar coefficients
\begin{equation}
h_{ij} \eqv e_i \dt \hu(e_j).  
\label{6defhij}
\end{equation}
The individual partial derivatives $\partial_{h_{ij}}$ are
assembled into a frame-free derivative by defining
\begin{equation}
\dhua \eqv a \dt e_j e_i \partial_{h_{ij}}.
\label{appres-dhu}
\end{equation}
The fundamental property of $\dhua$ is that
\begin{align}
\dhua \hu(b) \dt c &=  a \dt e_j e_i \partial_{h_{ij}} (h_{lk} b^k
c^l) \nn \\
&=  a \dt e_j \, e_i c^i b^j \nn \\
&= a \dt b \, c
\end{align}
which, together with Leibniz' rule, is sufficient to derive all the
required properties of the $\dhua$ operator.  For example, if $B$ is a
fixed bivector, 
\begin{align} 
\dhua \la \hu(b \wdg c) B \ra 
&= \dot{\partial}_{\hu(a)} \la \dhu(b) \hu(c) B \ra -
\dot{\partial}_{\hu(a)} \la \dhu (c) \hu(b) B \ra \nn \\ 
&= a \dt b \, \hu(c) \dt B - a \dt c \, \hu(b) \dt B \nn \\
&= \hu(a \dt (b \wdg c)) \dt B .
\end{align}
This result extends immediately to give
\begin{equation}
\dhua \la \hu(A_r) B_r \ra = \la \hu(a \dt A_r) B_r \ra_1.
\end{equation}
In particular, 
\begin{align} 
\dhua \det(\hu)
&= \dhua \la \hu(I) I^{-1} \ra \nn \\
&= \hu(a \dt I) I^{-1} \nn \\
&= \det(\hu) \, \ho^{-1}(a) ,
\label{appres-ddet}
\end{align}
where the definition of the inverse~\eqref{1fninv} has been employed.  The
derivation of~\eqref{appres-ddet} affords a remarkably direct proof of the
formula for the derivative of the determinant of a linear function.

The above results hold equally if $\hu$ is replaced throughout by its
adjoint $\ho$, which is the form of the derivative used throughout the
main text.  Note, however, that
\begin{align}  
\dhua \ho(b) &= \dhua \la \hu(c) b \ra \dc \nn \\
&= a \dt c \, b \dc \nn \\
&=  ba.
\end{align} 
Thus the derivatives of $\hu(b)$ and $\ho(b)$ give different results,
regardless of any symmetry properties of $\hu$.  This has immediate
implications for the symmetry (or lack of symmetry) of the functional
stress-energy tensors for certain fields.

We finally need some results for derivatives with respect to the
bivector-valued linear function $\Om(a)$.  The extensions are
straightforward, and we just give the required results:
\begin{align}
\partial_{\Omega(a)} \la \Omega(b)M \ra &= a \dt b \la M \ra_2 \\
\partial_{\Omega(b)_{,a}} \la c \dt \grad \Omega(d) M \ra 
&=  a \dt c \, b \dt d \, \la M \ra_2.
\end{align}

\section{The translation of tensor calculus}
\label{app-tens}

The reformulation of the gauge theory presented in this paper in terms
of conventional tensor calculus proceeds as follows.  A choice of
gauge is made and a set of scalar coordinates $\{x^\mu\}$ is
introduced.  We define the coordinate frame $\{e_\mu\}$,
\begin{equation}
e_\mu \eqv \deriv{x}{x^\mu},
\end{equation}
with the reciprocal frame denoted $\{e^\mu\}$.  In terms of this the
vector derivative is 
\begin{equation}
 \grad = e^\mu \deriv{}{x^\mu}.
\end{equation}
From the frame vectors we construct the vectors
\begin{equation}
g_\mu \eqv \hu^{-1}(e_\mu), \qquad
g^\mu \eqv \ho(e^\mu).
\end{equation}
The metric is then given by the $4\times 4$ matrix
\begin{equation}
g_{\mu \nu} \eqv g_\mu \dt g_\nu.
\end{equation}
If the $x$-dependence in $g_{\mu \nu}$ is replaced by dependence
solely on the coordinates $\{x^\mu\}$ then we recover Riemann--Cartan
geometry, where all relations are between coordinates and the concept
of a point as a vector is lost.

The connection is defined by (following the conventions
of~\cite{heh76,nak-geom}) 
\begin{equation}
\cld_\mu g_\nu = \Gam^\alp_{\mu \nu} g_\alp,
\end{equation}
where $\cld_\mu = g_\mu \dt \cld = \partial_\mu + \om(g_\mu) \times$\ .
We can therefore write
\begin{equation}
\Gam^\lam_{\mu \nu} = g^\lam \dt (\cld_\mu g_\nu).
\end{equation}
Since
\begin{equation}
\partial_\mu g_{\nu \lam} = (\cld_\mu g_\nu) \dt g_\lam + g_\nu \dt
(\cld_\mu g_\lam) ,
\end{equation}
we find that
\begin{equation}
\partial_\mu g_{\nu \lam} =  \Gam^\alp_{\mu \nu} g_{\alp \lam} +
\Gam^\alp_{\mu \lam} g_{\alp \nu},
\label{C-2}
\end{equation}
which recovers `metric compatibility' of the connection.  This is
nothing more than the statement that the $\cld_\mu$ operator satisfies
Leibniz' rule.  Equation~\eqref{C-2} can be inverted to show that the
connection contains a component given by the standard Christoffel
symbol.  The connection can then be written
\begin{equation}
\Gam^\nu_{\lam \mu} = \left\{ { }^\nu_{\lam \mu} \right\} - {K_{\lam
\mu}}^\nu \, , 
\end{equation}
where ${K_{\lam \mu}}^\nu$ is the contorsion tensor and is given by
\begin{equation}
{K_{\lam \mu}}^\nu = - {S_{\lam\mu}}^\nu +{{S_\lam}^\nu}_\mu -
{S^\nu}_{\lam\mu}.
\end{equation}
Here ${S_{\lam\mu}}^\nu$ is the torsion tensor, equal to the
antisymmetric part of the connection:
\begin{align}
{S_{\lam\mu}}^\nu 
&= \half (\Gam^\nu_{\lam \mu} - \Gam^\nu_{\mu \lam}) \nn \\
&= \half g^\nu \dt (\cld_\lam g_\mu - \cld_\mu g_\lam) \nn \\
&= - \half \bigl(g_\mu \dt (\cld_\lam g^\nu) - g_\lam \dt (\cld_\mu
g^\nu) \bigr)\nn \\
&= - \half (g_\mu \wdg g_\lam) \dt (\cld \wdg g^\nu) \nn \\
&= \half (g_\lam \wdg g_\mu) \dt \cls(g^\nu),
\end{align}
where $\cls(a)$ is the torsion bivector.  The contorsion
is formed from $\cls(a)$ by
\begin{align}
K_{\lam \mu \nu} 
&= -\half (g_\lam \wdg g_\mu)\dt\cls(g_\nu) +\half(g_\mu \wdg g_\nu)   
\dt\cls(g_\lam)  - \half (g_\nu \wdg g_\lam) \dt \cls(g_\mu) \nn \\
&= \half (g_\mu \wdg g_\nu)\dt\cls(g_\lam) - \half g_\lam \dt
(g_\mu \dt \cls(g_\nu) - g_\nu \dt \cls(g_\mu)) \nn \\
&= (g_\mu \wdg g_\nu) \dt \bigl( \cls(g_\lam) - \half g_\lam \dt (\da \wdg
\cls(a)) \bigr). 
\end{align}
For our gauge theory of gravity, the torsion is exclusively of the
type $\cls(a)=a\cdot\cls$, where $\cls$ is the spin trivector, in which
case
\begin{equation}
K_{\lam \mu \nu} = -S_{\lam \mu \nu} =- \half (g_\lam \wdg g_\mu \wdg
g_\nu) \dt \cls.
\end{equation}

If we now consider the covariant derivative of a covariant vector
$\cla=A^\alp g_\alp= A_\alp g^\alp$, we find that
\begin{align}
\cld_\mu \cla 
&= \cld_\mu (A^\alp g_\alp) \nn \\
&= (\partial_\mu A^\alp) g_\alp + A^\alp \Gam^\bet_{\mu \alp} g_\bet
\nn \\
&= (\partial_\mu A^\alp + \Gam^\alp_{\mu \bet} A^\bet) g_\alp,
\end{align}
so that the components of the vector $\cld_\mu \cla$ are those
expected for tensor calculus.  Obviously the fact that $A^\alp
g_\alp= A_\alp g^\alp$ implies that $A_\mu=A^\alp g_{\alp \mu}$, so
indices are raised and lowered in the expected manner.

For covariant quantities such as the Riemann tensor the translation
to tensor calculus is straightforward:
\begin{equation}
{R^\mu}_{\nu \rho \sigma} = (g^\mu \wdg g_\nu) \dt \clr(g_\sigma \wdg
g_\rho).
\end{equation}
The general scheme is that any covariant quantity in GTG can be
decomposed into tensor components by applying either the $\{g_\mu\}$
or $\{g^\mu\}$, or a mixture of both, to yield a tensor with the
appropriate number of upstairs and downstairs indices.  So, for
example, $\clf$ can be decomposed to $F_{\mu \nu}=\clf\cdot (g_\mu
\wdg g_\nu)$, ${F_\mu}^\nu=\clf\cdot (g_\mu \wdg g^\nu)$ or $F^{\mu
\nu}=\clf\cdot (g^\mu \wdg g^\nu)$.  Tensor calculus is poor at
revealing which, if any, of the components represent a physical
observable.  Such issues are much clearer in GTG, which focuses
attention on the single entity~$\clf$.

A vierbein ${e_\mu}^i$ (essentially an orthonormal tetrad) is given by
\begin{align}
{e_\mu}^i &= g_\mu \dt \gam^i \\
{e^\mu}_i &= g^\mu \dt \gam_i
\end{align}
where $\{\gam^i\}$ is a fixed orthonormal frame.  Any position
dependence in the $\{\gam^i\}$ is eliminated with a suitable rotor
transformation.  When matrix operators $\{\hgam^i\}$ are required
these are replaced by the $\{\gam^i\}$ frame vectors using the method
described in Appendix~\ref{app-dirac}.  In this way frame-free
vectors can be assembled.  For example, the Dirac
operator~\cite[Chapter 11]{goc-diff}
\begin{equation}
\not\!\!D |\psi \ra \eqv {e^\mu}_i\, \hgam^i \left(\deriv{}{x^\mu} +
\qrt \om_{jk\mu} \hgam^j\hgam^k \right) |\psi \ra
\end{equation} 
has the spacetime algebra equivalent
\begin{align}
g^\mu \dt \gam_i \, \gam^i \left( \deriv{}{x^\mu} + \qrt
\om(g_\mu)\dt(\gam_k \wdg \gam_j)  \gam^j \gam^k  \right) \psi \go 
&= \ho(\grad) \psi \go + \half g^\mu \om(g_\mu)  \psi \go \nn \\
&= D \psi \go.
\end{align}

The above relations enable many results from Riemann--Cartan geometry
to be carried over into our formalism, though the theory of gravity
presented here is restricted in the structures from Riemann--Cartan
geometry that it admits (the torsion is of trivector type, for
example).

A similar translation scheme is easily set up for the language of
differential forms, which is much closer to the spirit of geometric
algebra than tensor calculus.  Differential forms are scalar-valued
functions of an antisymmetrised set of vectors.  They can easily be
mapped to an equivalent multivector, and a full translation into
geometric algebra is quite straightforward~\cite{hes-gc}.  Here we
note in passing the geometric algebra equivalent of the Hodge dual of
a differential form, which is
\begin{equation}
{ }^* \alp_r \mapsto - \dhi \hu \ho(\tilde{A}_r) i
\end{equation}
where $A_r$ is the multivector equivalent of $\alp_r$.


\begin{thebibliography}{100}

\bibitem{uti56}
R.~Utiyama.
\newblock Invariant theoretical interpretation of interaction.
\newblock {\em Phys. Rev.}, 101(5):1597, 1956.

\bibitem{kib61}
T.W.B. Kibble.
\newblock Lorentz invariance and the gravitational field.
\newblock {\em J. Math. Phys.}, 2(3):212, 1961.

\bibitem{iva83}
D.~Ivanenko and G.~Sardanashvily.
\newblock The gauge treatment of gravity.
\newblock {\em Phys. Rep.}, 94(1):1, 1983.

\bibitem{car22}
E.~Cartan.
\newblock Sur un g{\'{e}}n{\'{e}}ralisation de la notion de courbure de
  {R}iemann et les espaces {\`{a}} torsion.
\newblock {\em C. R. Acad. Sci. (Paris)}, 174:593, 1922.

\bibitem{wey50}
H.~Weyl.
\newblock A remark on the coupling of gravitation and electron.
\newblock {\em Phys. Rev.}, 77(5):699, 1950.

\bibitem{sci64}
D.~Sciama.
\newblock The physical structure of general relativity.
\newblock {\em Rev. Mod. Phys.}, 36:463 and 1103, 1964.

\bibitem{weys47}
J.~Weyssenhoff and A.~Raabe.
\newblock Relativistic dynamics of spin-fluids and spin-particles.
\newblock {\em Acta Phys. Pol.}, 9:7, 1947.

\bibitem{bea63}
O.~{Costa de Beauregard}.
\newblock Translational inertial spin effect.
\newblock {\em Phys. Rev.}, 129(1):466, 1963.

\bibitem{heh76}
F.W. Hehl, P.~von~der Heyde, G.D. Kerlick, and J.M. Nester.
\newblock General relativity with spin and torsion: Foundations and prospects.
\newblock {\em Rev. Mod. Phys.}, 48:393, 1976.

\bibitem{egu80}
T.~Eguchi, P.B. Gilkey, and A.J. Hanson.
\newblock Gravitation, gauge theories and differential geometry.
\newblock {\em Phys. Rep.}, 66(6):213, 1980.

\bibitem{hes-gc}
D.~Hestenes and G.~Sobczyk.
\newblock {\em {C}lifford Algebra to Geometric Calculus}.
\newblock Reidel, Dordrecht, 1984.

\bibitem{DGL93-notreal}
S.F. Gull, A.N. Lasenby, and C.J.L. Doran.
\newblock Imaginary numbers are not real --- the geometric algebra of
  spacetime.
\newblock {\em Found. Phys.}, 23(9):1175, 1993.

\bibitem{hes86}
D.~Hestenes.
\newblock Curvature calculations with spacetime algebra.
\newblock {\em Int. J. Theor. Phys.}, 25(6):581, 1986.

\bibitem{hes-sta}
D.~Hestenes.
\newblock {\em Space-Time Algebra}.
\newblock {Gordon and Breach, New York}, 1966.

\bibitem{sob81}
G.~Sobczyk.
\newblock Space-time approach to curvature.
\newblock {\em J. Math. Phys.}, 22(2):333, 1981.

\bibitem{haw-large}
S.W. Hawking and G.F.R. Ellis.
\newblock {\em The Large Scale Structure of Space-Time}.
\newblock Cambridge University Press, 1973.

\bibitem{inv-rel}
R.~d'Inverno.
\newblock {\em Introducing {E}instein's Relativity}.
\newblock Oxford University Press, 1992.

\bibitem{kauf-front}
W.J. Kaufmann.
\newblock {\em The Cosmic Frontiers of General Relativity}.
\newblock Penguin Books, 1979.

\bibitem{haw-black}
S.W. Hawking.
\newblock {\em Black Holes and Baby Universes and Other Essays}.
\newblock Bantam Books, London, 1993.

\bibitem{hes75}
D.~Hestenes.
\newblock Observables, operators, and complex numbers in the {D}irac theory.
\newblock {\em J. Math. Phys.}, 16(3):556, 1975.

\bibitem{DGL93-states}
C.J.L. Doran, A.N. Lasenby, and S.F. Gull.
\newblock States and operators in the spacetime algebra.
\newblock {\em Found. Phys.}, 23(9):1239, 1993.

\bibitem{hes-nf1}
D.~Hestenes.
\newblock {\em New Foundations for Classical Mechanics (Second Edition)}.
\newblock Kluwer Academic Publishers, Dordrecht, 1999.

\bibitem{hes74a}
D.~Hestenes.
\newblock Proper particle mechanics.
\newblock {\em J. Math. Phys.}, 15(10):1768, 1974.

\bibitem{vol93}
T.G. Vold.
\newblock An introduction to geometric algebra with an application to rigid
  body mechanics.
\newblock {\em Am. J. Phys.}, 61(6):491, 1993.

\bibitem{hes74}
D.~Hestenes.
\newblock Proper dynamics of a rigid point particle.
\newblock {\em J. Math. Phys.}, 15(10):1778, 1974.

\bibitem{DGL93-paths}
S.F. Gull, A.N. Lasenby, and C.J.L. Doran.
\newblock Electron paths, tunnelling and diffraction in the spacetime algebra.
\newblock {\em Found. Phys.}, 23(10):1329, 1993.

\bibitem{DGL95-elphys}
C.J.L Doran, A.N. Lasenby, S.F. Gull, S.S. Somaroo, and A.D. Challinor.
\newblock Spacetime algebra and electron physics.
\newblock {\em Adv. Imag. \& Elect. Phys.}, 95:271, 1996.

\bibitem{vol93a}
T.G. Vold.
\newblock An introduction to geometric calculus and its application to
  electrodynamics.
\newblock {\em Am. J. Phys.}, 61(6):505, 1993.

\bibitem{DGL93-gras}
A.N. Lasenby, C.J.L. Doran, and S.F. Gull.
\newblock {G}rassmann calculus, pseudoclassical mechanics and geometric
  algebra.
\newblock {\em J. Math. Phys.}, 34(8):3683, 1993.

\bibitem{dor93-spin}
C.J.L. Doran, D.~Hestenes, F.~Sommen, and N.~van Acker.
\newblock Lie groups as spin groups.
\newblock {\em J.~Math. Phys.}, 34(8):3642, 1993.

\bibitem{DGL93-lft}
A.N. Lasenby, C.J.L. Doran, and S.F. Gull.
\newblock A multivector derivative approach to {L}agrangian field theory.
\newblock {\em Found. Phys.}, 23(10):1295, 1993.

\bibitem{DGL-polmvl}
C.J.L. Doran, A.N. Lasenby, and S.F. Gull.
\newblock Grassmann mechanics, multivector derivatives and geometric algebra.
\newblock In Z.~Oziewicz, B.~Jancewicz, and A.~Borowiec, editors, {\em Spinors,
  Twistors, {C}lifford Algebras and Quantum Deformations}, page 215. Kluwer
  Academic, Dordrecht, 1993.

\bibitem{DGL-polspin}
A.N. Lasenby, C.J.L. Doran, and S.F. Gull.
\newblock 2-spinors, twistors and supersymmetry in the spacetime algebra.
\newblock In Z.~Oziewicz, B.~Jancewicz, and A.~Borowiec, editors, {\em Spinors,
  Twistors, {C}lifford Algebras and Quantum Deformations}, page 233. Kluwer
  Academic, Dordrecht, 1993.

\bibitem{gap}
C.J.L Doran and A.N. Lasenby.
\newblock {\em Geometric Algebra for Physicists}.
\newblock Cambridge University Press, 2003.

\bibitem{hes-unified}
D.~Hestenes.
\newblock A unified language for mathematics and physics.
\newblock In J.S.R. Chisholm and A.K. Common, editors, {\em {C}lifford Algebras
  and their Applications in Mathematical Physics (1985)}, page~1. Reidel,
  Dordrecht, 1986.

\bibitem{hes91}
D.~Hestenes.
\newblock The design of linear algebra and geometry.
\newblock {\em Acta Appl. Math.}, 23:65, 1991.

\bibitem{DL-course}
C.J.L. Doran and A.N. Lasenby.
\newblock \textit{Physical {A}pplications of {G}eometric {A}lgebra}.
\newblock Cambridge Univesity lecture course. Lecture notes available from
  \texttt{http://www.mrao.cam.ac.uk/$\sim$clifford}.

\bibitem{cliffconf1}
{J.S.R. Chisholm and A.K. Common, eds.}
\newblock {\em {C}lifford Algebras and their Applications in Mathematical
  Physics (1985)}.
\newblock Reidel, Dordrecht, 1986.

\bibitem{cliffconf2}
{A. Micali, R. Boudet and J. Helmstetter, eds.}
\newblock {\em {C}lifford Algebras and their Applications in Mathematical
  Physics (1989)}.
\newblock Kluwer Academic, Dordrecht, 1991.

\bibitem{cliffconf3}
{F. Brackx and R. Delanghe and H. Serras, eds.}
\newblock {\em {C}lifford Algebras and their Applications in Mathematical
  Physics (1993)}.
\newblock Kluwer Academic, Dordrecht, 1993.

\bibitem{cli1878}
W.K. Clifford.
\newblock Applications of {G}rassmann's extensive algebra.
\newblock {\em Am. J. Math.}, 1:350, 1878.

\bibitem{gra1877}
H.~{G}rassmann.
\newblock Der ort der {H}amilton'schen quaternionen in der
  {A}us\-dehnung\-slehre.
\newblock {\em Math. Ann.}, 12:375, 1877.

\bibitem{DGL-grav-bel}
C.J.L. Doran, A.N. Lasenby, and S.F. Gull.
\newblock Gravity as a gauge theory in the spacetime algebra.
\newblock In F.~Brackx, R.~Delanghe, and H.~Serras, editors, {\em Clifford
  Algebras and their Applications in Mathematical Physics (1993)}, page 375.
  Kluwer Academic, Dordrecht, 1993.

\bibitem{nak-geom}
M.~Nakahara.
\newblock {\em Geometry, Topology and Physics}.
\newblock Adam Hilger, Bristol, 1990.

\bibitem{goc-diff}
M.~Gockeler and T.~Schucker.
\newblock {\em Differential Geometry, Gauge Theories, and Gravity}.
\newblock Cambridge University Press, 1987.

\bibitem{rau82}
R.T. Rauch.
\newblock Equivalence of an {$R+R^2$} theory of gravity to
  {E}instein--{C}artan--{S}ciama--{K}ibble theory in the presence of matter.
\newblock {\em Phys. Rev. D}, 26(4):931, 1982.

\bibitem{hec91}
R.D. Hecht, J.~Lemke, and R.P. Wallner.
\newblock Can {P}oincar\'{e} gauge theory be saved?
\newblock {\em Phys. Rev. D}, 44(8):2442, 1991.

\bibitem{kho92}
A.V. Khodunov and V.V. Zhytnikov.
\newblock Gravitational equations in space-time with torsion.
\newblock {\em J. Math. Phys.}, 33(10):3509, 1992.

\bibitem{DGL98-spintor}
C.J.L Doran, A.N. Lasenby, A.D. Challinor, and S.F Gull.
\newblock Effects of spin-torsion in gauge theory gravity.
\newblock {\em J. Math. Phys.}, 39(6):3303--3321, 1998.

\bibitem{fey-grav}
R.P. Feynman, F.B. Morinigo, and W.G. Wagner.
\newblock {\em Feynman Lectures on Gravitation}.
\newblock Addison--Wesley, Reading MA, 1995.

\bibitem{pen-I}
R.~Penrose and W.~Rindler.
\newblock {\em Spinors and space-time, Volume I: two-spinor calculus and
  relativistic fields}.
\newblock Cambridge University Press, 1984.

\bibitem{kra-exact}
D.~Kramer, H.~Stephani, M.~Mac{C}allum, and E.~Herlt.
\newblock {\em Exact Solutions of Einstein's Field Equations}.
\newblock Cambridge University Press, 1980.

\bibitem{dor-thesis}
C.J.L. Doran.
\newblock {\em Geometric Algebra and its Application to Mathematical Physics}.
\newblock PhD thesis, Cambridge University, 1994.

\bibitem{new65}
E.T. Newman and A.I. Janis.
\newblock Note on the {K}err spinning-particle metric.
\newblock {\em J.~Math. Phys.}, 6(4):915, 1965.

\bibitem{DL04-KS}
C.J.L. Doran and A.N. Lasenby.
\newblock Integral equations, {K}err--{S}child fields and
gravitational sources
\newblock gr-qc/0404081, 2004.

\bibitem{wah65}
F.B. Estabrook and H.D. Wahlquist.
\newblock Dyadic analysis of spacetime congruences.
\newblock {\em J. Math. Phys.}, 5:1629, 1965.

\bibitem{wah92}
H.D. Wahlquist.
\newblock The problem of exact interior solutions for rotating rigid bodies in
  general relativity.
\newblock {\em J. Math. Phys.}, 33(1):304, 1992.

\bibitem{DGL96-erice}
A.N. Lasenby, C.J.L. Doran, Y.~Dabrowski, and A.D. Challinor.
\newblock Rotating astrophysical systems and a gauge theory approach to
  gravity.
\newblock In N.~S{\'{a}}nchez and A.~Zichichi, editors, {\em Current Topics in
  Astrofundamental Physics, Erice 1996}, page 380. World Scientific, Singapore,
  1997.

\bibitem{DGL-erice}
A.N. Lasenby, C.J.L. Doran, and S.F. Gull.
\newblock Astrophysical and cosmological consequences of a gauge theory of
  gravity.
\newblock In N.~S{\'{a}}nchez and A.~Zichichi, editors, {\em Advances in
  Astrofundamental Physics, Erice 1994}, page 359. World Scientific, Singapore,
  1995.

\bibitem{DLkerr03}
C.J.L. Doran and A.N. Lasenby.
\newblock New techniques for analysing axisymmetric gravitational systems {I}.
  {V}acuum fields.
\newblock {\em Class. Quant. Gravity,}, 20(6):1077--1101, 2003.

\bibitem{DGL96-cylin}
C.J.L. Doran, A.N. Lasenby, and S.F. Gull.
\newblock Physics of rotating cylindrical strings.
\newblock {\em Phys. Rev. D}, 54(10):6021, 1996.

\bibitem{gaut84}
R.~Gautreau.
\newblock Curvature coordinates in cosmology.
\newblock {\em Phys. Rev. D}, 29(2):186, 1984.

\bibitem{gaut95}
R.~Gautreau and J.M. Cohen.
\newblock Gravitational collapse in a single coordinate system.
\newblock {\em Am. J. Phys.}, 63(11):991, 1995.

\bibitem{mart01}
K.~Martel and E.~Poisson.
\newblock Regular coordinate systems for {S}chwarzschild and other spherical
  spacetimes.
\newblock {\em Am. J. Phys.}, 69(4):476, 2001.

\bibitem{cho93}
M.W. Choptuik.
\newblock Universality and scaling in gravitational collapse of a massless
  scalar field.
\newblock {\em Phys. Rev. Lett.}, 70(1):9, 1993.

\bibitem{abr93}
A.M. Abrahams and C.R. Evans.
\newblock Critical behaviour and scaling in vacuum axisymmetric gravitational
  collapse.
\newblock {\em Phys. Rev. Lett.}, 70(20):2980, 1993.

\bibitem{gaut95a}
R.~Gautreau.
\newblock Light cones inside the {S}chwarzschild radius.
\newblock {\em Am. J. Phys.}, 63(5):431, 1995.

\bibitem{steph-grav}
H.~Stephani.
\newblock {\em General Relativity}.
\newblock Cambridge University Press, 1982.

\bibitem{lemait33}
G.~Lema{\^{i}}tre.
\newblock Spherical condensations in the expanding universe.
\newblock {\em Acad. Sci., Paris, Comptes Rend.}, 196:903, 1933.

\bibitem{kru60}
M.D. Kruskal.
\newblock Maximal extension of the {S}chwarzschild metric.
\newblock {\em Phys. Rev.}, 119:1743, 1960.

\bibitem{opp39}
J.R. Oppenheimer and H.~Snyder.
\newblock On continued gravitational contraction.
\newblock {\em Phys. Rev.}, 56:455, 1939.

\bibitem{mis64}
C.W. Misner and D.H. Sharp.
\newblock Relativistic equations for adiabatic, spherically symmetric
  gravitational collapse.
\newblock {\em Phys. Rev.}, 136(2B):571, 1964.

\bibitem{mis-grav}
C.W. Misner, K.S. Thorne, and J.A. Wheeler.
\newblock {\em Gravitation}.
\newblock W.H. Freeman and Company, San Francisco, 1973.

\bibitem{pan92}
M.~Panek.
\newblock Cosmic background radiation anisotropies from cosmic structures:
  Models based on the {T}olman solution.
\newblock {\em Ap.J.}, 388:225, 1992.

\bibitem{tol34}
R.C. Tolman.
\newblock Effect of inhomogeneity on cosmological models.
\newblock {\em Proc. Nat. Acad. Sci. U.S.}, 20:169, 1934.

\bibitem{bon47}
H.~Bondi.
\newblock Spherically symmetrical models in general relativity.
\newblock {\em Mon. Not. R. Astron. Soc.}, 107:410, 1947.

\bibitem{DDGHL-I}
A.N. Lasenby, C.J.L Doran, M.P. Hobson, Y.~Dabrowski, and A.D. Challinor.
\newblock Microwave background anisotropies and nonlinear structures {I}.
  {I}mproved theoretical models.
\newblock {\em Mon. Not. R. Astron. Soc.}, 302:748, 1999.

\bibitem{bek74}
J.~D. Bekenstein.
\newblock Generalized second law of thermodynamics in black-hole physics.
\newblock {\em Phys. Rev. D}, 9(12):3292, 1974.

\bibitem{nov-bh}
I.D. Novikov and V.P. Frolov.
\newblock {\em Physics of Black Holes}.
\newblock Kluwer Academic, Dordrecht, 1989.

\bibitem{nar-cosm}
J.V. Narlikar.
\newblock {\em Introduction to Cosmology}.
\newblock Cambridge University Press, 1993.

\bibitem{ellis93}
G.F.R. Ellis and T.~Rothman.
\newblock Lost horizons.
\newblock {\em Am. J. Phys.}, 61(10):883, 1993.

\bibitem{cop28}
E.T. Copson.
\newblock On electrostatics in a gravitational field.
\newblock {\em Proc. R. Soc. A}, 118:184, 1928.

\bibitem{lin76}
B.~Linet.
\newblock Electrostatics and magnetostatics in the {S}chwarzschild metric.
\newblock {\em J. Phys. A}, 9(7):1081, 1976.

\bibitem{han73}
R.~S. Hanni and R.~Ruffini.
\newblock Lines of force of a point charge near a {S}chwarzschild black hole.
\newblock {\em Phys. Rev. D.}, 8(10):3259, 1973.

\bibitem{tho-mem}
K.S. Thorne, R.H. Price, and D.A. Macdonald.
\newblock {\em Black Holes: The Membrane Paradigm}.
\newblock Yale University Press, 1986.

\bibitem{smi80}
A.G. Smith and C.M. Will.
\newblock Force on a static charge outside a {S}chwarzschild black hole.
\newblock {\em Phys. Rev. D}, 22(6):1276, 1980.

\bibitem{D00-kerr}
C.J.L. Doran.
\newblock New form of the {K}err solution.
\newblock {\em Phys. Rev. D}, 61(6):067503, 2000.

\bibitem{bstates}
A.N.~Lasenby \textit{et al.}
\newblock Bound states and decay times of fermions in a {S}chwarzschild black
  hole background.
\newblock gr-qc/0209090, 2002.

\bibitem{bjo-rel1}
J.D. Bjorken and S.D. Drell.
\newblock {\em Relativistic Quantum Mechanics, vol 1}.
\newblock McGraw-Hill, New York, 1964.

\bibitem{dam76}
T.~Damour and R.~Ruffini.
\newblock Black-hole evaporation in the {K}lein-{S}auter-{H}eisenberg-{E}uler
  formalism.
\newblock {\em Phys. Rev. D}, 14(2):332, 1976.

\bibitem{liu81}
{Zhao Zheng}, {Guei Yuan-xing}, and {Liu Liao}.
\newblock {H}awking evaporation of {D}irac particles.
\newblock {\em Chin. Phys.}, 1(4):934, 1981.

\bibitem{zha82}
{Zhao Zheng}, {Guei Yuan-xing}, and {Liu Liao}.
\newblock On the {H}awking evaporation of {D}irac particles in {K}err--{N}ewman
  space-time.
\newblock {\em Chin. Phys.}, 2(2):386, 1982.

\bibitem{mart77}
M.~Martellini and A.~Treves.
\newblock Comment on the {D}amour-{R}uffini treatment of black-hole
  evaporation.
\newblock {\em Phys. Rev. D}, 15(8):2415, 1977.

\bibitem{haw74}
S.~W. Hawking.
\newblock Black hole explosion?
\newblock {\em Nature}, 248:30, 1974.

\bibitem{DGL-cosm-bel}
A.N. Lasenby, C.J.L. Doran, and S.F. Gull.
\newblock Cosmological consequences of a flat-space theory of gravity.
\newblock In F.~Brackx, R.~Delanghe, and H.~Serras, editors, {\em Clifford
  Algebras and their Applications in Mathematical Physics (1993)}, page 387.
  Kluwer Academic, Dordrecht, 1993.

\bibitem{DGL97-selfcons}
A.D. Challinor, A.N. Lasenby, C.J.L Doran, and S.F Gull.
\newblock Massive, non-ghost solutions for the {D}irac field coupled
  self-consistently to gravity.
\newblock {\em General Rel. Grav.}, 29:1527, 1997.

\bibitem{ish74}
C.J. Isham and J.E. Nelson.
\newblock Quantization of a coupled {F}ermi field and {R}obertson-{W}alker
  metric.
\newblock {\em Phys. Rev. D}, 10(10):3226, 1974.

\bibitem{try73}
E.P. Tryon.
\newblock Is the {U}niverse a vacuum fluctuation?
\newblock {\em Nature}, 246:396, 1973.

\bibitem{try84}
E.P. Tryon.
\newblock What made the world?
\newblock {\em New Sci.}, 8 March:14, 1984.

\bibitem{mil34}
E.A. Milne and W.H. Mc{C}rea.
\newblock Newtonian {U}niverses and the curvature of space.
\newblock {\em Q. J. Maths.}, 5:73, 1934.

\bibitem{ellis-erice94}
G.F.R. Ellis.
\newblock The covariant and gauge invariant approach to perturbations in
  cosmology.
\newblock In N.~S{\'{a}}nchez and A.~Zichichi, editors, {\em Current Topics in
  Astrofundamental Physics: The Early Universe}, page~1. Kluwer Academic,
  Dordrecht, 1995.

\bibitem{itz-quant}
C.~Itzykson and J-B. Zuber.
\newblock {\em Quantum Field Theory}.
\newblock McGraw--Hill, New York, 1980.

\bibitem{gul-steps}
S.F. Gull.
\newblock Charged particles at potential steps.
\newblock In A.~Weingartshofer and D.~Hestenes, editors, {\em The Electron},
  page~37. Kluwer Academic, Dordrecht, 1991.

\end{thebibliography}
\end{document}